\documentclass[preprint, superscriptaddress, preprintnumbers,amsmath,amssymb]{revtex4}

\usepackage{booktabs}
\usepackage{color,graphicx}
\usepackage{amsmath}
\usepackage{dcolumn}
\usepackage{CJK}                 
\usepackage{bm}                  
\usepackage{soul}
\usepackage{enumerate}
\usepackage{multirow}
\usepackage{mathrsfs}
\usepackage{braket}
\usepackage{longtable}
\usepackage{pdflscape}

\usepackage[dvipdfm,
            pdfstartview=FitH,
            CJKbookmarks=true,
            bookmarksnumbered=true,
            bookmarksopen=true,
            colorlinks,
            pdfborder=001,
            linkcolor=blue,
            anchorcolor=blue,
            citecolor=blue
            ]{hyperref}

\usepackage{float}

\begin{document}

\title{Nuclear mass table in deformed relativistic Hartree-Bogoliubov theory in continuum: I. even-even nuclei}

\author{Kaiyuan Zhang}
\affiliation{State Key Laboratory of Nuclear Physics and Technology, School of Physics, Peking University, Beijing 100871, China}

\author{Myung-Ki Cheoun}
\affiliation{Department of Physics and Origin of Matter and Evolution of Galaxy (OMEG) Institute, Soongsil University, Seoul 156-743, Korea}

\author{Yong-Beom Choi}
\affiliation{Department of Physics, Pusan National University, Busan 46241, Korea}

\author{Pooi Seong Chong}
\affiliation{Department of Physics, The University of Hong Kong, Pokfulam 999077, Hong Kong, China}

\author{Jianmin Dong}
\affiliation{Institute of Modern Physics, Chinese Academy of Sciences, Lanzhou 730000, China}
\affiliation{School of Physics, University of Chinese Academy of Sciences, Beijing 100049, China}

\author{Zihao Dong}
\affiliation{State Key Laboratory of Nuclear Physics and Technology, School of Physics, Peking University, Beijing 100871, China}

\author{Xiaokai Du}
\affiliation{State Key Laboratory of Nuclear Physics and Technology, School of Physics, Peking University, Beijing 100871, China}

\author{Lisheng Geng}
\affiliation{School of Physics, Beihang University, Beijing 102206, China}
\affiliation{School of Physics and Microelectronics, Zhengzhou University, Zhengzhou 450001, China}

\author{Eunja Ha}
\affiliation{Department of General Education for Human Creativity, Hoseo University, Asan, Chungnam 336-851, Korea}

\author{Xiao-Tao He}
\affiliation{College of Materials Science and Technology, Nanjing University of Aeronautics and Astronautics, Nanjing 210016, China}

\author{Chan Heo}
\affiliation{Department of Physics, The University of Hong Kong, Pokfulam 999077, Hong Kong, China}

\author{Meng Chit Ho}
\affiliation{Department of Physics, The University of Hong Kong, Pokfulam 999077, Hong Kong, China}

\author{Eun Jin In}
\affiliation{Center for Exotic Nuclei Studies, Institute for Basic Science, Daejeon 34126, Korea}
\affiliation{Department of Energy Science, Sungkyunkwan University, Suwon 16419, Korea }

\author{Seonghyun Kim}
\affiliation{Department of Physics and Origin of Matter and Evolution of Galaxy (OMEG) Institute, Soongsil University, Seoul 156-743, Korea}

\author{Youngman Kim}
\affiliation{Rare Isotope Science Project, Institute for Basic Science, Daejeon 34000, Korea}

\author{Chang-Hwan Lee}
\affiliation{Department of Physics, Pusan National University, Busan 46241, Korea}

\author{Jenny Lee}
\affiliation{Department of Physics, The University of Hong Kong, Pokfulam 999077, Hong Kong, China}

\author{Hexuan Li}
\affiliation{State Key Laboratory of Nuclear Physics and Technology, School of Physics, Peking University, Beijing 100871, China}

\author{Zhipan Li}
\affiliation{School of Physical Science and Technology, Southwest University, Chongqing 400715, China}

\author{Tianpeng Luo}
\affiliation{State Key Laboratory of Nuclear Physics and Technology, School of Physics, Peking University, Beijing 100871, China}

\author{Jie Meng} \email{mengj@pku.edu.cn}
\affiliation{State Key Laboratory of Nuclear Physics and Technology, School of Physics, Peking University, Beijing 100871, China}

\author{Myeong-Hwan Mun}
\affiliation{Department of Physics and Origin of Matter and Evolution of Galaxy (OMEG) Institute, Soongsil University, Seoul 156-743, Korea}
\affiliation{Korea Institute of Science and Technology Information, Daejeon 34141, Korea}

\author{Zhongming Niu}
\affiliation{School of Physics and Optoelectronics Engineering, Anhui University, Hefei 230601, China}

\author{Cong Pan}
\affiliation{State Key Laboratory of Nuclear Physics and Technology, School of Physics, Peking University, Beijing 100871, China}

\author{Panagiota Papakonstantinou}
\affiliation{Rare Isotope Science Project, Institute for Basic Science, Daejeon 34000, Korea}

\author{Xinle Shang}
\affiliation{Institute of Modern Physics, Chinese Academy of Sciences, Lanzhou 730000, China}
\affiliation{School of Physics, University of Chinese Academy of Sciences, Beijing 100049, China}

\author{Caiwan Shen}
\affiliation{School of Science, Huzhou University, Huzhou 313000, China}

\author{Guofang Shen}
\affiliation{School of Physics, Beihang University, Beijing 102206, China}

\author{Wei Sun}
\affiliation{School of Physical Science and Technology, Southwest University, Chongqing 400715, China}

\author{Xiang-Xiang Sun}
\affiliation{School of Nuclear Science and Technology, University of Chinese Academy of Sciences, Beijing 100049, China}
\affiliation{CAS Key Laboratory of Theoretical Physics, Institute of Theoretical Physics, Chinese Academy of Sciences, Beijing 100190, China}

\author{Chi Kin Tam}
\affiliation{Department of Physics, The University of Hong Kong, Pokfulam 999077, Hong Kong, China}

\author{Thaivayongnou}
\affiliation{School of Physics, Beihang University, Beijing 102206, China}

\author{Chen Wang}
\affiliation{College of Materials Science and Technology, Nanjing University of Aeronautics and Astronautics, Nanjing 210016, China}

\author{Xingzhi Wang}
\affiliation{State Key Laboratory of Nuclear Physics and Technology, School of Physics, Peking University, Beijing 100871, China}

\author{Sau Hei Wong}
\affiliation{Department of Physics, The University of Hong Kong, Pokfulam 999077, Hong Kong, China}

\author{Jiawei Wu}
\affiliation{College of Materials Science and Technology, Nanjing University of Aeronautics and Astronautics, Nanjing 210016, China}

\author{Xinhui Wu}
\affiliation{State Key Laboratory of Nuclear Physics and Technology, School of Physics, Peking University, Beijing 100871, China}

\author{Xuewei Xia}
\affiliation{School of Physics and Electronic Engineering, Center for Computational Sciences, Sichuan Normal University, Chengdu 610068, China}

\author{Yijun Yan}
\affiliation{Institute of Modern Physics, Chinese Academy of Sciences, Lanzhou 730000, China}
\affiliation{School of Physics, University of Chinese Academy of Sciences, Beijing 100049, China}

\author{Ryan Wai-Yen Yeung}
\affiliation{Department of Physics, The University of Hong Kong, Pokfulam 999077, Hong Kong, China}

\author{To Chung Yiu}
\affiliation{Department of Physics, The University of Hong Kong, Pokfulam 999077, Hong Kong, China}

\author{Shuangquan Zhang}
\affiliation{State Key Laboratory of Nuclear Physics and Technology, School of Physics, Peking University, Beijing 100871, China}

\author{Wei Zhang}
\affiliation{School of Physics and Microelectronics, Zhengzhou University, Zhengzhou 450001, China}

\author{Xiaoyan Zhang}
\affiliation{School of Physics and Optoelectronics Engineering, Anhui University, Hefei 230601, China}

\author{Qiang Zhao}
\affiliation{Center for Exotic Nuclei Studies, Institute for Basic Science, Daejeon 34126, Korea}
\affiliation{State Key Laboratory of Nuclear Physics and Technology, School of Physics, Peking University, Beijing 100871, China}

\author{Shan-Gui Zhou}
\affiliation{CAS Key Laboratory of Theoretical Physics, Institute of Theoretical Physics, Chinese Academy of Sciences, Beijing 100190, China}
\affiliation{School of Physical Sciences, University of Chinese Academy of Sciences, Beijing 100049, China}
\affiliation{Center of Theoretical Nuclear Physics, National Laboratory of Heavy Ion Accelerator, Lanzhou 730000, China}
\affiliation{Synergetic Innovation Center for Quantum Effects and Application, Hunan Normal University, Changsha 410081, China}

\collaboration{DRHBc Mass Table Collaboration}

\begin{abstract}
  Ground-state properties of even-even nuclei with $8\le Z\le120$ from the proton drip line to the neutron drip line have been investigated using the deformed relativistic Hartree-Bogoliubov theory in continuum (DRHBc) with the density functional PC-PK1.
  With the effects of deformation and continuum included simultaneously, 2583 even-even nuclei are predicted to be bound.
  The calculated binding energies, two-nucleon separation energies, root-mean-square (rms) radii of neutron, proton, matter, and charge distributions, quadrupole deformations, and neutron and proton Fermi surfaces are tabulated and compared with available experimental data.
  The rms deviation from the 637 mass data is 1.518~MeV, providing one of the best microscopic descriptions for nuclear masses.
  The drip lines obtained from DRHBc calculations are compared with other calculations, including the spherical relativistic continuum Hartree-Bogoliubov (RCHB) and triaxial relativistic Hartree-Bogoliubov (TRHB) calculations with PC-PK1.
  The deformation and continuum effects on the limits of the nuclear landscape are discussed.
  Possible peninsulas consisting of bound nuclei beyond the two-neutron drip line are predicted.
  The systematics of the two-nucleon separation energies, two-nucleon gaps, rms radii, quadrupole deformations, potential energy curves, neutron densities, neutron mean-field potentials, and pairing energies in the DRHBc calculations are also discussed. In addition, the $\alpha$ decay energies extracted are in good agreement with available data.
\end{abstract}

\date{\today}

\maketitle
\tableofcontents


\section{Introduction}

The new generations of radioactive ion beam (RIB) facilities running and under construction worldwide, including the Cooler Storage Ring (CSR) at the Heavy Ion Research Facility in Lanzhou (HIRFL) in China~\cite{Zhan2010NPA}, the Radioactive Ion Beam Factory (RIBF) at RIKEN in Japan~\cite{Motobayashi2010NPA}, the Rare isotope Accelerator complex for ON-line experiments (RAON) in Korea~\cite{Tshoo2013NIMPRB}, the Facility for Antiproton and Ion Research (FAIR) in Germany~\cite{Sturm2010NPA}, the Second Generation System On-Line Production of Radioactive Ions (SPIRAL2) at GANIL in France~\cite{Gales2010NPA}, the Facility for Rare Isotope Beams (FRIB) in the USA~\cite{Thoennessen2010NPA}, etc. produce and study more and more nuclei far from the stability valley and, thus, continue to extend our knowledge of nuclear physics from stable nuclei to exotic ones~\cite{Baumann2012RPP}, which also leads to new insights into the origins of the chemical elements in stars and star explosions~\cite{Aprahamian2005PPNP,Schatz2013IJMS}.

Up to date, the existence of more than $3300$ nuclides has been confirmed~\cite{Thoennessen2016Book,DNP} and the masses of about $2500$ among them have been measured~\cite{AME2020(1),AME2020(2),AME2020(3)}.
Accurate nuclear masses~\cite{Lunney2003RMP,BLAUM20061} are of crucial importance not only in nuclear physics but also in other fields, such as particle physics, nuclear astrophysics, and neutrino physics.
However, most of neutron-rich nuclei far from the stability valley will remain beyond the experimental access in the foreseeable future.
For example, the proton drip line of neptunium ($Z=93$) has been reached~\cite{Zhang2019PRL}, but the neutron drip line is known only up to neon ($Z=10$)~\cite{Ahn2019PRL}.
Therefore, a reliable theoretical nuclear mass table is highly desired to further understand the nuclear landscape.

Lots of efforts have been made to predict nuclear masses and to explore great unknowns of the nuclear landscape~\cite{Niu2018SciBull}.
Precise descriptions of nuclear masses have been achieved with various macroscopic-microscopic models~\cite{Moller2016ADNDT,Aboussir1995ADNDT,Wang2014PLB,Zhang2014NPA}, among which the WS4 model fits its 18 parameters to 2353 available mass data and achieves the highest accuracy, 0.298 MeV~\cite{Wang2014PLB}.
A number of Skyrme~\cite{Samyn2002NPA,Stoitsov2003PRC,Goriely2009PRLSkyrme,Erler2012Nature,Goriely2013PRC,Goriely2013PRC(R)} or Gogny~\cite{Hilaire2007EPJA,Goriely2009PRLGogny,Delaroche2010PRC} Hartree-Fock-Bogoliubov (HFB) mass-table-type calculations have been performed based on the non-relativistic density functional theory.
The Skyrme mass model HFB-27*~\cite{Goriely2013PRC(R)} determines its 24 parameters in functionals and corrections by fitting 2353 nuclear masses, and achieves an accuracy of 0.512 MeV.
The Gogny mass model D1M~\cite{Goriely2009PRLGogny} includes the beyond-mean-field correlation energies and fits its 14 parameters to 2149 measured masses, and the final rms deviation is 0.798 MeV.

The covariant density functional theory (CDFT) has proven to be a powerful theory in nuclear physics which has been used to describe successfully a variety of nuclear phenomena~\cite{Ring1996PPNP,Vretenar2005PhysRep,Meng2006PPNP,Niksic2011PPNP,Meng2013FOP,Meng2015JPG,Zhou2016PhysScr,Meng2016Book,Shen2019PPNP}.
The CDFT has gained wide attention in recent years for many attractive advantages, such as the automatic inclusion of the nucleonic spin degree of freedom and the spin-orbital interaction~\cite{Ren2020PRC(R)}, the explanation of the pseudospin symmetry in nucleon spectrum~\cite{Ginocchio1997PRL,Meng1998Phys.Rev.C628,Meng1999PRC,Chen2003CPL,Ginocchio2005PhysRep,Liang2015PhysRep} and the spin symmetry in anti-nucleon spectrum~\cite{Zhou2003PRL,He2006EPJA,Liang2015PhysRep}, and the natural inclusion of the nuclear magnetism~\cite{Koepf1989NPA}, which plays an important role in nuclear magnetic moments~\cite{Yao2006Phys.Rev.C24307,Arima2011,Li2011Sci.ChinaPhys.Mech.Astron.204,Li2011Prog.Theor.Phys.1185,Li2018Front.Phys.Beijing132109} and nuclear rotations~\cite{Meng2013FOP,Konig1993PRL,Afanasjev2000NPA,PhysRevC.62.031302,PhysRevC.82.034329,Zhao2011PRL,Zhao2011PLB,Zhao2012Phys.Rev.C54310,Zhao2015PRL,Wang2017Phys.Rev.C54324,Wang2018Phys.Rev.64321,Ren2019SciChina,Ren2020NPA}.

Significant progresses on mass description have also been made based on the CDFT~\cite{Lalazissis1999ADNDT,Geng2005PTP,Meng2013FOP,Afanasjev2013PLB,Zhang2014FOP,Agbemava2014PRC,Afanasjev2015PRC,Lu2015PRC,Pena-Arteaga2016EPJA,Xia2018ADNDT,Yang2021PRC}.
The relativistic mean field (RMF) calculations including dynamical correlation energies describe the masses of 575 even-even nuclei with an accuracy of 1.14 MeV~\cite{Lu2015PRC}, based on the density functional PC-PK1 which has only 9 parameters in its Lagrangian and 2 ones for the pairing strength adjusted to the masses of 60 spherical nuclei~\cite{Zhao2010PRC}.

It is worth mentioning that the shell model and its generalizations can also be used to predict nuclear masses and other properties~\cite{Coraggio2009PPNP,Barrett2013PPNP,Otsuka2020RMP}.
The shell model strictly fulfills all conservation laws, accounts not only for pairing but for many possible residual interactions, and achieves success in correctly predicting the neutron drip line of oxygen isotopes~\cite{Hoffman2008PRL,Otsuka2010PRL}, etc.
Due to the computational demand, however, it is not realistic to apply the shell model to describe all the nuclei in the whole nuclear chart.

Based on the CDFT and with the pairing correlations and continuum effects properly treated, the relativistic continuum Hartree-Bogoliubov (RCHB) theory was developed in Refs.~\cite{Meng1996PRL,Meng1998NPA} with the relativistic Hartree-Bogoliubov (RHB) equations solved in coordinate space.
The RCHB theory has achieved great successes in the studies of both stable and exotic nuclei~\cite{Meng1996PRL,Meng1998PRL,Meng2002PRC(R),Zhang2003SciChina,Meng1998PLB,Meng2002PLB,Meng1998Phys.Rev.C628,Meng1999PRC,Zhang2005NPA,Lyu2003EPJA,Zhang2016CPC,Lim2016PRC}.
Based on the RCHB theory and the density functional PC-PK1, the first nuclear mass table including continuum effects has been constructed and the continuum effects on the limits of the nuclear landscape have been discussed~\cite{Xia2018ADNDT}.
It is demonstrated that the continuum effects are crucial for drip-line locations and $9035$ nuclei with $8\le Z\le 120$ are predicted to be bound, which remarkably extends the existing nuclear landscapes.

Except for doubly-magic nuclei, most nuclei in the nuclear chart deviate from the spherical shape.
Solving the deformed RHB equations in coordinate space is extremely difficult if not impossible~\cite{Zhou2000CPL}.
To provide a proper description of deformed exotic nuclei, the deformed relativistic Hartree-Bogoliubov theory in continuum (DRHBc) was developed~\cite{Zhou2010PRC(R),Li2012PRC}, with the deformed RHB equations solved in a Dirac Woods-Saxon basis~\cite{Zhou2003PRC}.
The DRHBc theory has been extended to the version with density-dependent meson-nucleon couplings~\cite{Chen2012PRC}, to incorporate the blocking effects of odd nucleon(s)~\cite{Li2012CPL}, and to explore the rotational excitations of exotic nuclei with the angular momentum projection~\cite{Sun2021SciBull,Sun2021arXiv}.
Inheriting the advantages of the RCHB theory and including the deformation degrees of freedom, the DRHBc theory was successful in studying deformed halos in magnesium isotopes and predicting an interesting shape decoupling between the core and the halo in $^{42,44}$Mg~\cite{Zhou2010PRC(R),Li2012PRC}, resolving the puzzles concerning the radius and configuration of valence neutrons in $^{22}$C~\cite{Sun2018PLB}, investigating the shell evolution of carbon isotopes and neutron halos in $^{15,19,22}$C~\cite{Sun2020NPA}, exploring particles in the classically forbidden regions for magnesium isotopes~\cite{Zhang2019PRC}, and interpreting the neutron halos in $^{17}$B~\cite{Yang2021PRL} and $^{19}$B~\cite{Sun2021PRC}.

The success of the RCHB mass table in extending the limit of nuclear existence and the advantage of the DRHBc theory for deformed nuclei stimulate the construction of an upgraded mass table including simultaneously the deformation and continuum effects by the DRHBc theory.
However, the DRHBc theory is numerically much more complicated than the RCHB theory.
Note that all the above mentioned DRHBc calculations focused upon light nuclei only~\cite{Zhou2010PRC(R),Li2012PRC,Li2012CPL,Chen2012PRC,Sun2018PLB,Sun2020NPA,Zhang2019PRC,Yang2021PRL,Sun2021PRC}.

In order to provide a unified description for all nuclei in the nuclear chart with the DRHBc theory, a DRHBc Mass Table Collaboration was established.
The strategy and techniques for the nuclear mass table by the DRHBc theory with point-coupling density functionals are presented in Ref.~\cite{Zhang2020PRC}.
During the construction of the DRHBc mass table, several important and interesting results have been reported~\cite{Pan2019IJMPE,In2021IJMPE,Zhang2021PRC(L),Pan2021PRC,He2021CPC}.
In Ref.~\cite{Pan2019IJMPE}, the dependence on the Legendre expansion of nuclear potentials and densities is investigated for light and heavy nuclei.
By comparing the neutron drip lines predicted by the DRHBc theory and the RCHB theory, the deformation effects on the location of neutron drip line have been studied from O to Ca isotopes in Ref.~\cite{In2021IJMPE}.
The predictive power of the DRHBc theory with PC-PK1 for the masses of superheavy nuclei has been shown by comparing with the latest AME2020 data, and an interesting peninsula consisting of bound superheavy nuclei beyond the two-neutron drip line has been predicted~\cite{Zhang2021PRC(L)}.
Similar peninsulas of stability have been found in the region of medium heavy nuclei~\cite{Pan2021PRC}, and the effects of higher order deformations such as $\beta_4$ and $\beta_6$ on this phenomenon have been investigated~\cite{He2021CPC}.

In this paper, we report the DRHBc mass table for even-even nuclei with $8\le Z\le120$.
The DRHBc theoretical framework is presented briefly in Sec.~\ref{theory}.
Numerical details in the construction of the DRHBc mass table for even-even nuclei are given in Sec.~\ref{numerical}.
Extensive results are compiled in Sec.~\ref{results}, including binding energies, two-nucleon separation energies, rms radii, quadrupole deformations, etc.
Finally, a summary is given in Sec.~\ref{summary}.


\section{Theoretical framework} \label{theory}

The details of the DRHBc theory with meson-exchange and point-coupling density functionals can be found in Refs.~\cite{Li2012PRC} and \cite{Zhang2020PRC}, respectively. In the following we introduce its theoretical framework briefly.

Starting from the Lagrangian density for point-coupling functionals, the energy density functional of the nuclear system can be constructed under the mean-field and no-sea approximations.
By minimizing the energy density functional with respect to the densities, one obtains the Dirac equation for nucleons within the relativistic
mean-field framework~\cite{Meng2016Book}.
Treating self-consistently the mean field and pairing correlations, the RHB equations for the nucleons read~\cite{Kucharek1991ZPA}
\begin{equation}\label{RHB}
\left(\begin{matrix}
h_D-\lambda_\tau & \Delta \\
-\Delta^* &-h_D^*+\lambda_\tau
\end{matrix}\right)\left(\begin{matrix}
U_k\\
V_k
\end{matrix}\right)=E_k\left(\begin{matrix}
U_k\\
V_k
\end{matrix}\right).
\end{equation}
For exotic nuclei with the Fermi energy very close to the continuum threshold, the pairing interaction can scatter nucleons from bound states to resonant ones in the continuum.
The density could become more diffuse due to this coupling to the continuum, and the position of the drip line might be influenced, which are the so-called continuum effects.
In order to consider such continuum effects, the deformed RHB equations in the DRHBc theory are solved in a Dirac Woods-Saxon basis, in which the radial wave functions have a proper asymptotic behavior for large $r$~\cite{Zhou2003PRC}.

In Eq.~(\ref{RHB}), $\lambda_\tau$ is the Fermi energy ($\tau = n/p$ for neutrons or protons), $E_k$ and $(U_k, V_k)^{\rm T}$ the quasiparticle energy and wave function, and $h_D$ the Dirac Hamiltonian,
\begin{equation}
h_D(\bm{r})=\bm{\alpha}\cdot\bm{p}+V(\bm{r})+\beta[M+S(\bm{r})],
\end{equation}
with the scalar $S(\bm r)$ and vector $V(\bm r)$ potentials.
The pairing potential $\Delta$ reads
\begin{equation}\label{Delta}
\Delta(\bm r_1,\bm r_2) = V^{\mathrm{pp}}(\bm r_1,\bm r_2)\kappa(\bm r_1,\bm r_2),
\end{equation}
with a density-dependent force of zero range,
\begin{equation}\label{pair}
V^{\mathrm{pp}}(\bm r_1,\bm r_2)= V_0 \frac{1}{2}(1-P^\sigma)\delta(\bm r_1-\bm r_2)\left(1-\frac{\rho(\bm r_1)}{\rho_{\mathrm{sat}}}\right),
\end{equation}
and the pairing tensor $\kappa$~\cite{Peter1980Book}.

For axially deformed nuclei, the potentials and densities are expanded in terms of the Legendre polynomials,
\begin{equation}\label{legendre}
f(\bm r)=\sum_{\lambda} f_{\lambda}(r)P_{\lambda}(\cos\theta),~~\lambda=0,2,4,\cdots,
\end{equation}
where $\lambda$ is restricted to be even numbers due to spatial reflection symmetry.

Under the mean-field approximation, many-body correlations are taken into account with the symmetry breaking, e.g., the loss of translational invariance and rotational invariance~\cite{Peter1980Book}.
The deformed state thus obtained does not have a good angular momentum quantum number, and should be regarded as the intrinsic ground state.
One can restore the broken symmetries by using beyond-mean-field techniques such as the angular momentum projection and the generator coordinate method.
In the present DRHBc theory, beyond-mean-field correlations are taken into account by the microscopic calculation of center-of-mass and rotational correction energies, and the details can be found in Ref.~\cite{Zhang2020PRC}.


\section{Numerical details}\label{numerical}

In Ref.~\cite{Zhang2020PRC}, systematic numerical convergence checks from light to heavy nuclei for the DRHBc calculations have been performed, and numerical details for the DRHBc mass table of even-even nuclei have been suggested.
Numerical details used in the present study are almost the same as those suggested.

In the following we summarize the numerical details in the present DRHBc mass table for even-even nuclei.
\begin{itemize}
  \item [$\bullet$] The density functional PC-PK1~\cite{Zhao2010PRC}, which provides one of the best density-functional descriptions for nuclear masses~\cite{Zhao2012Phys.Rev.C64324,Zhang2014FOP,Lu2015PRC,Zhang2021PRC(L)}, is employed.
  \item [$\bullet$] The pairing strength $V_0=-325~\mathrm{MeV~fm}^3$ and the saturation density $\rho_{\mathrm{sat}}=0.152~\mathrm{fm}^{-3}$ in Eq.~(\ref{pair}) together with a pairing window of $100$ MeV, which reproduce well the odd-even mass differences for calcium and lead isotopes~\cite{Zhang2020PRC}.
  \item [$\bullet$] The energy cutoff for the Dirac Woods-Saxon basis $E^+_{\mathrm{cut}}=300$ MeV, which guarantees the convergence accuracy of 0.01~MeV for the total energies of doubly magic nuclei $^{40}$Ca, $^{100}$Sn, and $^{208}$Pb, reproducing exactly the results from RCHB calculations~\cite{Zhang2020PRC}.
  \item [$\bullet$] The angular momentum cutoff for the Dirac Woods-Saxon basis $J_{\max}=\frac{23}{2}~\hbar$, which guarantees the convergence accuracy of $0.01\%$ for the total energy of the deformed heavy nucleus $^{300}$Th~\cite{Zhang2020PRC}.
  \item [$\bullet$] The number of the Dirac Woods-Saxon basis states in the Dirac sea is the same as that in the Fermi sea~\cite{Zhou2003PRC,Zhou2010PRC(R),Li2012PRC}.
  \item [$\bullet$] The Legendre expansion truncations in Eq.~(\ref{legendre}) are chosen as $\lambda_{\max}=6$ and $8$ for nuclei with $8\le Z\le 70$ and $72\le Z\le 100$, respectively. This guarantees the convergence accuracy of $0.03\%$ for the total energies of $^{20}$Ne and $^{112}$Mo and of $0.01\%$ for the total energy of $^{300}$Th, when their deformations are constrained to be $\beta_2=0.6$~\cite{Zhang2020PRC}. For superheavy nuclei with $102\le Z\le 120$, $\lambda_{\max}=10$ is adopted~\cite{Pan2019IJMPE}.
\end{itemize}

\section{Results and Discussion}\label{results}

\subsection{Overview of nuclear masses and two-nucleon separation energies}

\subsubsection{Nuclear masses}

\begin{figure}[htbp]
  \centering
  \includegraphics[width=1.0\textwidth]{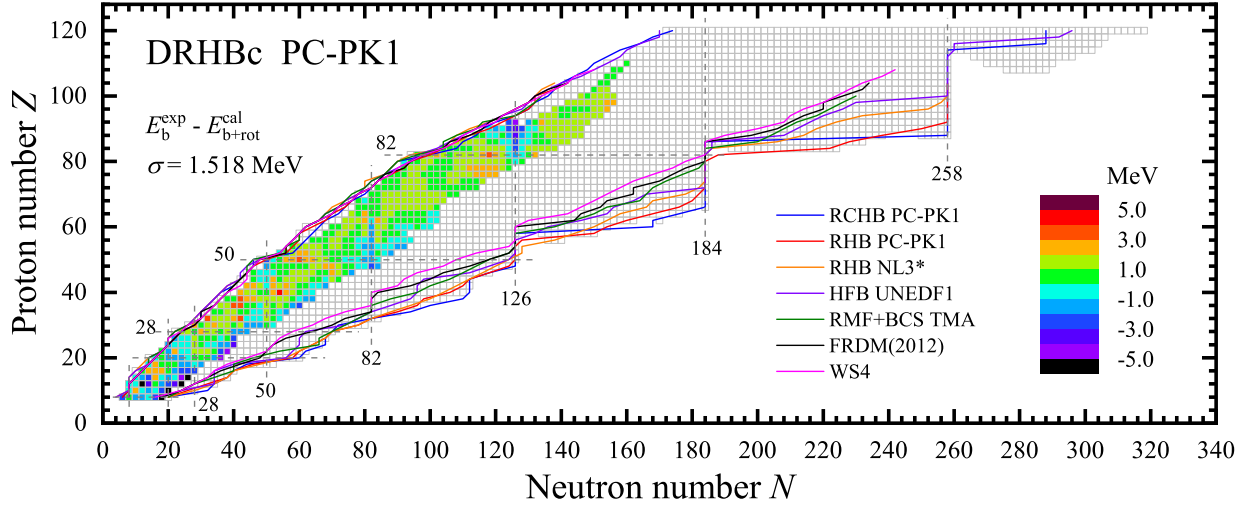}
  \caption{(Color online) 2583 bound even-even nuclei from O ($Z = 8$) to $Z = 120$ predicted by the DRHBc theory with PC-PK1. For the 637 even-even nuclei with mass measured, the binding energy differences between the data~\cite{AME2020(3)} and the DRHBc calculations (with rotational correction energy included) are scaled by colors. The nucleon drip lines predicted by different mass tables, RCHB with PC-PK1~\cite{Xia2018ADNDT}, RHB with PC-PK1~\cite{Yang2021PRC} and with NL3*~\cite{Agbemava2014PRC}, HFB with UNEDF1~\cite{Erler2012Nature}, RMF+BCS with TMA~\cite{Geng2005PTP}, FRDM(2012)~\cite{Moller2016ADNDT}, and WS4~\cite{Wang2014PLB}, are plotted for comparison.}
\label{fig1}
\end{figure}

We perform systematic calculations for all even-even nuclei from $Z = 8$ to $Z = 120$ from the proton drip
line to the neutron drip line.
In Table~\ref{tab2}, the ground-state properties of these nuclei are summarized.
The mass number $A$, neutron number $N$, binding energy $E_{\mathrm{b}}^{\mathrm{cal}}$, binding energy including rotational correction $E^{\mathrm{cal}}_{\mathrm{b}+\mathrm{rot}}$, neutron rms radius $R_n$, proton rms radius $R_p$, matter rms radius $R_m$, charge radius $R_{\mathrm{ch}}$, neutron quadrupole deformation $\beta_{2n}$, proton quadrupole deformation $\beta_{2p}$, total quadrupole deformation $\beta_{2}$, neutron Fermi surface $\lambda_n$, and proton Fermi surface $\lambda_p$ are listed.
The two-neutron separation energy $S_{2n}$ and the two-proton separation energy $S_{2p}$ for each nucleus are also provided.
The available experimental binding energies~\cite{AME2020(3)} and charge radii~\cite{Angeli2013ADNDT,Li2021ADNDT} are shown for comparison.
2583 even-even nuclei from O ($Z = 8$) to $Z = 120$ are predicted to be bound by the DRHBc theory with the relativistic
density functional PC-PK1.
For guidance, some unbound nuclei are listed and underlined in Table~\ref{tab2} as well.

In Fig.~\ref{fig1}, the nuclear landscape for even-even nuclei from O ($Z = 8$) to $Z = 120$ explored by the DRHBc theory with PC-PK1 is shown, where the squares represent bound nuclei.
Among these 2583 bound even-even nuclei, the masses of 637 nuclei have been measured experimentally~\cite{AME2020(3)}.
The binding energy differences $E^{\mathrm{exp}}_{\mathrm{b}}-E^{\mathrm{cal}}_{\mathrm{b}+\mathrm{rot}}$ for these measured nuclei scaled by colors are plotted.
It is found that the overall agreement between experimental and calculated binding energies is quite good, and the rms deviation for these 637 nuclei is $\sigma = 1.518$ MeV.

\subsubsection{Two-nucleon separation energies}

The two-neutron separation energy $S_{2n}$ and the two-proton separation energy $S_{2p}$ are respectively defined as
\begin{gather}
S_{2n}(Z,N) = E_{\mathrm{b}}(Z,N) - E_{\mathrm{b}}(Z,N-2),\\
S_{2p}(Z,N) = E_{\mathrm{b}}(Z,N) - E_{\mathrm{b}}(Z-2,N),
\end{gather}
where $E_{\mathrm{b}}(Z,N)$ is the binding energy of the nucleus with the proton number $Z$ and the neutron number $N$.
These quantities provide information on whether one nucleus is stable against two-nucleon emissions, and thus define the two-nucleon drip lines.
In this work, a nucleus is bound only if its two-nucleon and multi-nucleon separation energies are positive, i.e., it is stable against two- and multi-nucleon emissions.
For each isotopic chain, the location where $S_{2n}=0$ or $S_{2p}=0$ defines the two-neutron drip line or the two-proton drip line.
Because this paper concerns only even-even nuclei, in the following we refer to the two-nucleon drip line as the nucleon drip line, unless otherwise specified.

\begin{figure}[htbp]
  \centering
  \includegraphics[width=1.0\textwidth]{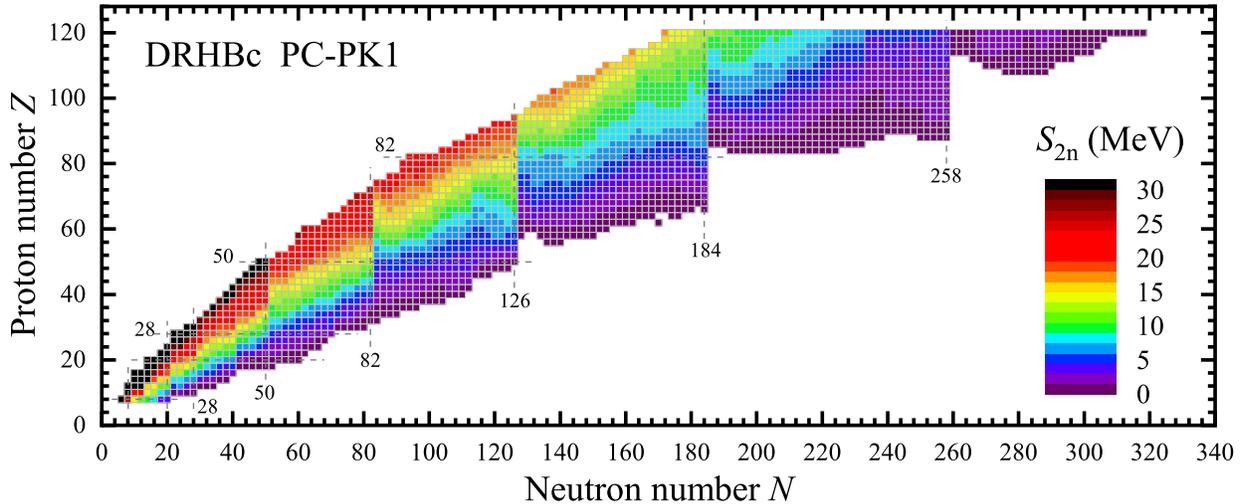}
  \caption{(Color online) Two-neutron separation energies of bound even-even nuclei with $8 \le Z \le 120$ in the DRHBc calculations with PC-PK1 scaled by colors.}
\label{fig2}
\end{figure}

In Fig.~\ref{fig2}, the two-neutron separation energies $S_{2n}$ of the bound even-even nuclei predicted by the DRHBc theory with PC-PK1 are shown.
From a global view, $S_{2n}$ is large near the proton drip line and close to zero near the neutron drip line.
For a given isotopic (isotonic) chain, $S_{2n}$ decreases (increases) with the increasing neutron (proton) number.
There are 41 nuclei with the predicted two-neutron separation energies larger than 30 MeV, and most of them are located at the proton-rich side of nuclear
landscape with $Z \le 50$.
There are 103 nuclei with $S_{2n}$ in the range of 21-30 MeV, mainly located in the proton-rich region of the nuclear
chart; 501 nuclei with $S_{2n}$ in the range of 12-21 MeV, most of which are near the valley of stability; 1168 nuclei with $S_{2n}$ in the
range of 3-12 MeV which lie on the neutron-rich region mostly.
In addition, there are 770 nuclei with $S_{2n}$ less than 3 MeV, and most of them are located in the region far from the stability line, and are even approaching the neutron drip line.

It should be noted that, there are 253 weakly bound nuclei with $S_{2n} \le 1$ MeV in the DRHBc calculations.
They are extremely neutron-rich, and many of them lie even beyond the neutron drip lines predicted by the other nuclear mass models.
For these weakly bound nuclei, as the neutron Fermi surface is close to the continuum threshold, pairing correlations could scatter the nucleons from bound states to resonant ones in the continuum and, thus, provide a significant coupling between the continuum and bound states, which might affect the location of the drip line~\cite{Xia2018ADNDT}.
In addition, the nearly vanishing $S_{2n}$ around the neutron drip line might be regarded as a sign of the neutron halo or giant halo~\cite{Meng2006PPNP}, which requires a detailed analysis of neutron radii and single-particle levels.

The two-neutron separation energy $S_{2n}$ is a widely used probe of the neutron shell structure.
In general, for a given isotopic chain, $S_{2n}$ decreases smoothly with the neutron number, except at a magic number where $S_{2n}$ drops significantly.
An abrupt decline of $S_{2n}$ indicates the occurrence of neutron shell closure.
It can be seen in Fig.~\ref{fig2} that the significant drops exist at the traditional magic numbers $N = 20, 28, 50, 82$, and $126$, which demonstrates that these shell closures are well reproduced by the DRHBc theory.
Apart from this, dramatic declines of $S_{2n}$ can be found at $N = 184$ and $258$, which indicates that the neutron numbers $184$ and $258$ may be neutron magic numbers in the superheavy mass region~\cite{Zhang2005NPA,Li2014PLB}.

\begin{figure}[htbp]
  \centering
  \includegraphics[width=1.0\textwidth]{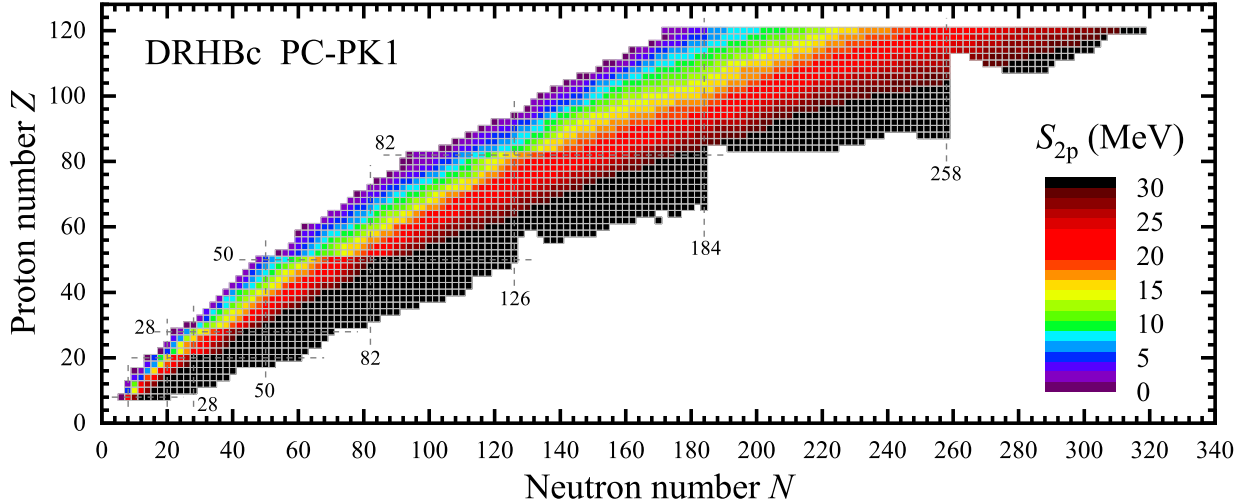}
  \caption{(Color online) Two-proton separation energies of bound even-even nuclei with $8 \le Z \le 120$ in the DRHBc calculations with PC-PK1 scaled by colors.}
\label{fig3}
\end{figure}

The two-proton separation energies $S_{2p}$ of the bound even-even nuclei predicted by the DRHBc theory with PC-PK1 are scaled by colors in Fig.~\ref{fig3}.
$S_{2p}$ increases with the neutron number for a given isotopic chain, while decreases with the proton number for a given isotonic chain.
In the present study, there are 232 nuclei with $S_{2p} \ge 40$ MeV located near the neutron drip line, 611 neutron-rich nuclei with 30 MeV $\le S_{2p} < 40$ MeV, 754 nuclei with 20 MeV $\le S_{2p} < 30$ MeV near the $\beta$-stability line, 566 nuclei with 10 MeV $\le S_{2p} < $ 20 MeV, and 420 nuclei with  $S_{2p} < $ 10 MeV near the proton drip line.
16 nuclei have $S_{2p} \le 1$ MeV, which is much less than the number of the nuclei with $S_{2n} \le 1$ MeV.
This is consistent with the fact that the number of halo nuclei observed on the proton-rich side is much less than that on the neutron-rich side, due to the existence of the Coulomb barrier.
Moreover, although the proton Fermi surface is approaching zero for neutron-deficient nuclei, the contributions from the continuum are suppressed due to the Coulomb barrier in comparison with that for the neutron-rich nuclei.
This also partially explains why the proton drip lines predicted by different models are very close.
At the proton magic numbers 20, 28, 50, and 82, some abrupt changes of $S_{2p}$ are exhibited, which indicates that the DRHBc theory reproduces the traditional proton closed shells.
The closed shells will be discussed further down in terms of two-nucleon gaps.

\subsubsection{Comparison with other predictions}

\begin{table}
\centering
\caption{The rms deviations of binding energies, of two-neutron separation energies, and of two-proton separation energies for the DRHBc calculations with PC-PK1 with respect to the AME2020 data~\cite{AME2020(3)} in the unit of MeV. The results of other relativistic and non-relativistic density functional calculations are also listed for comparison.}
\begin{tabular}{cccccccc}
\hline
Model & Symmetry & Density functional & ~~$\sigma(E_{\mathrm{b}})$~~ & ~~$\sigma(S_{2n})$~~ & ~~$\sigma(S_{2p})$~~ & Data numbers & Reference \\
\hline
DRHBc$^{\mathrm{w/o}~E_{\mathrm{rot}}}$ & Axial & PC-PK1 & \textbf{2.744} & \textbf{1.067} & \textbf{0.959} & 637 & This work \\
DRHBc$^{\mathrm{w/} ~~E_{\mathrm{rot}}}$ & Axial & PC-PK1 & \textbf{1.518} & \textbf{1.104} & \textbf{1.095} & 637 & This work \\
RCHB & Spherical & PC-PK1 & 8.036 & 1.573 & 1.587 & 630 & \cite{Xia2018ADNDT} \\
RHB$^{\mathrm{w/o} ~E_{\mathrm{corr}}}$ & Triaxial & PC-PK1 & 2.635 & 1.064 & 0.929 & 628 & \cite{Yang2021PRC} \\
RHB$^{\mathrm{w/} ~~E_{\mathrm{corr}}}$ & Triaxial & PC-PK1 & 1.335 & 0.751 & 0.755 & 628 & \cite{Yang2021PRC} \\
RHB & Axial & DD-ME2 & 2.377 & 1.007 & 0.878 & 636 & \cite{Agbemava2014PRC} \\
RHB & Axial & DD-ME$\delta$ & 2.309 & 1.035 & 1.041 & 634 & \cite{Agbemava2014PRC} \\
RHB & Axial & DD-PC1 & 2.019 & 1.074 & 0.900 & 636 & \cite{Agbemava2014PRC} \\
RHB & Axial & NL3* & 2.907 & 1.082 & 1.189 & 638 & \cite{Agbemava2014PRC} \\
RMF+BCS & Axial & TMA & 2.113 & 0.967 & 1.150 & 642 & \cite{Geng2005PTP} \\
\hline
HFB & Axial & SkM & 7.269 & 1.223 & 1.823 & 624 & \cite{Erler2012Nature} \\
HFB & Axial & SLy4 & 5.344 & 0.996 & 0.897 & 631 & \cite{Erler2012Nature} \\
HFB & Axial & SV-min & 3.426 & 0.790 & 0.817 & 629 & \cite{Erler2012Nature} \\
HFB & Axial & UNEDF1 & 1.926 & 0.747 & 0.780 & 632 & \cite{Erler2012Nature} \\
\hline
\end{tabular}
\label{tab1}
\end{table}

For a quantitative comparison with previous works, the rms deviations of binding energies $\sigma(E_{\mathrm{b}})$, two-neutron separation energies $\sigma(S_{2n})$, and two-proton separation energies $\sigma(S_{2p})$ for the present calculations with respect to the data available from AME2020~\cite{AME2020(3)} are listed in Table~\ref{tab1}, together with some previous density functional calculations.
Because the number of bound nuclei may differ by models, the data numbers for extracting the rms deviations are also listed in Table~\ref{tab1}.
It is seen that the accuracy of the DRHBc calculations with PC-PK1 is significantly improved by including the rotational correction energies, which has been shown for Nd isotopes in Ref.~\cite{Zhang2020PRC}.
By comparing the DRHBc results with the spherical RCHB results, we find that the inclusion of deformation degrees of freedom is very important for the description of nuclear masses.
By comparing the DRHBc results with the triaxial RHB (TRHB) results, including triaxial deformation degree of freedom and considering the dynamical correlation energies might further improve the accuracy.
By comparing the DRHBc results with those from other relativistic and nonrelativistic density functional calculations, we find that the present PC-PK1 calculations including rotational correction energies provide a better description for nuclear masses.

The two-nucleon separation energies are described with an accuracy around 1~MeV for most density functional calculations as shown in Table~\ref{tab1}.
Because of the assumed spherical symmetry, the description accuracy of the RCHB theory for two-nucleon separation energies is about 1.5~MeV, slightly larger than the others.
Different from the TRHB calculations with PC-PK1, the inclusion of the rotational correction energies does not improve the global description of two-nucleon separation energies in the DRHBc calculations.
This is because the cranking approximation used to obtain the rotational correction energy in the present DRHBc calculations~\cite{Zhang2020PRC} is not suitable for spherical or nearly spherical nuclei.
Using the collective Hamiltonian method to estimate the beyond-mean-field correlation energies in the DRHBc theory is expected, and such work is in progress.
Following the works in Ref.~\cite{Yang2021PRC}, the accuracies for nuclear masses and two-nucleon separation energies can be expected to be further improved to less than 1.4 MeV and 0.8 MeV respectively by including the beyond-mean-field correlation energies from the collective Hamiltonian method.

\subsection{The limits of the nuclear landscape}

In Fig.~\ref{fig1}, the nucleon drip lines predicted by other mass tables, RCHB with PC-PK1~\cite{Xia2018ADNDT}, RHB with PC-PK1~\cite{Yang2021PRC} and with NL3*~\cite{Agbemava2014PRC}, HFB with UNEDF1~\cite{Erler2012Nature}, RMF+BCS with TMA~\cite{Geng2005PTP}, FRDM(2012)~\cite{Moller2016ADNDT}, and WS4~\cite{Wang2014PLB} have been plotted.
At present, the experimental proton-rich border of the nuclear territory has been reached up to neptunium ($Z=93$)~\cite{Zhang2019PRL}.
Due to the Coulomb repulsive interaction among protons, the proton drip line does not lie so far away from the valley of stability.
Moreover, the proton continuum is effectively shifted up in energy due to the Coulomb barrier.
Therefore, the proton drip lines obtained by different models are close and roughly consistent with the experimental observations.

On the neutron-rich side, however, the neutron-rich boundary is known only up to neon ($Z = 10$) experimentally~\cite{Ahn2019PRL}.
In contrast to the proton drip line, the locations of neutron drip lines from various mass tables obviously differ with each other, and the differences increase with the proton number.
Compared with the neutron drip line in the spherical RCHB theory, the inclusion of the deformation effects does not necessarily extend the drip line.
The deformation effects on the location of neutron drip line have been studied from O to Ca isotopes in Ref.~\cite{In2021IJMPE}.
It is found that the direction of the change in the neutron drip line depends on the evolution of deformation towards the drip line: the drip line extends to the more neutron-rich side if the deformation increases towards the drip line, and \emph{vice versa}.
This mechanism is found to be valid for heavier Sm, Gd, and Dy isotopic chains in Ref.~\cite{Pan2021PRC}.

There are almost no triaxially deformed even-even nuclei near the neutron drip line~\cite{Yang2021PRC}.
However, as shown in Fig.~\ref{fig1}, the neutron drip lines predicted by the DRHBc and TRHB calculations are not the same.
For several isotopic chains, e.g. $38\le Z\le 42$, the DRHBc calculations predict a more extended neutron drip line than the TRHB calculations, where the harmonic oscillator basis is used, in which the continuum effects are not taken into account well.
It is shown in Ref.~\cite{Xia2018ADNDT} that the coupling between the continuum and bound states is relevant to the extension of the neutron drip line.
For the Pb ($Z=82$) isotopic chain, the neutron drip line locates at the predicted shell closure $N=184$ in the DRHBc calculations, while it extends slightly to the more neutron-rich region in the TRHB calculations.
This may be caused by the adopted different pairing interactions, i.e., the $\delta$ pairing in DRHBc and the separable pairing~\cite{Tian2009PLB} in TRHB.
In addition, the isospin-violating nucleon-nucleon interaction may influence the location of the neutron drip line as well~\cite{Dong2018PRC(R),Dong2019NPA}.

Compared with other density functional calculations, the DRHBc calculations with PC-PK1 generally predict a more extended neutron drip line with a few exceptions.
This is mainly because of the proper treatment of the continuum in the DRHBc theory and the adopted density functional.
On the other hand, compared with density functional calculations, the neutron drip lines predicted by the macroscopic-microscopic mass models FRDM~\cite{Moller2016ADNDT} and WS4~\cite{Wang2014PLB} are systematically closer to the valley of stability.
It has been shown in Ref.~\cite{Zhang2021PRC(L)} that, compared with the microscopic DRHBc calculations with PC-PK1, these macroscopic-microscopic mass models have a different isospin dependence in describing superheavy nuclear masses.

As shown in Fig.~\ref{fig1}, there exist some bound nuclei beyond the primary neutron drip line in the regions of $50\le Z\le 70$, $80\le Z\le90$, and $100\le Z\le 120$, forming peninsulas of stability adjacent to the nuclear mainland.
Near these peninsulas, there are several multi-neutron emitters that have positive $S_{2n}$ but negative multi-neutron separation energies.
To explore the underlying mechanism behind this interesting phenomenon, the neutron separation energies, Fermi surfaces, quadrupole deformations, and single-particle spectra in the canonical basis have been investigated for $50\le Z\le 70$ and $100\le Z\le 120$~\cite{Zhang2021PRC(L),Pan2021PRC,He2021CPC}.
It is found that the deformation plays a decisive role in the formation of these stability peninsulas, and meanwhile the pairing correlations and continuum effects also influence them in a self-consistent way.
Furthermore, it is shown in Refs.~\cite{He2021CPC,Wang2021CPC} that the effects of higher order deformations such as $\beta_4$ and $\beta_6$ are very important.
The decay rates of multi-neutron radioactivity in Ba and Sm isotopic chains are estimated by using the direct decay model in Ref.~\cite{Pan2021PRC}.

\subsection{Two-nucleon gaps and possible magic numbers}

The two-neutron gap $\delta_{2n}$ and the two-proton gap $\delta_{2p}$ are respectively defined as
\begin{gather}
\delta_{2n}(Z,N) = S_{2n}(Z,N) - S_{2n}(Z,N+2),\\
\delta_{2p}(Z,N) = S_{2p}(Z,N) - S_{2p}(Z+2,N).
\end{gather}
A peak of the two-nucleon gaps implicates the drastic change of the two-nucleon separation energies, which can be used as one of the signatures for magic numbers~\cite{Zhang2005NPA,Li2014PLB}.
Compared with two-nucleon separation energies, two-nucleon gaps are more intuitive in indicating possible magic numbers and clearer in exploring possible subshells.

\begin{figure*}[htbp]
  \centering
  \includegraphics[width=1.0\textwidth]{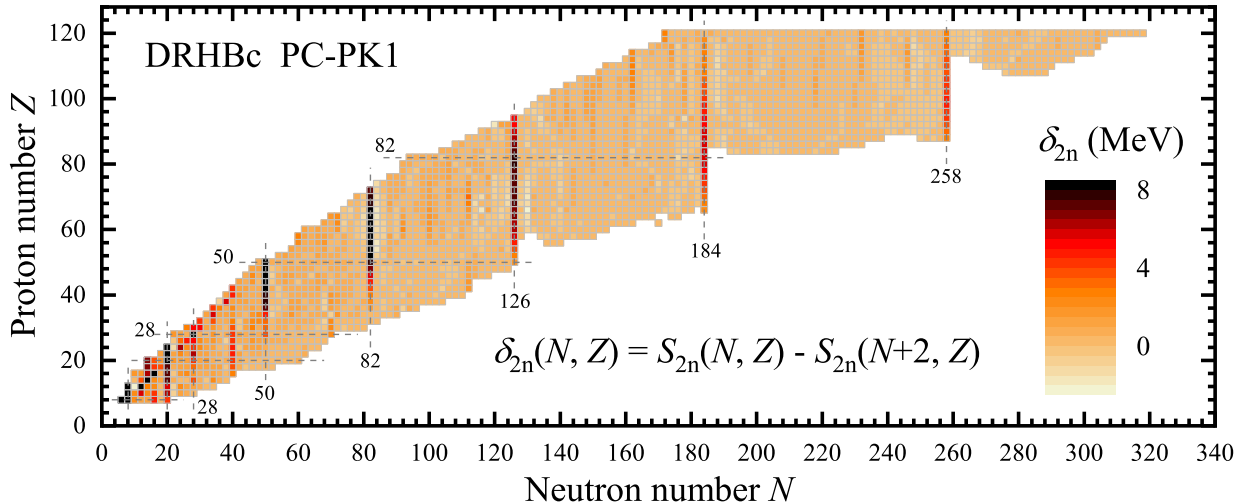}
  \caption{(Color online) Two-neutron gaps $\delta_{2n}$ of bound even-even nuclei with $8 \le Z \le 120$ in the DRHBc calculations with PC-PK1 scaled by colors.}
\label{fig4}
\end{figure*}

Figure \ref{fig4} shows the two-neutron gaps $\delta_{2n}$ of bound even-even nuclei with $8 \le Z \le 120$ in the DRHBc calculations with PC-PK1.
Peaks of $\delta_{2n}$ at neutron numbers $N = 8, 20, 28, 50, 82, 126, 184$ and $258$ can be clearly seen in Fig.~\ref{fig4}, which leads to the same conclusion for shell closures as in Fig.~\ref{fig2}.
It is noted that the peaks at $N = 28, 50,$ and $80$ become weaker or even disappear near the neutron drip lines for $Z \approx$ 10, 20, and 32, respectively, which suggests the quenching or even collapse of the traditional neutron shell closures 28, 50, and 82 in the neutron-rich region far from the stability valley.
There are also some hints at spherical or deformed subshells, for example, $N=40$ in the $Z\approx20$ region and $N=162$ in the $Z\approx 110$ region.
Further confirmation of such subshells needs detailed analysis for deformation and the evolution of single-neutron levels with deformation from constrained calculations.

\begin{figure*}[htbp]
  \centering
  \includegraphics[width=1.0\textwidth]{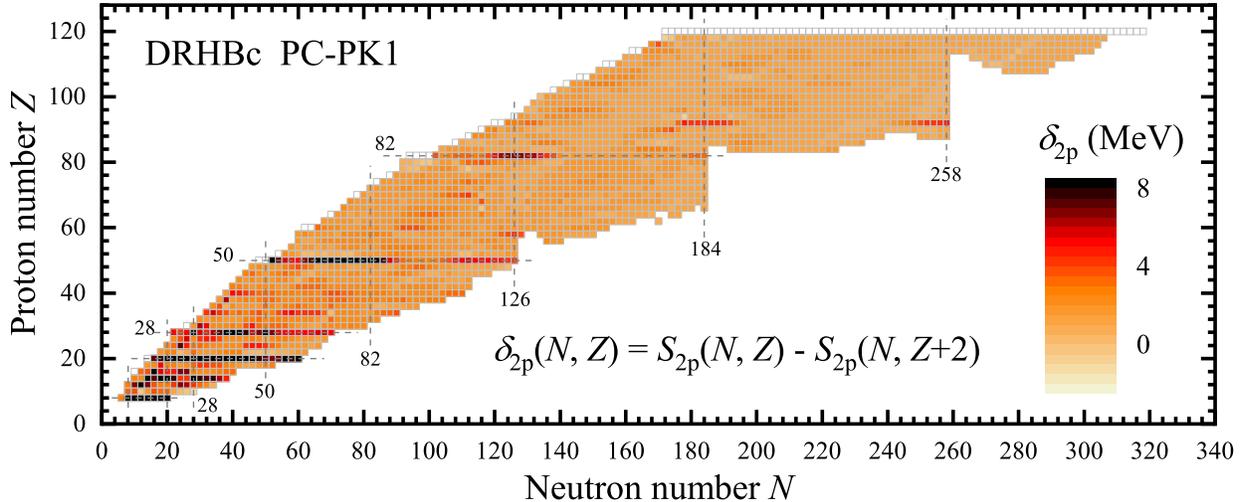}
  \caption{(Color online) Two-proton gaps $\delta_{2p}$ of bound even-even nuclei with $8 \le Z \le 120$ in the DRHBc calculations with PC-PK1 scaled by colors.}
\label{fig5}
\end{figure*}

Figure \ref{fig5} shows the two-proton gaps $\delta_{2p}$ of bound even-even nuclei with $8 \le Z \le 120$ in the DRHBc calculations with PC-PK1.
The traditional magic numbers $Z= 8, 20, 28, 50$ and $82$ are obvious.
The $\delta_{2p}$ for $Z=120$ has not been extracted since the $Z=122$ isotopes have not been calculated in this work.
It is noted that $Z=120$ was predicted to be the next proton magic number in many density functional calculations~\cite{Zhang2005NPA,Li2014PLB}.
Peaks of the $\delta_{2p}$ can be also found for several $Z=14$ isotopes and at $Z = 92$ near the regions of $N\approx 184$ and $258$.
$Z=14$ is regarded as an oblate subshell from the level scheme of the Nilsson model for light nuclei~\cite{Peter1980Book}, and here the $Z=14$ isotopes with large $\delta_{2p}$ are generally oblate in their ground states, see Figs.~\ref{fig9} and \ref{fig10}(a).
However, $Z=92$ has been considered as a pseudo shell in the previous relativistic mean-field calculations~\cite{Geng2006CPL}.

\subsection{Rms radii}

\subsubsection{Charge radii}

\begin{figure*}[htbp]
  \centering
  \includegraphics[width=1.0\textwidth]{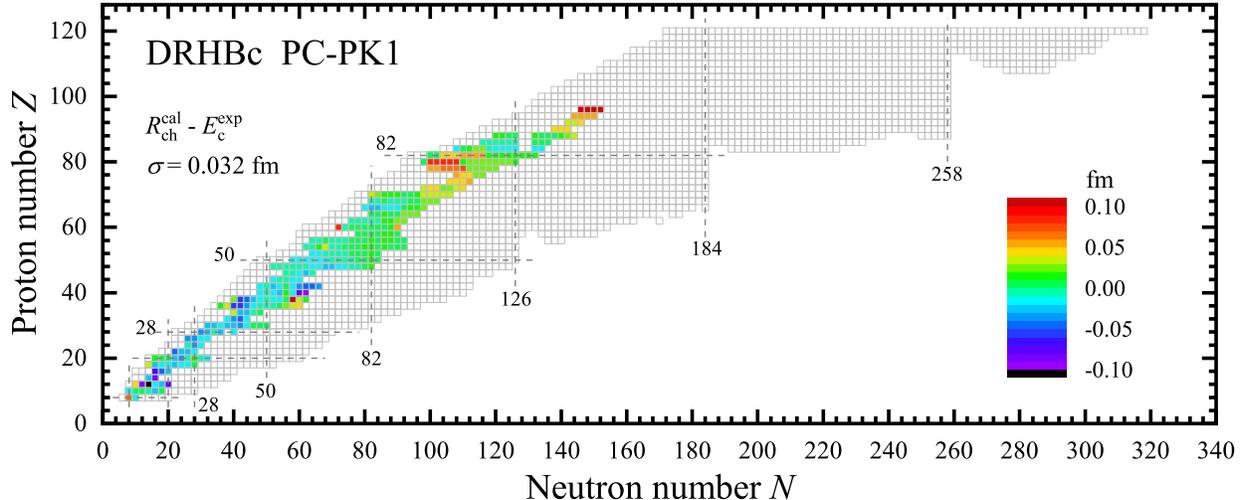}
  \caption{(Color online) For the 369 even-even nuclei ($8 \le Z \le 120$) with charge radius measured, the deviations between the DRHBc calculations with PC-PK1 and the data are scaled by colors.}
\label{fig6}
\end{figure*}

The nuclear charge radius is one of the most important nuclear observables that estimates the size of nuclei.
In Fig.~\ref{fig6}, the deviations of charge radii between the DRHBc calculations and the experimental data are scaled by colors for the 369 even-even nuclei with the charge radius measured~\cite{Angeli2013ADNDT,Li2021ADNDT}.
One can see that the DRHBc calculations provide, in general, a good description of the data, and deviations are mostly in the range of $-0.05$ to $0.05$~fm.
Compared with the rms deviation $\sigma = 0.036$~fm in the RCHB calculations, the rms deviation has been reduced to $\sigma = 0.032$~fm in the DRHBc calculations by the inclusion of deformation.
However, for several light nuclei with $Z<20$, there exist discrepancies between theory and experiment.
For some Pt ($Z=78$) and Hg ($Z=80$) isotopes around $N\approx104$, the DRHBc theory overestimates the experimental values.
This overestimation can be understood, as their ground states are predicted to be prolate with large quadrupole deformation $\beta_2\approx 0.3$ (see Fig.~\ref{fig9}) which is in contradiction with experimental observations~\cite{Pritychenko2016ADNDT}.
Finally, a systematic overestimation is seen for the Cm ($Z=96$) isotopic chain, which is related to the unusual behavior of charge radii in the U-Pu-Cm isotopes~\cite{Angeli2013ADNDT}.
The increase of proton number in an isotonic chain generally leads to an increase of the charge radius, but the experimental charge radii of Cm ($Z=96$) isotopes are found lower than those of the Pu ($Z=94$) and U ($Z=92$) isotones.
This is the only case of such inversion in the nuclear chart and considered to be a challenge for the current density functional theory~\cite{Agbemava2014PRC}.
Machine learning has recently been applied to improve the description accuracy of charge radii based on the density functional theory~\cite{Ma2020PRC} and the empirical formula~\cite{Dong2021arXiv}.

\subsubsection{Neutron radii}

\begin{figure*}[htbp]
  \centering
  \includegraphics[width=1.0\textwidth]{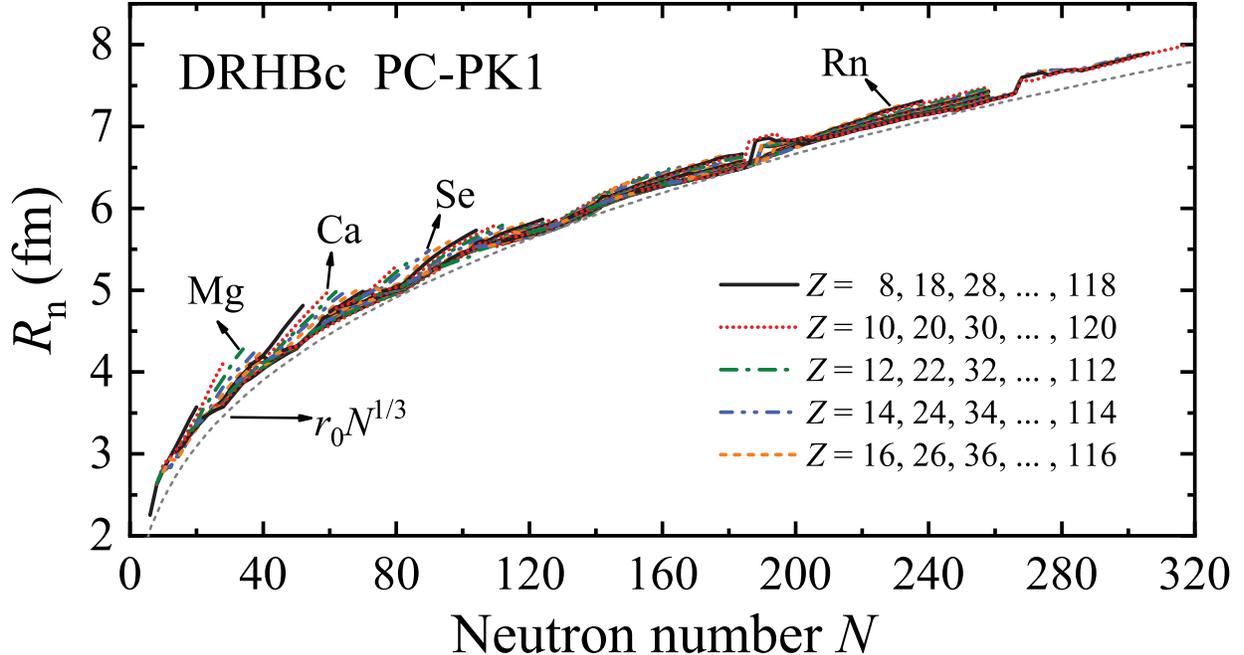}
  \caption{(Color online) Neutron rms radii for even-even nuclei with $8 \le Z \le 120$ from the DRHBc calculations with PC-PK1 as a function of the neutron number. The empirical formula $r_0 N^{1/3}$ with $r_0 = 1.140$~fm determined from the neutron rms radius of $^{208}$Pb is plotted for guidance.}
\label{fig7}
\end{figure*}

In Fig.~\ref{fig7}, the calculated neutron rms radii for even-even nuclei with $8 \le Z \le 120$ are shown as a function of the neutron number.
In addition, the empirical formula $R_n = r_0N^{1/3}$ is shown for guidance with $r_0 = 1.140$~fm determined by the $R_n$ of $^{208}$Pb.
Except for extremely neutron-rich nuclei, the systematic trend of the neutron radii reasonably follows the simple empirical formula.
Pronounced deviations of the DRHBc calculations from the empirical formula can be found in some extremely neutron-rich nuclei near the drip line, e.g., the near-drip-line Mg and Ca nuclei.
Such deviations can be regarded as one of the signals for the halo or giant halo phenomena.
$^{42,44}$Mg have been predicted to be deformed halo nuclei by the DRHBc theory~\cite{Zhou2010PRC(R),Li2012PRC}, and giant halos have been predicted to exist in neutron-rich Ca isotopes near the drip line by the RCHB theory~\cite{Meng2002PRC(R),Zhang2003SciChina,Terasaki2006PRC} together with the Skyrme HFB theory~\cite{Terasaki2006PRC}.

\begin{figure*}[htbp]
  \centering
  \includegraphics[width=1.0\textwidth]{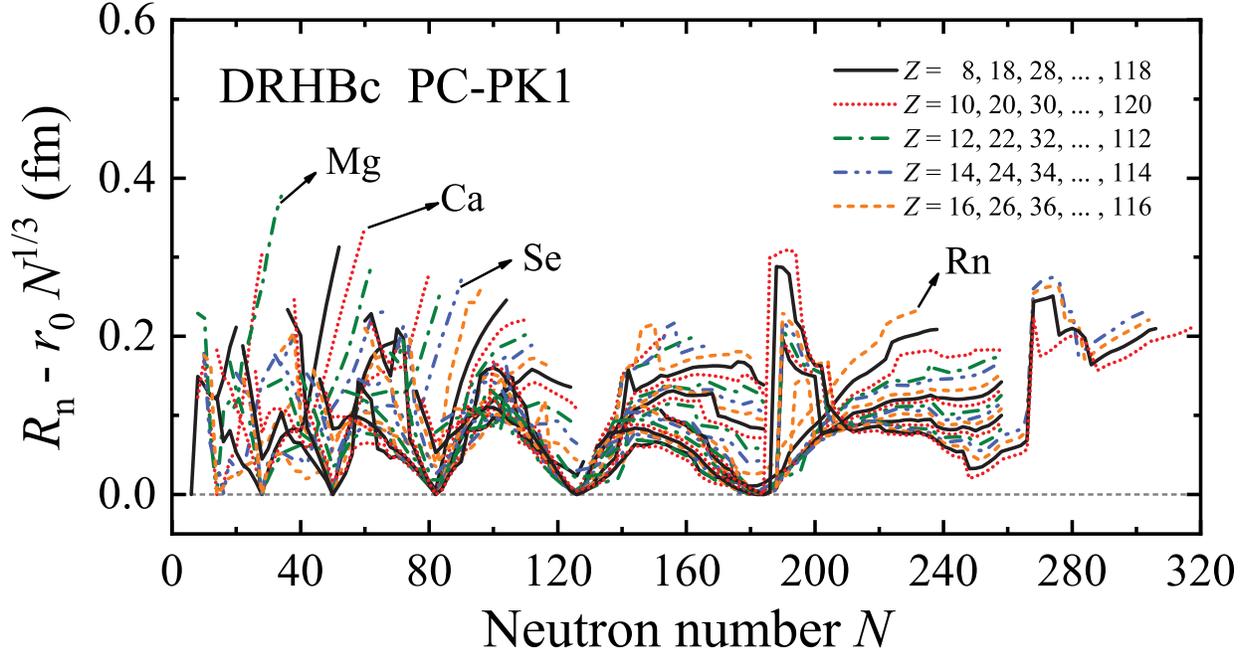}
  \caption{(Color online) Deviations of the DRHBc calculated neutron rms radii from the empirical formula $r_0 N^{1/3}$ for even-even nuclei with $8 \le Z \le 120$, in which the smallest ratio $R_n/N^{1/3}$ is chosen as $r_0$ for each isotopic chain to ensure non-negative values.}
\label{fig8}
\end{figure*}

To achieve deeper insight into the deviations from the empirical formula, the differences of the neutron rms radii between the DRHBc calculations and the empirical formula are shown in Fig.~\ref{fig8} for even-even nuclei with $8 \le Z \le 120$, in which the smallest ratio $R_n/N^{1/3}$ in each isotopic chain is chosen as $r_0$ to ensure non-negative values.
In general, along one isotopic chain, the $R_n-r_0 N^{1/3}$ value first decreases and then increases with the neutron number, and the zero deviation appears near the most stable isotope.
It is found that the zero deviation concentrates near $N=20,28,50,82,126,184$, and $258$, which reflects the effects of neutron shell closures.
There are some sudden increases of $R_n-r_0 N^{1/3}$ at $N\approx 190$ and $270$, which is related to the sudden shape changes from nearly spherical to strongly prolate for $Z\gtrsim 110$ isotopic chains and will be discussed later in Fig.~\ref{fig9}.
Near the neutron drip line, one can see pronounced deviations in some isotopic chains, such as Mg, Ca, Se, and Rn, which indicate the possible existence of the halo or giant halo phenomena.
A further study to explore the possible halo structures in these nuclei needs careful analysis on the neutron separation energies, single-neutron levels near the Fermi surface, and their components, which is beyond the scope of the present work.

\subsection{Quadrupole deformation and potential energy curves}

\subsubsection{Quadrupole deformation}

\begin{figure*}[htbp]
  \centering
  \includegraphics[width=1.0\textwidth]{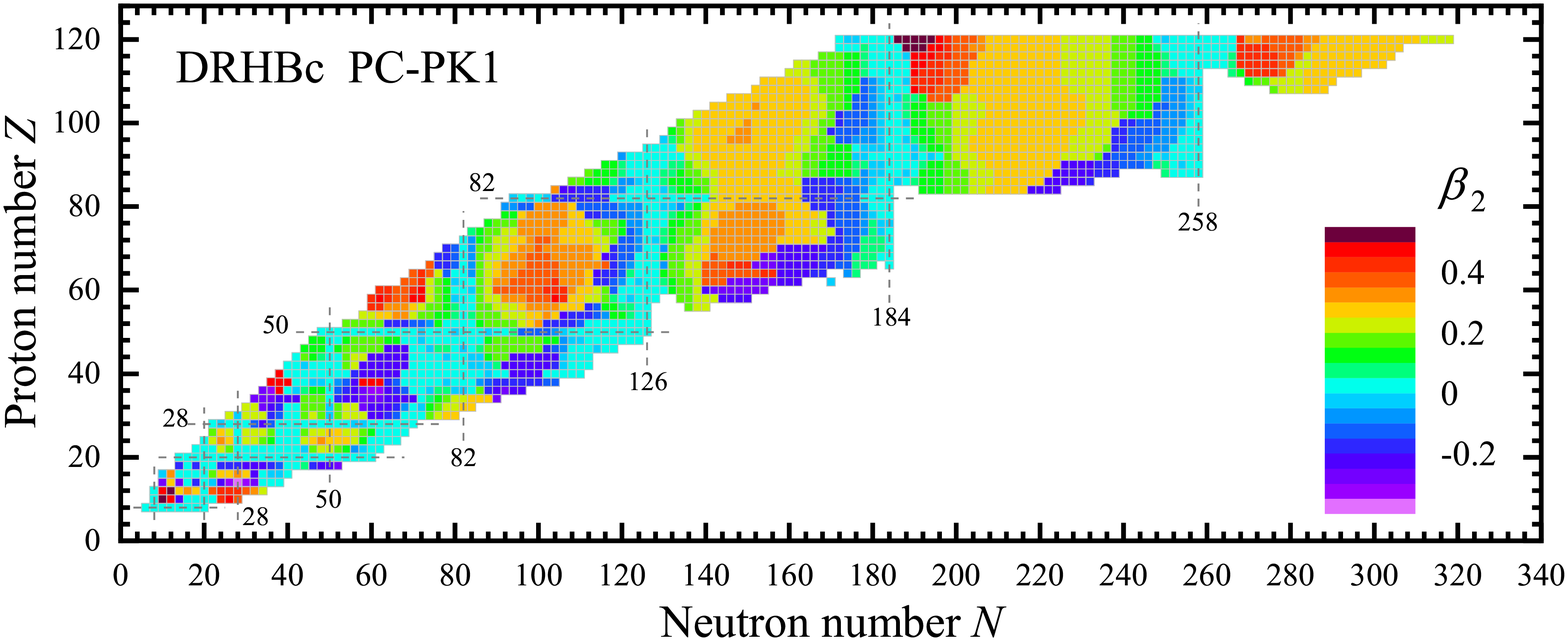}
  \caption{(Color online) Quadrupole deformations from the DRHBc calculation with PC-PK1 for bound even-even nuclei with $8 \le Z \le 120$ scaled by colors.}
\label{fig9}
\end{figure*}

In Fig.~\ref{fig9}, the ground-state quadrupole deformations $\beta_2$ given by the DRHBc calculations are presented.
542 nuclei are spherical near the closed shells.
The number of nuclei with prolate ground-state deformations $\beta_2>0$ is 1547, while that of oblate nuclei with $\beta_2 < 0$ is 478.
In general, the nuclear shape evolution in a isotopic chain between two closed shells is parabolic, i.e., $|\beta_2|$ gradually increases from zero to a certain value in the mid-shell region, and then decreases to zero, for instance, see $Z=58, 82\le N\le 126$.
Some sudden shape changes from $\beta_2\approx 0$ to $\beta_2 \gtrsim 0.4$ are found after the predicted shell closures $N=184$ and $258$ in the superheavy $Z\gtrsim 110$ isotopic chains.
This is due to the competition between the nearly spherical minimum and the prolate one, which is clearly shown in their potential energy curves, e.g., see Fig.~\ref{fig10}(c).
For these nuclei, the prolate minimum becomes the global minimum, i.e., the ground state, after adding a few neutrons to the closed shells $N=184$ and $258$.
Consequently, their rms neutron radii increase dramatically relative to the nearly spherical neighbors, as shown in Fig.~\ref{fig8}.
It should be noted that in some regions with prolate-oblate shape transitions, the triaxial deformation may play an important role.
For example, several cases can be found in the light mass region with $Z<20$, which are also shown more clearly in Fig.~\ref{fig10}(a) and in Fig.~1 of Ref.~\cite{In2021IJMPE}.
Actually, the effects of triaxiality have been investigated in this region, e.g. on the ground states and low-energy collective states in Mg isotopes~\cite{Yao2010Phys.Rev.C44311,Yao2011Phys.Rev.C14308}.
In addition, many nuclei in the heavier mass region with $N=114$ to 120 and $Z=54$ to 78 are predicted to be triaxially deformed in TRHB calculations with PC-PK1~\cite{Yang2021PRC}.

\subsubsection{Evolutions of potential energy curves}

\begin{figure*}[htbp]
  \centering
  \includegraphics[width=1.0\textwidth]{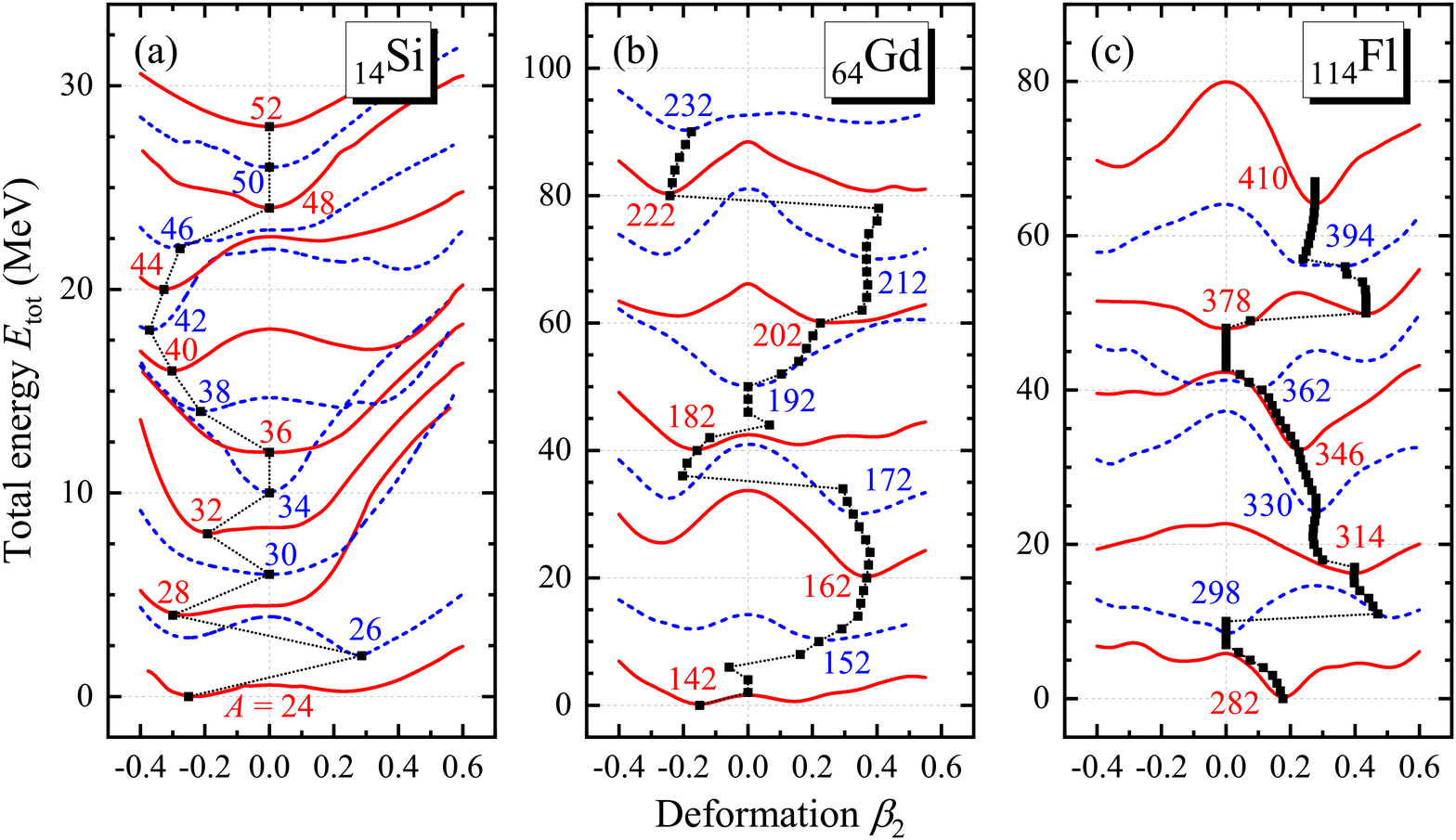}
  \caption{(Color online) Evolution of potential energy curves of $^{24,36,\cdots,52}$Si (a), $^{142,152,\cdots,232}$Gd (b), and $^{282,298,\cdots,410}$Fl (c) from the constrained DRHBc calculations with PC-PK1. For clarity reasons, in each panel, the PEC of the lightest isotope ($^{24}$Si, $^{142}$Gd, and $^{282}$Fl, respectively) is renormalized to its ground state (filled square), and other PECs are shifted upward one by one by $2$ MeV for Si and Gd and by 1 MeV for Fl, per increasing 2 neutrons. The ground-state deformations are denoted by squares.}
\label{fig10}
\end{figure*}

To understand the shape evolution better, Fig.~\ref{fig10} shows the potential energy curves (PECs) of $^{24,26,\cdots,52}$Si, $^{142,152,\cdots,232}$Gd, and $^{282,298,\cdots,410}$Fl as examples.
The evolution of ground-state deformation is also depicted for guidance.
The global minima of these PECs obtained from constrained calculations are consistent with the ground states from unconstrained ones, which guarantees the self-consistency of the present DRHBc calculations.

In the Si isotopic chain, the steep PEC of $^{34}$Si reflects clearly the effects of the closed shell $N=20$.
The maximal $|\beta_2|$ for $^{42}$Si and its rather flat PEC between $\beta_2= -0.3$ and $0.1$ indicate the disappearance of the traditional magic number $N=28$.
In fact, as shown in Fig.~\ref{fig9}, all the $N=28$ even-even isotones from $Z=10$ to $18$ are predicted to be deformed in their ground states, suggesting the collapse of $N=28$ shell closure in this neutron-rich region.
Further exploration of the possible occurrence of new magic numbers $N=32$ and $34$ in Si isotopes requires the analysis of neutron separation energies, single-neutron levels, the energy of the first excited state, etc.

In the Gd isotopic chain, the shape evolution of nearly parabolic type is found between $N=82$ and $126$ as well as between $N=126$ and $184$, while the prolate-oblate shape transitions also exist near the regions of $N = 110$ and $N = 160$.
The TRHB calculations with PC-PK1~\cite{Yang2021PRC} predict several triaxially deformed nuclei in the former region, and a similar prolate-oblate shape transition in the latter one.
Therefore, the shape coexistence in nuclei near the region of $Z\approx64,N\approx160$ is expected.

In the Fl chain, except for a smooth shape evolution with the neutron number approaching the predicted shell closure $N=258$, the above mentioned sudden changes of ground-state deformation from $\beta_2\approx 0$ to $\beta_2 \gtrsim 0.4$ after $N=184$ and $258$ are displayed.
The nearly degenerate spherical and prolate minima are clearly revealed in the PECs of $^{298}$Fl and $^{378}$Fl.
The competition between these two minima after the closed shells leads to the sudden change of ground-state deformation.

\subsection{Neutron density distributions}

\begin{figure*}[htbp]
  \centering
  \includegraphics[width=1.0\textwidth]{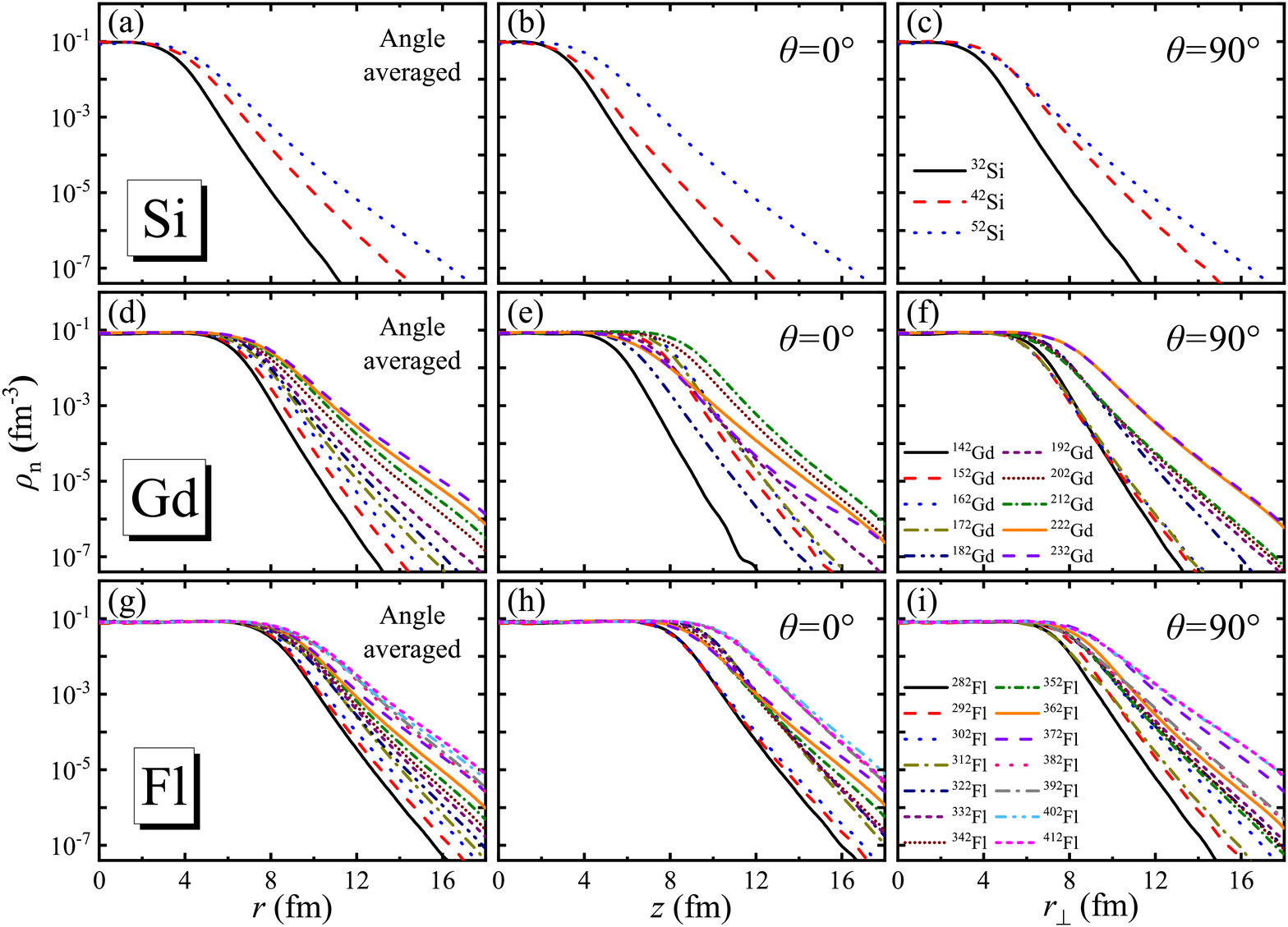}
  \caption{(Color online) Angle averaged neutron density distribution (Angle averaged), the neutron density distribution along the symmetry axis $z$ ($\theta=0^\circ$), and that perpendicular to the symmetry axis with $r_\perp = \sqrt{x^2 + y^2}$ ($\theta=90^\circ$), for selected even-even isotopes $^{32,42,52}$Si (a,b,c), $^{142,152,\cdots,232}$Gd (d,e,f), and $^{282,292,\cdots,412}$Fl (g,h,i) in the DRHBc calculations with PC-PK1.}
\label{fig11}
\end{figure*}

To investigate the evolution of the neutron densities, Fig.~\ref{fig11} shows the neutron density profiles of selected even-even isotopes $^{32,42,52}$Si, $^{142,152,\cdots,232}$Gd, and $^{282,292,\cdots,412}$Fl.
The angle averaged neutron density distributions, the neutron density distributions along the symmetry axis ($\theta=0^\circ$), and those perpendicular to the symmetry axis ($\theta=90^\circ$) are depicted in the left, middle, and right panels, respectively.
As seen in panels (a, d, g), there is a global trend that the angle averaged density distributions are extended further with the neutron number.
The surface expands outward rapidly, while the internal density distribution changes slightly.
Comparing the $\theta=0^\circ$ and $90^\circ$ parts, one can find that the neutron density distributions manifest not only the diffuseness with the increasing neutron number but also the deformation effects.
Due to the oblate deformations of $^{222}$Gd and $^{232}$Gd, their densities along the symmetry axis are less than those of $^{202}$Gd and $^{212}$Gd at $6\lesssim z\le18$~fm, as exhibited in Fig.~\ref{fig11}(e).
On the contrary, their densities perpendicular to the symmetry axis shown in Fig.~\ref{fig11}(f) are obviously larger than others at $z\gtrsim6$~fm.
More specifically, the sudden shape transitions from near spherical to large prolate in Fl isotopes are clearly reflected by the dramatic changes in density distributions along the symmetry axis, i.e., from $^{302}$Fl to $^{312}$Fl and from $^{372}$Fl to $^{382}$Fl, as shown in Fig.~\ref{fig11}(h).
Because of the spherical shape, the density distribution of $^{372}$Fl is angular independent and larger than those of the heavier and prolate deformed $^{382,392}$Fl at $\theta=90^\circ$ in the region of large $r_\perp$.
For the nuclei near the neutron drip line, the diffuse neutron density distribution is an indicator for the neutron skin or halo.
The further distinction between the neutron skin and halo requires a careful analysis of single-neutron levels, their components, and their contributions to the total neutron density, as was done for the neutron-rich $^{214}$Nd nucleus in Ref.~\cite{Zhang2020PRC}.

\subsection{Neutron potential and diffuseness}

\begin{figure*}[htbp]
  \centering
  \includegraphics[width=1.0\textwidth]{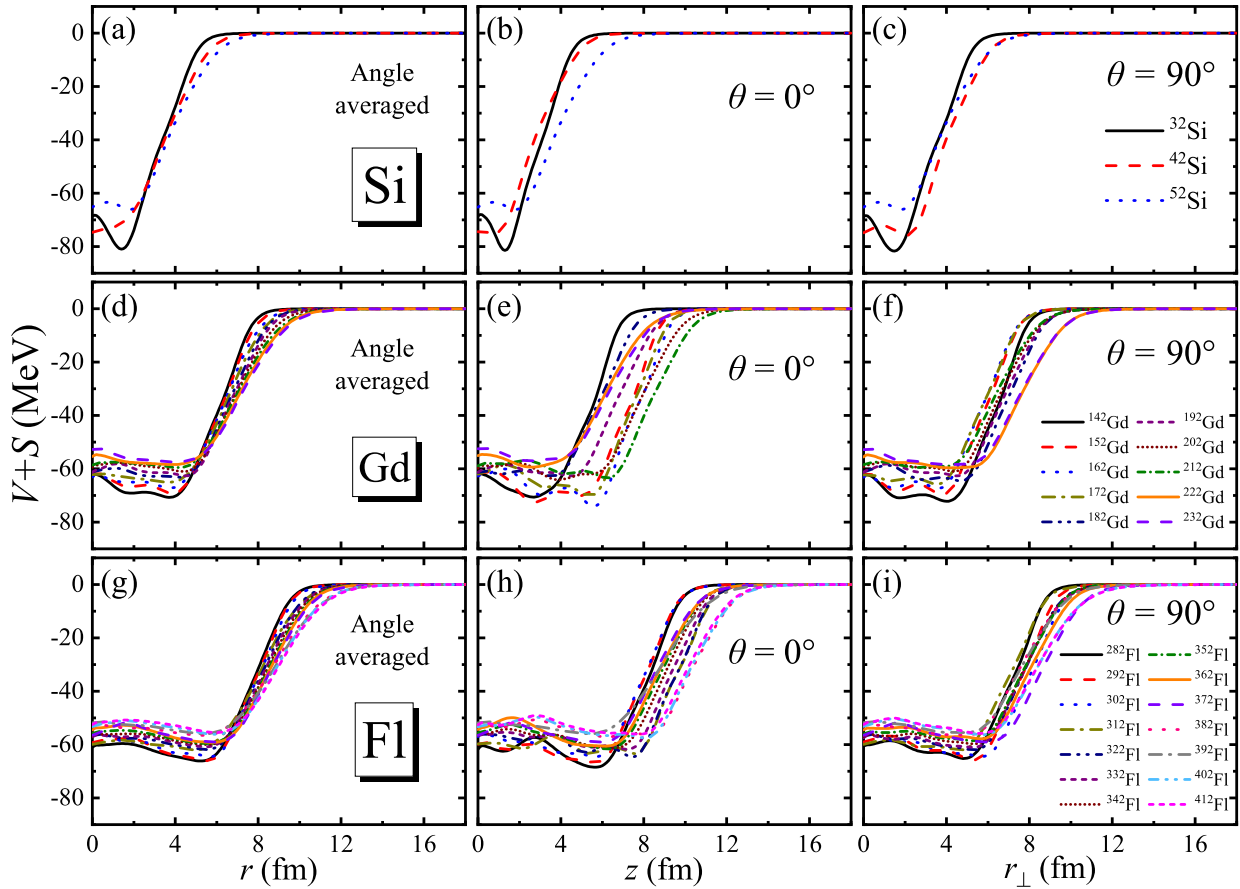}
  \caption{(Color online) Same as Fig.~\ref{fig11}, but for the neutron mean-field potential $V+S$.}
\label{fig12}
\end{figure*}

To examine the isospin dependence of the mean-field potentials, the neutron vector plus scalar potentials $V(\bm r) + S(\bm r)$ for selected even-even isotopes $^{32,42,52}$Si, $^{142,152,\cdots,232}$Gd, and $^{282,292,\cdots,412}$Fl are reported in Fig.~\ref{fig12}, in terms of the angle averaged potential and those along ($\theta=0^\circ$) and perpendicular to ($\theta=90^\circ$) the symmetry axis.
Generally, the depth of the potential rises with the neutron number, except for some fluctuations due to the shell structure and deformation effects.
Intriguingly, a central depletion is found in the neutron potential for $^{32}$Si in upper panels, which might be related to its neutron density, since the potentials and densities are connected in a self-consistent way.
By plotting the neutron density of $^{32}$Si in the linear coordinate, a similar central depletion can be found.
By the further analysis of the components of the single-neutron levels, the level mainly occupied by the two valance neutrons has about $93\%$ $1d_{3/2}$ and $7\%$ $1d_{5/2}$ components, while the level just below it, which should be $2s_{1/2}$ in the spherical limit, is now mixed with around $40\%$ $1d$ components due to the deformation effects.
The loss of $2s$ component leads to the central depletion in the neutron density and is to some extent responsible for the central depletion in the potential.

At the surface, the potentials extend outward and the diffuseness increases generally with the neutron number.
Consequently, the potentials for the nuclei near the neutron drip line become highly diffused.
This will dramatically influence the weakly bound orbitals and resonant ones near the threshold, especially those with low-$l$ components, and may lead to halos in drip-line nuclei.

\subsection{Pairing energies}

\begin{figure*}[htbp]
  \centering
  \includegraphics[width=1.0\textwidth]{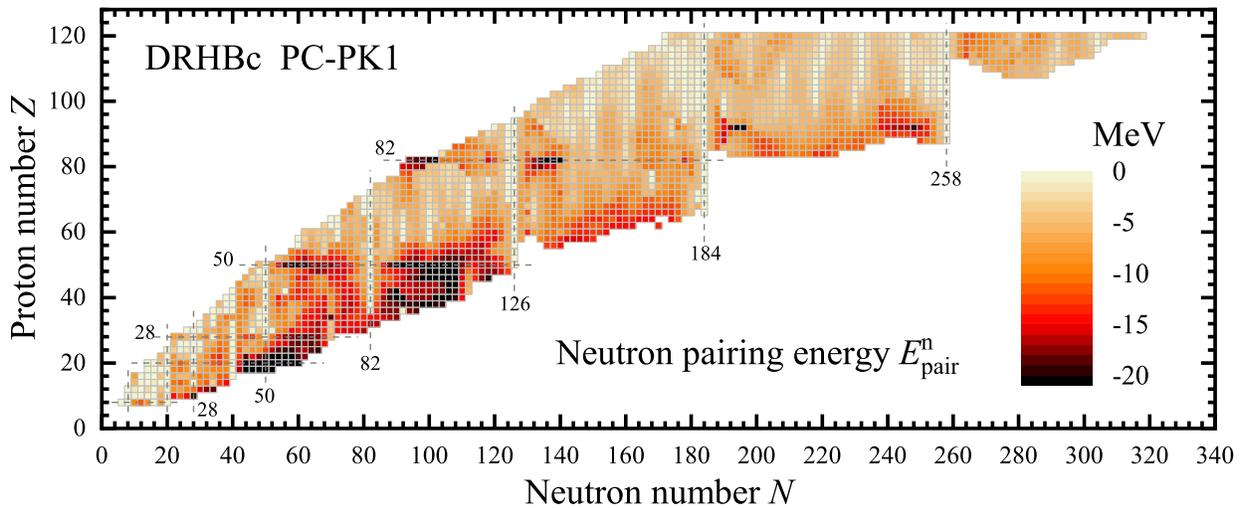}
  \caption{(Color online) Neutron pairing energies of bound even-even nuclei with $8 \le Z \le 120$ in the DRHBc calculations with PC-PK1 scaled by colors.}
\label{fig13}
\end{figure*}

To examine the pairing correlations globally, we investigate the pairing energies for even-even nuclei in the DRHBc calculations with PC-PK1.
In Fig.~\ref{fig13}, the neutron pairing energies $E_{\mathrm{pair}}^{n}$ of bound even-even nuclei with $8 \le Z \le 120$ are scaled by colors.
One can see that the neutron pairing energies approach zero or even vanish for the nuclei near the closed shells $N = 8, 20, 28, 50, 82$ and $126$, and in general the maximum values appear in the middle of the shells.
The neutron pairing energies vanish at $N = 184$ and $258$, which agrees with the above discussion for shell closures by $S_{2n}$ and $\delta_{2n}$.
The very small $E_{\mathrm{pair}}^{n}$ for light nuclei with $Z<20$ and $N<20$ may be due to the low density of single-particle levels.
In other regions, the vanishing $E_{\mathrm{pair}}^{n}$ along a part of an isotonic chain may be related to spherical or deformed subshells, e.g. $N = 40$ in the $Z \approx 20$ region and $N = 162$ in the $Z \approx 110$ region, as shown in Fig.~\ref{fig4} for $\delta_{2n}$.
Finally, for the isotopic chains with $Z$ around 10, 20, and 32, $E_{\mathrm{pair}}^{n}$ for the nuclei near the neutron drip line are not small, which suggests the disappearance of the traditional neutron magic numbers 28, 50, and 82 in these neutron-rich exotic nuclei.
This is also consistent with the weak $\delta_{2n}$ in these regions in Fig.~\ref{fig4}.

\begin{figure*}[htbp]
  \centering
  \includegraphics[width=1.0\textwidth]{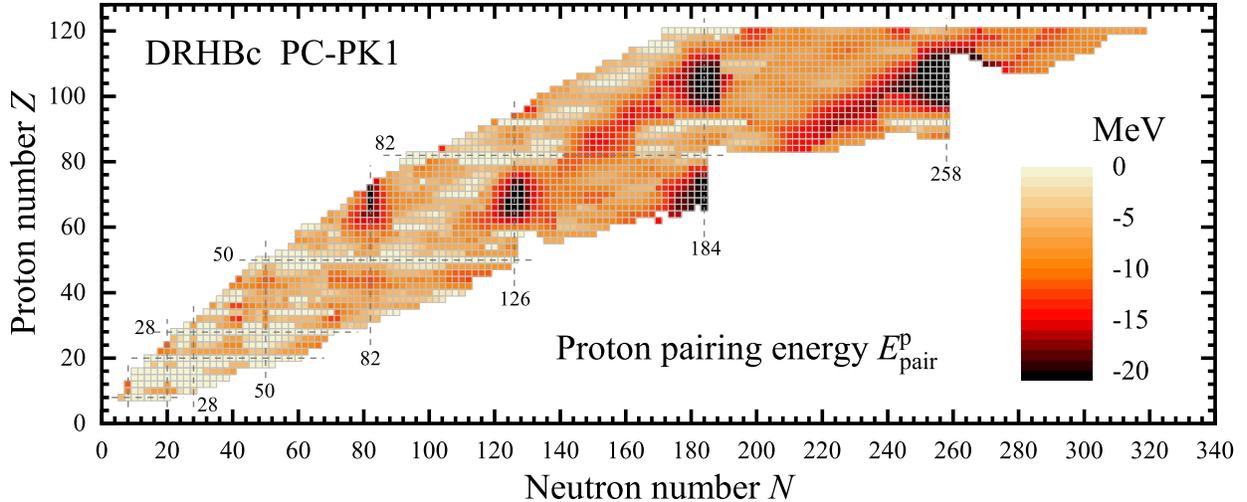}
  \caption{(Color online) Proton pairing energies of bound even-even nuclei with $8 \le Z \le 120$ in the DRHBc calculations with PC-PK1 scaled by colors.}
\label{fig14}
\end{figure*}

The proton pairing energies $E_{\mathrm{pair}}^{p}$ of bound even-even nuclei with $8 \le Z \le 120$ are presented in Fig.~\ref{fig14}.
Many features similar to the neutron pairing energies can be found, such as the nearly vanishing $E_{\mathrm{pair}}^{p}$ near the closed shells $Z = 8, 20, 28, 50,$ and $82$ and the very small $E_{\mathrm{pair}}^{p}$ for light nuclei with $Z<20$ and $N<20$.
$E_{\mathrm{pair}}^{p}$ also approach zero or even vanish for the nuclei near the proton drip line of $Z=120$, which is predicted to be the next proton magic number by many relativistic density functionals~\cite{Zhang2005NPA,Li2014PLB}.
There are also small $E_{\mathrm{pair}}^{p}$ along a part of an isotopic chain, e.g. near $N=60$ at $Z=34$ and near $N=184, 258$ at $Z=92$.
The former might indicate $Z=34$ as a deformed proton subshell, whose confirmation requires a further investigation of the evolution of single-proton levels with the deformation.
The latter corresponds to $Z=92$, and from Fig.~\ref{fig9} these nuclei with small $E_{\mathrm{pair}}^{p}$ are spherical.
However, no experimental evidence indicates $Z=92$ as a magic number, and it has been considered as a pseudo shell in relativistic density functionals TMA, NL3, PKDD, and DD-ME2~\cite{Geng2006CPL}.

\subsection{Alpha decay energies}

\begin{figure*}[htbp]
  \centering
  \includegraphics[width=1.0\textwidth]{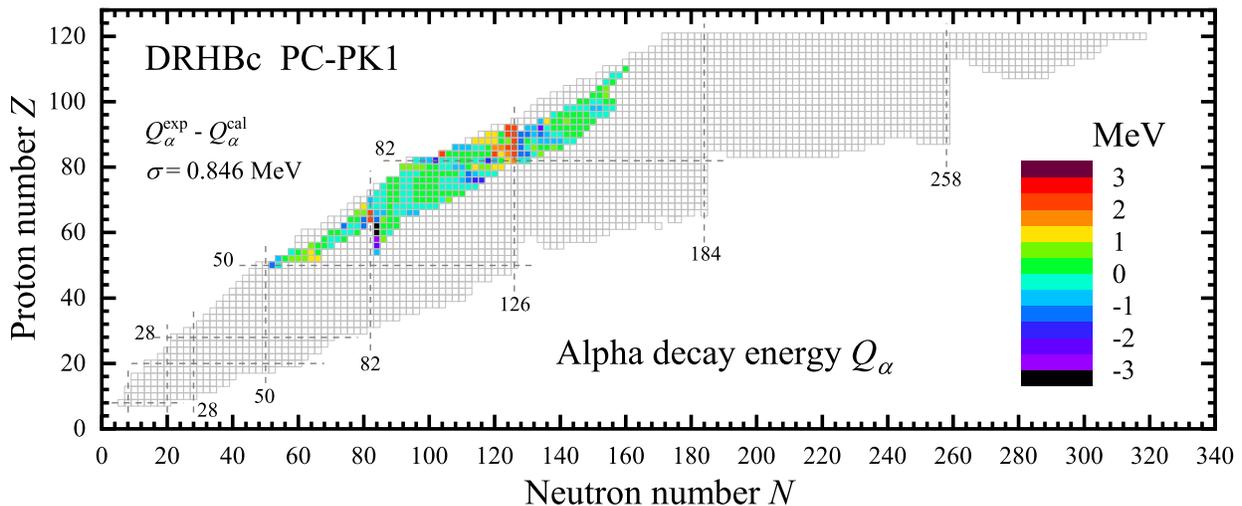}
  \caption{(Color online) $\alpha$ decay energy $Q_\alpha$ differences between the data~\cite{AME2020(3)} and the DRHBc calculations with PC-PK1 for nuclei with $Q_\alpha > 0$, scaled by colors.}
\label{fig15}
\end{figure*}

By using the binding energies in the DRHBc calculations with PC-PK1, we can extract the $\alpha$ decay energies $Q_\alpha$ via,
\begin{equation}
Q_\alpha = E_{\mathrm{b}}(Z-2,N-2)+E_{\mathrm{b}}(2,2)-E_{\mathrm{b}}(Z,N).
\end{equation}
In Fig.~\ref{fig15}, the differences of $Q_\alpha$ between our calculations and the data~\cite{AME2020(3)} for nuclei with $Q_\alpha>0$ are scaled by colors.
The number of data is 296 and the obtained rms deviation is $\sigma=0.846$~MeV, which is comparable to some other representative mass tables, such as $\sigma=0.901$~MeV in RHB + DD-PC1 calculations and $\sigma=0.939$~MeV in RHB + DD-ME2 calculations~\cite{Agbemava2014PRC}.
Compared with $\sigma\approx2$~MeV in the RCHB + PC-PK1 calculations~\cite{Zhang2016CPC}, the description of $Q_\alpha$ has been improved by the inclusion of deformation degrees of freedom.
In the TRHB + PC-PK1 calculations, the description of $Q_\alpha$ is improved a lot from $\sigma=0.989$~MeV to $0.552$~MeV by including the beyond-mean-field correlation energies from the collective Hamiltonian method, which is suitable for both nearly spherical and deformed nuclei.
We note that many large deviations in the DRHBc calculations appear near the shell closures $N=82$ and $126$, which is caused by the overestimation of the binding energies for the nuclei with $N=82$ and $126$ and the underestimation for their nearly spherical neighbors, as shown in Fig.~\ref{fig1}.
The strength of shell closures is generally overestimated by the CDFT~\cite{Meng2016Book}, and the cranking approximation used to obtain the rotational correction energy in the DRHBc calculations is not suitable for nearly spherical nuclei.
Therefore, the ongoing work implementing the collective Hamiltonian method in the DRHBc theory is expected to improve the description of $Q_\alpha$, especially near the closed shells.
Similar to the study on the $\alpha$ decay energies from the RCHB calculations in Ref.~\cite{Zhang2016CPC}, the DRHBc calculated $Q_\alpha$ can be investigated systematically to explore the shell effects and predict new magic numbers.
Furthermore, combining the density functional theory with the Wentzel-Kramers-Brillouin (WKB) method, the $\alpha$ decay half-lives can be estimated without introducing any adjustable parameter~\cite{Pei2007PRC}.
It would be interesting to systematically study $\alpha$ decay energies and estimate the $\alpha$ decay half-lives with the DRHBc theory in the near future.

\section{Summary}\label{summary}

In summary, we have performed systematic studies of all even-even nuclei from $Z=8$ to $Z=120$ by using the DRHBc theory with the density functional PC-PK1.
The calculated binding energies, two-nucleon separation energies, rms radii of neutron, proton, matter, and charge distributions, quadrupole deformations, and neutron and proton Fermi surfaces are tabulated.

With the deformation degrees of freedom and continuum effects included, there are 2583 even-even nuclei in total, which are predicted to be bound from $Z = 8$ to $Z = 120$.
For the experimental binding energies of 637 even-even nuclei, the description accuracy is $1.518$ MeV, which is one of the highest accuracies in the microscopic description of nuclear masses, and shows the importance of deformation effects by comparing with the spherical RCHB mass table.
The proton drip line predicted by the DRHBc theory is close to those predicted by other models, because on the proton-rich side the continuum effects are not pronounced owing to the Coulomb barrier.
By contrast, the locations of neutron drip lines from various models obviously differ from each other, especially in the heavy mass region.
Compared with the spherical RCHB calculations with PC-PK1, it is found that, the drip line extends to the more neutron-rich side if the deformation increases towards the drip line, and \emph{vice versa}~\cite{In2021IJMPE,Pan2021PRC}.
Compared with the TRHB calculations with PC-PK1, almost no even-even nuclei near the neutron drip line are triaxially deformed, and the continuum effects may be responsible for the extension of the neutron drip line in several isotopic chains.
Compared with other density functional calculations, the DRHBc calculations with PC-PK1 generally predict a more extended neutron drip line with a few exceptions.
This is mainly because of the proper treatment of the continuum in the DRHBc theory and the adopted density functional.
In addition, there exist some bound nuclei beyond the primary neutron drip line in several regions, forming peninsulas of stability adjacent to the nuclear mainland.
The deformation plays a decisive role in the formation of these stability peninsulas, and the pairing correlations and continuum effects also influence them in a self-consistent way~\cite{Pan2021PRC,Zhang2021PRC(L),He2021CPC}.

The two-neutron and two-proton separation energies are extracted and systematically discussed over the nuclear chart.
The traditional magic numbers are reproduced by the two-nucleon separation energies and two-nucleon gaps, and new magic numbers $N=184$ and $258$ are predicted.
Furthermore, possible spherical or deformed subshells are indicated by the two-nucleon gaps.

For the experimental charge radii of 369 even-even nuclei, the description accuracy is $0.032$ fm, which has been improved by the inclusion of deformation in comparison with the spherical RCHB calculations.
The comparison of the neutron rms radii of even-even nuclei with $8 \le Z \le 120$ between the DRHBc calculations and the empirical formula are performed.
The systematic trend of the neutron radii in the DRHBc calculations follows the simple empirical formula quite well with a few exceptions.
Some sudden increases of neutron radii reflect the deformation effects, e.g., the sudden shape change from $\beta_2\approx 0$ to $\beta_2 \gtrsim 0.4$ in $Z\gtrsim 110$ isotopic chains.
The pronounced deviations of the DRHBc calculations from the empirical formula for some extremely neutron-rich nuclei might be a signal for the halo or giant halo phenomena.

The quadrupole deformations for even-even nuclei with $8 \le Z \le 120$ given by the DRHBc calculations are presented and discussed.
The evolution of ground-state deformations is understood with the help of the potential energy curves from constrained calculations.
For some regions with prolate-oblate shape transitions, the possible triaxial deformation and shape coexistence are discussed in combination with the results from TRHB calculations with PC-PK1~\cite{Yang2021PRC}.

The angle averaged neutron density distributions as well as those along and perpendicular to the symmetry axis for selected even-even Si, Gd, and Fl isotopes are depicted.
The diffuseness with the increasing neutron number and the deformation effects on the neutron density distributions are discussed.
Similar to the neutron density, the neutron mean-field potentials for selected even-even Si, Gd, and Fl isotopes are discussed.
In general, the depth of the potential rises with the neutron number, except for some fluctuations due to the shell structure and deformation effects.
At the surface, the potentials extend outward and the diffuseness increases generally with the neutron number.
Taking $^{32}$Si as an example, the central depletion in its neutron potential is investigated by analyzing the components of single-neutron levels near the Fermi surface.

The pairing energies of the even-even nuclei over the nuclear landscape are presented.
The pairing energies approach zero or even vanish for the nuclei near the closed shells, and in general they have maximum values for nuclei in the middle of the shells.
New magic numbers $N=184$ and $258$ and $Z=120$ for superheavy nuclei and some subshells are predicted.
The collapse of traditional shell closures near the neutron drip lines of some isotopic chains is discussed.

Finally, the $\alpha$ decay energies $Q_\alpha$ are extracted for the nuclei with $Q_\alpha>0$ and compared with 296 available data.
The obtained rms deviation is $\sigma = 0.846$~MeV, which has been reduced by deformation effects in comparison with the RCHB calculations~\cite{Zhang2016CPC}.
A future work combining the DRHBc theory and WKB method to estimate the $\alpha$ decay half-lives is expected.

The successful exploration of the even-even nuclei in the nuclear chart by using the DRHBc theory with the relativistic density functional PC-PK1 demonstrates the importance of the simultaneous consideration of the deformation and continuum effects.
Works in constructing a DRHBc mass table for odd-mass and odd-odd nuclei are in progress.

\begin{acknowledgments}

This work was partly supported by the National Natural Science Foundation of China (Grants No.~11935003, No.~11875075, No.~11975031, No.~12141501, No.~12070131001, No.~12075085, No.~12047568, No.~11790325, No.~11790323, No.~11735003, No.~11975041, No.~12047503, No.~11975237, and No.~11961141004), the National Key R\&D Program of China (Contracts No.~2017YFE0116700 and No.~2018YFA0404400), the State Key Laboratory of Nuclear Physics and Technology, Peking University (Grant No.~NPT2020ZZ01), the Strategic Priority Research Program of Chinese Academy of Sciences (Grants No.~XDB34010000 and No.~XDPB15), the High-performance Computing Platform of Peking University, the High-performance Computing Platform of Anhui University, the Rare Isotope Science Project of Institute for Basic Science, funded by Ministry of Science and ICT (MSICT), and National Research Foundation of Korea (2013M7A1A1075764).
This work was partly supported by the National Research Foundation of Korea (Grant Nos.~NRF-2018R1D1A1B05048026, NRF-2020R1A2C3006177, NRF-2020K1A3A7A09080134, NRF-2021R1F1A1060066, and NRF-2021R1A6A1A03043957) and the National Supercomputing Center with supercomputing resources including technical support (KSC-2020-CRE-0329, KSC-2021-CRE-0126, and KSC-2021-CRE-0272).
YBC and CHL were supported by  National Research Foundation of Korea (NRF) grants funded by the Korea government (Ministry of Science and ICT and Ministry of Education) (No.~2016R1A5A1013277 and No.~2018R1D1A1B07048599).
QZ was supported by Institute for Basic Science (Grant IBS-R031-D1).
The results from ITP side are obtained on the High-performance Computing Cluster of ITP-CAS and the ScGrid of the Supercomputing Center, Computer Network Information Center of Chinese Academy of Sciences.

\end{acknowledgments}


\begin{thebibliography}{133}%
\makeatletter
\providecommand \@ifxundefined [1]{%
 \@ifx{#1\undefined}
}%
\providecommand \@ifnum [1]{%
 \ifnum #1\expandafter \@firstoftwo
 \else \expandafter \@secondoftwo
 \fi
}%
\providecommand \@ifx [1]{%
 \ifx #1\expandafter \@firstoftwo
 \else \expandafter \@secondoftwo
 \fi
}%
\providecommand \natexlab [1]{#1}%
\providecommand \enquote  [1]{``#1''}%
\providecommand \bibnamefont  [1]{#1}%
\providecommand \bibfnamefont [1]{#1}%
\providecommand \citenamefont [1]{#1}%
\providecommand \href@noop [0]{\@secondoftwo}%
\providecommand \href [0]{\begingroup \@sanitize@url \@href}%
\providecommand \@href[1]{\@@startlink{#1}\@@href}%
\providecommand \@@href[1]{\endgroup#1\@@endlink}%
\providecommand \@sanitize@url [0]{\catcode `\\12\catcode `\$12\catcode
  `\&12\catcode `\#12\catcode `\^12\catcode `\_12\catcode `\%12\relax}%
\providecommand \@@startlink[1]{}%
\providecommand \@@endlink[0]{}%
\providecommand \url  [0]{\begingroup\@sanitize@url \@url }%
\providecommand \@url [1]{\endgroup\@href {#1}{\urlprefix }}%
\providecommand \urlprefix  [0]{URL }%
\providecommand \Eprint [0]{\href }%
\providecommand \doibase [0]{http://dx.doi.org/}%
\providecommand \selectlanguage [0]{\@gobble}%
\providecommand \bibinfo  [0]{\@secondoftwo}%
\providecommand \bibfield  [0]{\@secondoftwo}%
\providecommand \translation [1]{[#1]}%
\providecommand \BibitemOpen [0]{}%
\providecommand \bibitemStop [0]{}%
\providecommand \bibitemNoStop [0]{.\EOS\space}%
\providecommand \EOS [0]{\spacefactor3000\relax}%
\providecommand \BibitemShut  [1]{\csname bibitem#1\endcsname}%
\let\auto@bib@innerbib\@empty
\bibitem [{\citenamefont {Zhan}\ \emph {et~al.}(2010)\citenamefont {Zhan},
  \citenamefont {Xu}, \citenamefont {Xiao}, \citenamefont {Xia}, \citenamefont
  {Zhao},\ and\ \citenamefont {Yuan}}]{Zhan2010NPA}%
  \BibitemOpen
  \bibfield  {author} {\bibinfo {author} {\bibfnamefont {W.}~\bibnamefont
  {Zhan}}, \bibinfo {author} {\bibfnamefont {H.}~\bibnamefont {Xu}}, \bibinfo
  {author} {\bibfnamefont {G.}~\bibnamefont {Xiao}}, \bibinfo {author}
  {\bibfnamefont {J.}~\bibnamefont {Xia}}, \bibinfo {author} {\bibfnamefont
  {H.}~\bibnamefont {Zhao}}, \ and\ \bibinfo {author} {\bibfnamefont
  {Y.}~\bibnamefont {Yuan}},\ }\href {\doibase 10.1016/j.nuclphysa.2010.01.126}
  {\bibfield  {journal} {\bibinfo  {journal} {Nucl. Phys. A}\ }\textbf
  {\bibinfo {volume} {834}},\ \bibinfo {pages} {694c } (\bibinfo {year}
  {2010})}\BibitemShut {NoStop}%
\bibitem [{\citenamefont {Motobayashi}(2010)}]{Motobayashi2010NPA}%
  \BibitemOpen
  \bibfield  {author} {\bibinfo {author} {\bibfnamefont {T.}~\bibnamefont
  {Motobayashi}},\ }\href {\doibase 10.1016/j.nuclphysa.2010.01.128} {\bibfield
   {journal} {\bibinfo  {journal} {Nucl. Phys. A}\ }\textbf {\bibinfo {volume}
  {834}},\ \bibinfo {pages} {707c } (\bibinfo {year} {2010})}\BibitemShut
  {NoStop}%
\bibitem [{\citenamefont {Tshoo}\ \emph {et~al.}(2013)\citenamefont {Tshoo},
  \citenamefont {Kim}, \citenamefont {Kwon}, \citenamefont {Woo}, \citenamefont
  {Kim}, \citenamefont {Kim}, \citenamefont {Kang}, \citenamefont {Park},
  \citenamefont {Park}, \citenamefont {Yoon}, \citenamefont {Kim},
  \citenamefont {Lee}, \citenamefont {Seo}, \citenamefont {Hwang},
  \citenamefont {Yun}, \citenamefont {Jeon},\ and\ \citenamefont
  {Kim}}]{Tshoo2013NIMPRB}%
  \BibitemOpen
  \bibfield  {author} {\bibinfo {author} {\bibfnamefont {K.}~\bibnamefont
  {Tshoo}}, \bibinfo {author} {\bibfnamefont {Y.}~\bibnamefont {Kim}}, \bibinfo
  {author} {\bibfnamefont {Y.}~\bibnamefont {Kwon}}, \bibinfo {author}
  {\bibfnamefont {H.}~\bibnamefont {Woo}}, \bibinfo {author} {\bibfnamefont
  {G.}~\bibnamefont {Kim}}, \bibinfo {author} {\bibfnamefont {Y.}~\bibnamefont
  {Kim}}, \bibinfo {author} {\bibfnamefont {B.}~\bibnamefont {Kang}}, \bibinfo
  {author} {\bibfnamefont {S.}~\bibnamefont {Park}}, \bibinfo {author}
  {\bibfnamefont {Y.-H.}\ \bibnamefont {Park}}, \bibinfo {author}
  {\bibfnamefont {J.}~\bibnamefont {Yoon}}, \bibinfo {author} {\bibfnamefont
  {J.}~\bibnamefont {Kim}}, \bibinfo {author} {\bibfnamefont {J.}~\bibnamefont
  {Lee}}, \bibinfo {author} {\bibfnamefont {C.}~\bibnamefont {Seo}}, \bibinfo
  {author} {\bibfnamefont {W.}~\bibnamefont {Hwang}}, \bibinfo {author}
  {\bibfnamefont {C.}~\bibnamefont {Yun}}, \bibinfo {author} {\bibfnamefont
  {D.}~\bibnamefont {Jeon}}, \ and\ \bibinfo {author} {\bibfnamefont
  {S.}~\bibnamefont {Kim}},\ }\href {\doibase 10.1016/j.nimb.2013.05.058}
  {\bibfield  {journal} {\bibinfo  {journal} {Nucl. Instrum. Methods Phys. Res.
  B}\ }\textbf {\bibinfo {volume} {317}},\ \bibinfo {pages} {242} (\bibinfo
  {year} {2013})}\BibitemShut {NoStop}%
\bibitem [{\citenamefont {Sturm}\ \emph {et~al.}(2010)\citenamefont {Sturm},
  \citenamefont {Sharkov},\ and\ \citenamefont {Stcker}}]{Sturm2010NPA}%
  \BibitemOpen
  \bibfield  {author} {\bibinfo {author} {\bibfnamefont {C.}~\bibnamefont
  {Sturm}}, \bibinfo {author} {\bibfnamefont {B.}~\bibnamefont {Sharkov}}, \
  and\ \bibinfo {author} {\bibfnamefont {H.}~\bibnamefont {Stcker}},\ }\href
  {\doibase 10.1016/j.nuclphysa.2010.01.124} {\bibfield  {journal} {\bibinfo
  {journal} {Nucl. Phys. A}\ }\textbf {\bibinfo {volume} {834}},\ \bibinfo
  {pages} {682c } (\bibinfo {year} {2010})}\BibitemShut {NoStop}%
\bibitem [{\citenamefont {Gales}(2010)}]{Gales2010NPA}%
  \BibitemOpen
  \bibfield  {author} {\bibinfo {author} {\bibfnamefont {S.}~\bibnamefont
  {Gales}},\ }\href {\doibase 10.1016/j.nuclphysa.2010.01.130} {\bibfield
  {journal} {\bibinfo  {journal} {Nucl. Phys. A}\ }\textbf {\bibinfo {volume}
  {834}},\ \bibinfo {pages} {717c } (\bibinfo {year} {2010})}\BibitemShut
  {NoStop}%
\bibitem [{\citenamefont {Thoennessen}(2010)}]{Thoennessen2010NPA}%
  \BibitemOpen
  \bibfield  {author} {\bibinfo {author} {\bibfnamefont {M.}~\bibnamefont
  {Thoennessen}},\ }\href {\doibase 10.1016/j.nuclphysa.2010.01.125} {\bibfield
   {journal} {\bibinfo  {journal} {Nucl. Phys. A}\ }\textbf {\bibinfo {volume}
  {834}},\ \bibinfo {pages} {688c } (\bibinfo {year} {2010})}\BibitemShut
  {NoStop}%
\bibitem [{\citenamefont {Baumann}\ \emph {et~al.}(2012)\citenamefont
  {Baumann}, \citenamefont {Spyrou},\ and\ \citenamefont
  {Thoennessen}}]{Baumann2012RPP}%
  \BibitemOpen
  \bibfield  {author} {\bibinfo {author} {\bibfnamefont {T.}~\bibnamefont
  {Baumann}}, \bibinfo {author} {\bibfnamefont {A.}~\bibnamefont {Spyrou}}, \
  and\ \bibinfo {author} {\bibfnamefont {M.}~\bibnamefont {Thoennessen}},\
  }\href
  {https://iopscience.iop.org/article/10.1088/0034-4885/75/3/036301/meta}
  {\bibfield  {journal} {\bibinfo  {journal} {Rep. Prog. Phys.}\ }\textbf
  {\bibinfo {volume} {75}},\ \bibinfo {pages} {036301} (\bibinfo {year}
  {2012})}\BibitemShut {NoStop}%
\bibitem [{\citenamefont {Aprahamian}\ \emph {et~al.}(2005)\citenamefont
  {Aprahamian}, \citenamefont {Langanke},\ and\ \citenamefont
  {Wiescher}}]{Aprahamian2005PPNP}%
  \BibitemOpen
  \bibfield  {author} {\bibinfo {author} {\bibfnamefont {A.}~\bibnamefont
  {Aprahamian}}, \bibinfo {author} {\bibfnamefont {K.}~\bibnamefont
  {Langanke}}, \ and\ \bibinfo {author} {\bibfnamefont {M.}~\bibnamefont
  {Wiescher}},\ }\href {\doibase 10.1016/j.ppnp.2004.09.002} {\bibfield
  {journal} {\bibinfo  {journal} {Prog. Part. Nucl. Phys.}\ }\textbf {\bibinfo
  {volume} {54}},\ \bibinfo {pages} {535} (\bibinfo {year} {2005})}\BibitemShut
  {NoStop}%
\bibitem [{\citenamefont {Schatz}(2013)}]{Schatz2013IJMS}%
  \BibitemOpen
  \bibfield  {author} {\bibinfo {author} {\bibfnamefont {H.}~\bibnamefont
  {Schatz}},\ }\href {\doibase https://doi.org/10.1016/j.ijms.2013.03.016}
  {\bibfield  {journal} {\bibinfo  {journal} {Int. J. Mass Spectrometry}\
  }\textbf {\bibinfo {volume} {349-350}},\ \bibinfo {pages} {181} (\bibinfo
  {year} {2013})}\BibitemShut {NoStop}%
\bibitem [{\citenamefont {Thoennessen}(2016)}]{Thoennessen2016Book}%
  \BibitemOpen
  \bibfield  {author} {\bibinfo {author} {\bibfnamefont {M.}~\bibnamefont
  {Thoennessen}},\ }\href@noop {} {\emph {\bibinfo {title} {The Discovery of
  Isotopes}}}\ (\bibinfo  {publisher} {Springer, New York},\ \bibinfo {year}
  {2016})\BibitemShut {NoStop}%
\bibitem [{DNP(2021)}]{DNP}%
  \BibitemOpen
  \href {https://people.nscl.msu.edu/~thoennes/isotopes/} {\enquote {\bibinfo
  {title} {{Discovery of Nuclides Project}},}\ } (\bibinfo {year}
  {2021})\BibitemShut {NoStop}%
\bibitem [{\citenamefont {Kondev}\ \emph {et~al.}(2021)\citenamefont {Kondev},
  \citenamefont {Wang}, \citenamefont {Huang}, \citenamefont {Naimi},\ and\
  \citenamefont {Audi}}]{AME2020(1)}%
  \BibitemOpen
  \bibfield  {author} {\bibinfo {author} {\bibfnamefont {F.}~\bibnamefont
  {Kondev}}, \bibinfo {author} {\bibfnamefont {M.}~\bibnamefont {Wang}},
  \bibinfo {author} {\bibfnamefont {W.}~\bibnamefont {Huang}}, \bibinfo
  {author} {\bibfnamefont {S.}~\bibnamefont {Naimi}}, \ and\ \bibinfo {author}
  {\bibfnamefont {G.}~\bibnamefont {Audi}},\ }\href
  {https://iopscience.iop.org/article/10.1088/1674-1137/abddae/meta} {\bibfield
   {journal} {\bibinfo  {journal} {Chin. Phys. C}\ }\textbf {\bibinfo {volume}
  {45}},\ \bibinfo {pages} {030001} (\bibinfo {year} {2021})}\BibitemShut
  {NoStop}%
\bibitem [{\citenamefont {Huang}\ \emph {et~al.}(2021)\citenamefont {Huang},
  \citenamefont {Wang}, \citenamefont {Kondev}, \citenamefont {Audi},\ and\
  \citenamefont {Naimi}}]{AME2020(2)}%
  \BibitemOpen
  \bibfield  {author} {\bibinfo {author} {\bibfnamefont {W.}~\bibnamefont
  {Huang}}, \bibinfo {author} {\bibfnamefont {M.}~\bibnamefont {Wang}},
  \bibinfo {author} {\bibfnamefont {F.}~\bibnamefont {Kondev}}, \bibinfo
  {author} {\bibfnamefont {G.}~\bibnamefont {Audi}}, \ and\ \bibinfo {author}
  {\bibfnamefont {S.}~\bibnamefont {Naimi}},\ }\href
  {https://iopscience.iop.org/article/10.1088/1674-1137/abddb0/meta} {\bibfield
   {journal} {\bibinfo  {journal} {Chin. Phys. C}\ }\textbf {\bibinfo {volume}
  {45}},\ \bibinfo {pages} {030002} (\bibinfo {year} {2021})}\BibitemShut
  {NoStop}%
\bibitem [{\citenamefont {Wang}\ \emph
  {et~al.}(2021{\natexlab{a}})\citenamefont {Wang}, \citenamefont {Huang},
  \citenamefont {Kondev}, \citenamefont {Audi},\ and\ \citenamefont
  {Naimi}}]{AME2020(3)}%
  \BibitemOpen
  \bibfield  {author} {\bibinfo {author} {\bibfnamefont {M.}~\bibnamefont
  {Wang}}, \bibinfo {author} {\bibfnamefont {W.}~\bibnamefont {Huang}},
  \bibinfo {author} {\bibfnamefont {F.}~\bibnamefont {Kondev}}, \bibinfo
  {author} {\bibfnamefont {G.}~\bibnamefont {Audi}}, \ and\ \bibinfo {author}
  {\bibfnamefont {S.}~\bibnamefont {Naimi}},\ }\href
  {https://iopscience.iop.org/article/10.1088/1674-1137/abddaf/meta} {\bibfield
   {journal} {\bibinfo  {journal} {Chin. Phys. C}\ }\textbf {\bibinfo {volume}
  {45}},\ \bibinfo {pages} {030003} (\bibinfo {year}
  {2021}{\natexlab{a}})}\BibitemShut {NoStop}%
\bibitem [{\citenamefont {Lunney}\ \emph {et~al.}(2003)\citenamefont {Lunney},
  \citenamefont {Pearson},\ and\ \citenamefont {Thibault}}]{Lunney2003RMP}%
  \BibitemOpen
  \bibfield  {author} {\bibinfo {author} {\bibfnamefont {D.}~\bibnamefont
  {Lunney}}, \bibinfo {author} {\bibfnamefont {J.}~\bibnamefont {Pearson}}, \
  and\ \bibinfo {author} {\bibfnamefont {C.}~\bibnamefont {Thibault}},\ }\href
  {\doibase 10.1103/RevModPhys.75.1021} {\bibfield  {journal} {\bibinfo
  {journal} {Rev. Mod. Phys.}\ }\textbf {\bibinfo {volume} {75}},\ \bibinfo
  {pages} {1021} (\bibinfo {year} {2003})}\BibitemShut {NoStop}%
\bibitem [{\citenamefont {Blaum}(2006)}]{BLAUM20061}%
  \BibitemOpen
  \bibfield  {author} {\bibinfo {author} {\bibfnamefont {K.}~\bibnamefont
  {Blaum}},\ }\href {\doibase 10.1016/j.physrep.2005.10.011} {\bibfield
  {journal} {\bibinfo  {journal} {Phys. Rep.}\ }\textbf {\bibinfo {volume}
  {425}},\ \bibinfo {pages} {1 } (\bibinfo {year} {2006})}\BibitemShut
  {NoStop}%
\bibitem [{\citenamefont {Zhang}\ \emph
  {et~al.}(2019{\natexlab{a}})\citenamefont {Zhang}, \citenamefont {Gan},
  \citenamefont {Yang}, \citenamefont {Ma}, \citenamefont {Huang},
  \citenamefont {Yang}, \citenamefont {Zhang}, \citenamefont {Tian},
  \citenamefont {Wang}, \citenamefont {Sun}, \citenamefont {Lu}, \citenamefont
  {Zhang}, \citenamefont {Zhou}, \citenamefont {Wang}, \citenamefont {Wu},
  \citenamefont {Duan}, \citenamefont {Huang}, \citenamefont {Liu},
  \citenamefont {Ren}, \citenamefont {Zhou}, \citenamefont {Zhou},
  \citenamefont {Xu}, \citenamefont {Tsyganov}, \citenamefont {Voinov},\ and\
  \citenamefont {Polyakov}}]{Zhang2019PRL}%
  \BibitemOpen
  \bibfield  {author} {\bibinfo {author} {\bibfnamefont {Z.~Y.}\ \bibnamefont
  {Zhang}}, \bibinfo {author} {\bibfnamefont {Z.~G.}\ \bibnamefont {Gan}},
  \bibinfo {author} {\bibfnamefont {H.~B.}\ \bibnamefont {Yang}}, \bibinfo
  {author} {\bibfnamefont {L.}~\bibnamefont {Ma}}, \bibinfo {author}
  {\bibfnamefont {M.~H.}\ \bibnamefont {Huang}}, \bibinfo {author}
  {\bibfnamefont {C.~L.}\ \bibnamefont {Yang}}, \bibinfo {author}
  {\bibfnamefont {M.~M.}\ \bibnamefont {Zhang}}, \bibinfo {author}
  {\bibfnamefont {Y.~L.}\ \bibnamefont {Tian}}, \bibinfo {author}
  {\bibfnamefont {Y.~S.}\ \bibnamefont {Wang}}, \bibinfo {author}
  {\bibfnamefont {M.~D.}\ \bibnamefont {Sun}}, \bibinfo {author} {\bibfnamefont
  {H.~Y.}\ \bibnamefont {Lu}}, \bibinfo {author} {\bibfnamefont {W.~Q.}\
  \bibnamefont {Zhang}}, \bibinfo {author} {\bibfnamefont {H.~B.}\ \bibnamefont
  {Zhou}}, \bibinfo {author} {\bibfnamefont {X.}~\bibnamefont {Wang}}, \bibinfo
  {author} {\bibfnamefont {C.~G.}\ \bibnamefont {Wu}}, \bibinfo {author}
  {\bibfnamefont {L.~M.}\ \bibnamefont {Duan}}, \bibinfo {author}
  {\bibfnamefont {W.~X.}\ \bibnamefont {Huang}}, \bibinfo {author}
  {\bibfnamefont {Z.}~\bibnamefont {Liu}}, \bibinfo {author} {\bibfnamefont
  {Z.~Z.}\ \bibnamefont {Ren}}, \bibinfo {author} {\bibfnamefont {S.~G.}\
  \bibnamefont {Zhou}}, \bibinfo {author} {\bibfnamefont {X.~H.}\ \bibnamefont
  {Zhou}}, \bibinfo {author} {\bibfnamefont {H.~S.}\ \bibnamefont {Xu}},
  \bibinfo {author} {\bibfnamefont {Y.~S.}\ \bibnamefont {Tsyganov}}, \bibinfo
  {author} {\bibfnamefont {A.~A.}\ \bibnamefont {Voinov}}, \ and\ \bibinfo
  {author} {\bibfnamefont {A.~N.}\ \bibnamefont {Polyakov}},\ }\href {\doibase
  10.1103/PhysRevLett.122.192503} {\bibfield  {journal} {\bibinfo  {journal}
  {Phys. Rev. Lett.}\ }\textbf {\bibinfo {volume} {122}},\ \bibinfo {pages}
  {192503} (\bibinfo {year} {2019}{\natexlab{a}})}\BibitemShut {NoStop}%
\bibitem [{\citenamefont {Ahn}\ \emph {et~al.}(2019)\citenamefont {Ahn},
  \citenamefont {Fukuda}, \citenamefont {Geissel}, \citenamefont {Inabe},
  \citenamefont {Iwasa}, \citenamefont {Kubo}, \citenamefont {Kusaka},
  \citenamefont {Morrissey}, \citenamefont {Murai}, \citenamefont {Nakamura},
  \citenamefont {Ohtake}, \citenamefont {Otsu}, \citenamefont {Sato},
  \citenamefont {Sherrill}, \citenamefont {Shimizu}, \citenamefont {Suzuki},
  \citenamefont {Takeda}, \citenamefont {Tarasov}, \citenamefont {Ueno},
  \citenamefont {Yanagisawa},\ and\ \citenamefont {Yoshida}}]{Ahn2019PRL}%
  \BibitemOpen
  \bibfield  {author} {\bibinfo {author} {\bibfnamefont {D.~S.}\ \bibnamefont
  {Ahn}}, \bibinfo {author} {\bibfnamefont {N.}~\bibnamefont {Fukuda}},
  \bibinfo {author} {\bibfnamefont {H.}~\bibnamefont {Geissel}}, \bibinfo
  {author} {\bibfnamefont {N.}~\bibnamefont {Inabe}}, \bibinfo {author}
  {\bibfnamefont {N.}~\bibnamefont {Iwasa}}, \bibinfo {author} {\bibfnamefont
  {T.}~\bibnamefont {Kubo}}, \bibinfo {author} {\bibfnamefont {K.}~\bibnamefont
  {Kusaka}}, \bibinfo {author} {\bibfnamefont {D.~J.}\ \bibnamefont
  {Morrissey}}, \bibinfo {author} {\bibfnamefont {D.}~\bibnamefont {Murai}},
  \bibinfo {author} {\bibfnamefont {T.}~\bibnamefont {Nakamura}}, \bibinfo
  {author} {\bibfnamefont {M.}~\bibnamefont {Ohtake}}, \bibinfo {author}
  {\bibfnamefont {H.}~\bibnamefont {Otsu}}, \bibinfo {author} {\bibfnamefont
  {H.}~\bibnamefont {Sato}}, \bibinfo {author} {\bibfnamefont {B.~M.}\
  \bibnamefont {Sherrill}}, \bibinfo {author} {\bibfnamefont {Y.}~\bibnamefont
  {Shimizu}}, \bibinfo {author} {\bibfnamefont {H.}~\bibnamefont {Suzuki}},
  \bibinfo {author} {\bibfnamefont {H.}~\bibnamefont {Takeda}}, \bibinfo
  {author} {\bibfnamefont {O.~B.}\ \bibnamefont {Tarasov}}, \bibinfo {author}
  {\bibfnamefont {H.}~\bibnamefont {Ueno}}, \bibinfo {author} {\bibfnamefont
  {Y.}~\bibnamefont {Yanagisawa}}, \ and\ \bibinfo {author} {\bibfnamefont
  {K.}~\bibnamefont {Yoshida}},\ }\href {\doibase
  10.1103/PhysRevLett.123.212501} {\bibfield  {journal} {\bibinfo  {journal}
  {Phys. Rev. Lett.}\ }\textbf {\bibinfo {volume} {123}},\ \bibinfo {pages}
  {212501} (\bibinfo {year} {2019})}\BibitemShut {NoStop}%
\bibitem [{\citenamefont {Niu}\ \emph {et~al.}(2018)\citenamefont {Niu},
  \citenamefont {Liang}, \citenamefont {Sun}, \citenamefont {Niu},
  \citenamefont {Guo},\ and\ \citenamefont {Meng}}]{Niu2018SciBull}%
  \BibitemOpen
  \bibfield  {author} {\bibinfo {author} {\bibfnamefont {Z.}~\bibnamefont
  {Niu}}, \bibinfo {author} {\bibfnamefont {H.}~\bibnamefont {Liang}}, \bibinfo
  {author} {\bibfnamefont {B.}~\bibnamefont {Sun}}, \bibinfo {author}
  {\bibfnamefont {Y.}~\bibnamefont {Niu}}, \bibinfo {author} {\bibfnamefont
  {J.}~\bibnamefont {Guo}}, \ and\ \bibinfo {author} {\bibfnamefont
  {J.}~\bibnamefont {Meng}},\ }\href {\doibase
  https://doi.org/10.1016/j.scib.2018.05.009} {\bibfield  {journal} {\bibinfo
  {journal} {Sci. Bull.}\ }\textbf {\bibinfo {volume} {63}},\ \bibinfo {pages}
  {759} (\bibinfo {year} {2018})}\BibitemShut {NoStop}%
\bibitem [{\citenamefont {M\"oller}\ \emph {et~al.}(2016)\citenamefont
  {M\"oller}, \citenamefont {Sierk}, \citenamefont {Ichikawa},\ and\
  \citenamefont {Sagawa}}]{Moller2016ADNDT}%
  \BibitemOpen
  \bibfield  {author} {\bibinfo {author} {\bibfnamefont {P.}~\bibnamefont
  {M\"oller}}, \bibinfo {author} {\bibfnamefont {A.}~\bibnamefont {Sierk}},
  \bibinfo {author} {\bibfnamefont {T.}~\bibnamefont {Ichikawa}}, \ and\
  \bibinfo {author} {\bibfnamefont {H.}~\bibnamefont {Sagawa}},\ }\href
  {\doibase 10.1016/j.adt.2015.10.002Get} {\bibfield  {journal} {\bibinfo
  {journal} {Atom. Data Nucl. Data Tabl.}\ }\textbf {\bibinfo {volume}
  {109-110}},\ \bibinfo {pages} {1 } (\bibinfo {year} {2016})}\BibitemShut
  {NoStop}%
\bibitem [{\citenamefont {Aboussir}\ \emph {et~al.}(1995)\citenamefont
  {Aboussir}, \citenamefont {Pearson}, \citenamefont {Dutta},\ and\
  \citenamefont {Tondeur}}]{Aboussir1995ADNDT}%
  \BibitemOpen
  \bibfield  {author} {\bibinfo {author} {\bibfnamefont {Y.}~\bibnamefont
  {Aboussir}}, \bibinfo {author} {\bibfnamefont {J.}~\bibnamefont {Pearson}},
  \bibinfo {author} {\bibfnamefont {A.}~\bibnamefont {Dutta}}, \ and\ \bibinfo
  {author} {\bibfnamefont {F.}~\bibnamefont {Tondeur}},\ }\href {\doibase
  10.1016/S0092-640X(95)90014-4} {\bibfield  {journal} {\bibinfo  {journal}
  {Atom. Data Nucl. Data Tabl.}\ }\textbf {\bibinfo {volume} {61}},\ \bibinfo
  {pages} {127 } (\bibinfo {year} {1995})}\BibitemShut {NoStop}%
\bibitem [{\citenamefont {Wang}\ \emph {et~al.}(2014)\citenamefont {Wang},
  \citenamefont {Liu}, \citenamefont {Wu},\ and\ \citenamefont
  {Meng}}]{Wang2014PLB}%
  \BibitemOpen
  \bibfield  {author} {\bibinfo {author} {\bibfnamefont {N.}~\bibnamefont
  {Wang}}, \bibinfo {author} {\bibfnamefont {M.}~\bibnamefont {Liu}}, \bibinfo
  {author} {\bibfnamefont {X.}~\bibnamefont {Wu}}, \ and\ \bibinfo {author}
  {\bibfnamefont {J.}~\bibnamefont {Meng}},\ }\href {\doibase
  10.1016/j.physletb.2014.05.049} {\bibfield  {journal} {\bibinfo  {journal}
  {Phys. Lett. B}\ }\textbf {\bibinfo {volume} {734}},\ \bibinfo {pages} {215 }
  (\bibinfo {year} {2014})}\BibitemShut {NoStop}%
\bibitem [{\citenamefont {Zhang}\ \emph
  {et~al.}(2014{\natexlab{a}})\citenamefont {Zhang}, \citenamefont {Dong},
  \citenamefont {Ma}, \citenamefont {Royer}, \citenamefont {Li},\ and\
  \citenamefont {Zhang}}]{Zhang2014NPA}%
  \BibitemOpen
  \bibfield  {author} {\bibinfo {author} {\bibfnamefont {H.}~\bibnamefont
  {Zhang}}, \bibinfo {author} {\bibfnamefont {J.}~\bibnamefont {Dong}},
  \bibinfo {author} {\bibfnamefont {N.}~\bibnamefont {Ma}}, \bibinfo {author}
  {\bibfnamefont {G.}~\bibnamefont {Royer}}, \bibinfo {author} {\bibfnamefont
  {J.}~\bibnamefont {Li}}, \ and\ \bibinfo {author} {\bibfnamefont
  {H.}~\bibnamefont {Zhang}},\ }\href {\doibase
  10.1016/j.nuclphysa.2014.05.019} {\bibfield  {journal} {\bibinfo  {journal}
  {Nucl. Phys. A}\ }\textbf {\bibinfo {volume} {929}},\ \bibinfo {pages} {38 }
  (\bibinfo {year} {2014}{\natexlab{a}})}\BibitemShut {NoStop}%
\bibitem [{\citenamefont {Samyn}\ \emph {et~al.}(2002)\citenamefont {Samyn},
  \citenamefont {Goriely}, \citenamefont {Heenen}, \citenamefont {Pearson},\
  and\ \citenamefont {Tondeur}}]{Samyn2002NPA}%
  \BibitemOpen
  \bibfield  {author} {\bibinfo {author} {\bibfnamefont {M.}~\bibnamefont
  {Samyn}}, \bibinfo {author} {\bibfnamefont {S.}~\bibnamefont {Goriely}},
  \bibinfo {author} {\bibfnamefont {P.-H.}\ \bibnamefont {Heenen}}, \bibinfo
  {author} {\bibfnamefont {J.}~\bibnamefont {Pearson}}, \ and\ \bibinfo
  {author} {\bibfnamefont {F.}~\bibnamefont {Tondeur}},\ }\href {\doibase
  10.1016/S0375-9474(01)01316-1} {\bibfield  {journal} {\bibinfo  {journal}
  {Nucl. Phys. A}\ }\textbf {\bibinfo {volume} {700}},\ \bibinfo {pages} {142 }
  (\bibinfo {year} {2002})}\BibitemShut {NoStop}%
\bibitem [{\citenamefont {Stoitsov}\ \emph {et~al.}(2003)\citenamefont
  {Stoitsov}, \citenamefont {Dobaczewski}, \citenamefont {Nazarewicz},
  \citenamefont {Pittel},\ and\ \citenamefont {Dean}}]{Stoitsov2003PRC}%
  \BibitemOpen
  \bibfield  {author} {\bibinfo {author} {\bibfnamefont {M.~V.}\ \bibnamefont
  {Stoitsov}}, \bibinfo {author} {\bibfnamefont {J.}~\bibnamefont
  {Dobaczewski}}, \bibinfo {author} {\bibfnamefont {W.}~\bibnamefont
  {Nazarewicz}}, \bibinfo {author} {\bibfnamefont {S.}~\bibnamefont {Pittel}},
  \ and\ \bibinfo {author} {\bibfnamefont {D.~J.}\ \bibnamefont {Dean}},\
  }\href {\doibase 10.1103/PhysRevC.68.054312} {\bibfield  {journal} {\bibinfo
  {journal} {Phys. Rev. C}\ }\textbf {\bibinfo {volume} {68}},\ \bibinfo
  {pages} {054312} (\bibinfo {year} {2003})}\BibitemShut {NoStop}%
\bibitem [{\citenamefont {Goriely}\ \emph
  {et~al.}(2009{\natexlab{a}})\citenamefont {Goriely}, \citenamefont {Chamel},\
  and\ \citenamefont {Pearson}}]{Goriely2009PRLSkyrme}%
  \BibitemOpen
  \bibfield  {author} {\bibinfo {author} {\bibfnamefont {S.}~\bibnamefont
  {Goriely}}, \bibinfo {author} {\bibfnamefont {N.}~\bibnamefont {Chamel}}, \
  and\ \bibinfo {author} {\bibfnamefont {J.~M.}\ \bibnamefont {Pearson}},\
  }\href {\doibase 10.1103/PhysRevLett.102.152503} {\bibfield  {journal}
  {\bibinfo  {journal} {Phys. Rev. Lett.}\ }\textbf {\bibinfo {volume} {102}},\
  \bibinfo {pages} {152503} (\bibinfo {year} {2009}{\natexlab{a}})}\BibitemShut
  {NoStop}%
\bibitem [{\citenamefont {Erler}\ \emph {et~al.}(2012)\citenamefont {Erler},
  \citenamefont {Birge}, \citenamefont {Kortelainen}, \citenamefont
  {Nazarewicz}, \citenamefont {Olsen}, \citenamefont {Perhac},\ and\
  \citenamefont {Stoitsov}}]{Erler2012Nature}%
  \BibitemOpen
  \bibfield  {author} {\bibinfo {author} {\bibfnamefont {J.}~\bibnamefont
  {Erler}}, \bibinfo {author} {\bibfnamefont {N.}~\bibnamefont {Birge}},
  \bibinfo {author} {\bibfnamefont {M.}~\bibnamefont {Kortelainen}}, \bibinfo
  {author} {\bibfnamefont {W.}~\bibnamefont {Nazarewicz}}, \bibinfo {author}
  {\bibfnamefont {E.}~\bibnamefont {Olsen}}, \bibinfo {author} {\bibfnamefont
  {A.~M.}\ \bibnamefont {Perhac}}, \ and\ \bibinfo {author} {\bibfnamefont
  {M.}~\bibnamefont {Stoitsov}},\ }\href {\doibase 10.1038/nature11188}
  {\bibfield  {journal} {\bibinfo  {journal} {Nature}\ }\textbf {\bibinfo
  {volume} {486}},\ \bibinfo {pages} {509} (\bibinfo {year}
  {2012})}\BibitemShut {NoStop}%
\bibitem [{\citenamefont {Goriely}\ \emph
  {et~al.}(2013{\natexlab{a}})\citenamefont {Goriely}, \citenamefont {Chamel},\
  and\ \citenamefont {Pearson}}]{Goriely2013PRC}%
  \BibitemOpen
  \bibfield  {author} {\bibinfo {author} {\bibfnamefont {S.}~\bibnamefont
  {Goriely}}, \bibinfo {author} {\bibfnamefont {N.}~\bibnamefont {Chamel}}, \
  and\ \bibinfo {author} {\bibfnamefont {J.~M.}\ \bibnamefont {Pearson}},\
  }\href {\doibase 10.1103/PhysRevC.88.024308} {\bibfield  {journal} {\bibinfo
  {journal} {Phys. Rev. C}\ }\textbf {\bibinfo {volume} {88}},\ \bibinfo
  {pages} {024308} (\bibinfo {year} {2013}{\natexlab{a}})}\BibitemShut
  {NoStop}%
\bibitem [{\citenamefont {Goriely}\ \emph
  {et~al.}(2013{\natexlab{b}})\citenamefont {Goriely}, \citenamefont {Chamel},\
  and\ \citenamefont {Pearson}}]{Goriely2013PRC(R)}%
  \BibitemOpen
  \bibfield  {author} {\bibinfo {author} {\bibfnamefont {S.}~\bibnamefont
  {Goriely}}, \bibinfo {author} {\bibfnamefont {N.}~\bibnamefont {Chamel}}, \
  and\ \bibinfo {author} {\bibfnamefont {J.~M.}\ \bibnamefont {Pearson}},\
  }\href {\doibase 10.1103/PhysRevC.88.061302} {\bibfield  {journal} {\bibinfo
  {journal} {Phys. Rev. C}\ }\textbf {\bibinfo {volume} {88}},\ \bibinfo
  {pages} {061302(R)} (\bibinfo {year} {2013}{\natexlab{b}})}\BibitemShut
  {NoStop}%
\bibitem [{\citenamefont {Hilaire}\ and\ \citenamefont
  {Girod}(2007)}]{Hilaire2007EPJA}%
  \BibitemOpen
  \bibfield  {author} {\bibinfo {author} {\bibfnamefont {S.}~\bibnamefont
  {Hilaire}}\ and\ \bibinfo {author} {\bibfnamefont {M.}~\bibnamefont
  {Girod}},\ }\href
  {https://link.springer.com/article/10.1140/epja/i2007-10450-2} {\bibfield
  {journal} {\bibinfo  {journal} {Eur. Phys. J. A}\ }\textbf {\bibinfo {volume}
  {33}},\ \bibinfo {pages} {237} (\bibinfo {year} {2007})}\BibitemShut
  {NoStop}%
\bibitem [{\citenamefont {Goriely}\ \emph
  {et~al.}(2009{\natexlab{b}})\citenamefont {Goriely}, \citenamefont {Hilaire},
  \citenamefont {Girod},\ and\ \citenamefont {P\'eru}}]{Goriely2009PRLGogny}%
  \BibitemOpen
  \bibfield  {author} {\bibinfo {author} {\bibfnamefont {S.}~\bibnamefont
  {Goriely}}, \bibinfo {author} {\bibfnamefont {S.}~\bibnamefont {Hilaire}},
  \bibinfo {author} {\bibfnamefont {M.}~\bibnamefont {Girod}}, \ and\ \bibinfo
  {author} {\bibfnamefont {S.}~\bibnamefont {P\'eru}},\ }\href {\doibase
  10.1103/PhysRevLett.102.242501} {\bibfield  {journal} {\bibinfo  {journal}
  {Phys. Rev. Lett.}\ }\textbf {\bibinfo {volume} {102}},\ \bibinfo {pages}
  {242501} (\bibinfo {year} {2009}{\natexlab{b}})}\BibitemShut {NoStop}%
\bibitem [{\citenamefont {Delaroche}\ \emph {et~al.}(2010)\citenamefont
  {Delaroche}, \citenamefont {Girod}, \citenamefont {Libert}, \citenamefont
  {Goutte}, \citenamefont {Hilaire}, \citenamefont {P\'eru}, \citenamefont
  {Pillet},\ and\ \citenamefont {Bertsch}}]{Delaroche2010PRC}%
  \BibitemOpen
  \bibfield  {author} {\bibinfo {author} {\bibfnamefont {J.~P.}\ \bibnamefont
  {Delaroche}}, \bibinfo {author} {\bibfnamefont {M.}~\bibnamefont {Girod}},
  \bibinfo {author} {\bibfnamefont {J.}~\bibnamefont {Libert}}, \bibinfo
  {author} {\bibfnamefont {H.}~\bibnamefont {Goutte}}, \bibinfo {author}
  {\bibfnamefont {S.}~\bibnamefont {Hilaire}}, \bibinfo {author} {\bibfnamefont
  {S.}~\bibnamefont {P\'eru}}, \bibinfo {author} {\bibfnamefont
  {N.}~\bibnamefont {Pillet}}, \ and\ \bibinfo {author} {\bibfnamefont {G.~F.}\
  \bibnamefont {Bertsch}},\ }\href {\doibase 10.1103/PhysRevC.81.014303}
  {\bibfield  {journal} {\bibinfo  {journal} {Phys. Rev. C}\ }\textbf {\bibinfo
  {volume} {81}},\ \bibinfo {pages} {014303} (\bibinfo {year}
  {2010})}\BibitemShut {NoStop}%
\bibitem [{\citenamefont {Ring}(1996)}]{Ring1996PPNP}%
  \BibitemOpen
  \bibfield  {author} {\bibinfo {author} {\bibfnamefont {P.}~\bibnamefont
  {Ring}},\ }\href {\doibase 10.1016/0146-6410(96)00054-3} {\bibfield
  {journal} {\bibinfo  {journal} {Prog. Part. Nucl. Phys.}\ }\textbf {\bibinfo
  {volume} {37}},\ \bibinfo {pages} {193 } (\bibinfo {year}
  {1996})}\BibitemShut {NoStop}%
\bibitem [{\citenamefont {Vretenar}\ \emph {et~al.}(2005)\citenamefont
  {Vretenar}, \citenamefont {Afanasjev}, \citenamefont {Lalazissis},\ and\
  \citenamefont {Ring}}]{Vretenar2005PhysRep}%
  \BibitemOpen
  \bibfield  {author} {\bibinfo {author} {\bibfnamefont {D.}~\bibnamefont
  {Vretenar}}, \bibinfo {author} {\bibfnamefont {A.~V.}\ \bibnamefont
  {Afanasjev}}, \bibinfo {author} {\bibfnamefont {G.~A.}\ \bibnamefont
  {Lalazissis}}, \ and\ \bibinfo {author} {\bibfnamefont {P.}~\bibnamefont
  {Ring}},\ }\href {\doibase 10.1016/j.physrep.2004.10.001} {\bibfield
  {journal} {\bibinfo  {journal} {Phys. Rep.}\ }\textbf {\bibinfo {volume}
  {409}},\ \bibinfo {pages} {101 } (\bibinfo {year} {2005})}\BibitemShut
  {NoStop}%
\bibitem [{\citenamefont {Meng}\ \emph {et~al.}(2006)\citenamefont {Meng},
  \citenamefont {Toki}, \citenamefont {Zhou}, \citenamefont {Zhang},
  \citenamefont {Long},\ and\ \citenamefont {Geng}}]{Meng2006PPNP}%
  \BibitemOpen
  \bibfield  {author} {\bibinfo {author} {\bibfnamefont {J.}~\bibnamefont
  {Meng}}, \bibinfo {author} {\bibfnamefont {H.}~\bibnamefont {Toki}}, \bibinfo
  {author} {\bibfnamefont {S.~G.}\ \bibnamefont {Zhou}}, \bibinfo {author}
  {\bibfnamefont {S.~Q.}\ \bibnamefont {Zhang}}, \bibinfo {author}
  {\bibfnamefont {W.~H.}\ \bibnamefont {Long}}, \ and\ \bibinfo {author}
  {\bibfnamefont {L.~S.}\ \bibnamefont {Geng}},\ }\href {\doibase
  10.1016/j.ppnp.2005.06.001} {\bibfield  {journal} {\bibinfo  {journal} {Prog.
  Part. Nucl. Phys.}\ }\textbf {\bibinfo {volume} {57}},\ \bibinfo {pages}
  {470} (\bibinfo {year} {2006})}\BibitemShut {NoStop}%
\bibitem [{\citenamefont {Niksic}\ \emph {et~al.}(2011)\citenamefont {Niksic},
  \citenamefont {Vretenar},\ and\ \citenamefont {Ring}}]{Niksic2011PPNP}%
  \BibitemOpen
  \bibfield  {author} {\bibinfo {author} {\bibfnamefont {T.}~\bibnamefont
  {Niksic}}, \bibinfo {author} {\bibfnamefont {D.}~\bibnamefont {Vretenar}}, \
  and\ \bibinfo {author} {\bibfnamefont {P.}~\bibnamefont {Ring}},\ }\href
  {\doibase 10.1016/j.ppnp.2011.01.055} {\bibfield  {journal} {\bibinfo
  {journal} {Prog. Part. Nucl. Phys.}\ }\textbf {\bibinfo {volume} {66}},\
  \bibinfo {pages} {519 } (\bibinfo {year} {2011})}\BibitemShut {NoStop}%
\bibitem [{\citenamefont {Meng}\ \emph {et~al.}(2013)\citenamefont {Meng},
  \citenamefont {Peng}, \citenamefont {Zhang},\ and\ \citenamefont
  {Zhao}}]{Meng2013FOP}%
  \BibitemOpen
  \bibfield  {author} {\bibinfo {author} {\bibfnamefont {J.}~\bibnamefont
  {Meng}}, \bibinfo {author} {\bibfnamefont {J.}~\bibnamefont {Peng}}, \bibinfo
  {author} {\bibfnamefont {S.~Q.}\ \bibnamefont {Zhang}}, \ and\ \bibinfo
  {author} {\bibfnamefont {P.~W.}\ \bibnamefont {Zhao}},\ }\href {\doibase
  10.1007/s11467-013-0287-y} {\bibfield  {journal} {\bibinfo  {journal} {Front.
  Phys.}\ }\textbf {\bibinfo {volume} {8}},\ \bibinfo {pages} {55} (\bibinfo
  {year} {2013})}\BibitemShut {NoStop}%
\bibitem [{\citenamefont {Meng}\ and\ \citenamefont
  {Zhou}(2015)}]{Meng2015JPG}%
  \BibitemOpen
  \bibfield  {author} {\bibinfo {author} {\bibfnamefont {J.}~\bibnamefont
  {Meng}}\ and\ \bibinfo {author} {\bibfnamefont {S.~G.}\ \bibnamefont
  {Zhou}},\ }\href {\doibase 10.1088/0954-3899/42/9/093101} {\bibfield
  {journal} {\bibinfo  {journal} {J. Phys. G}\ }\textbf {\bibinfo {volume}
  {42}},\ \bibinfo {pages} {093101} (\bibinfo {year} {2015})}\BibitemShut
  {NoStop}%
\bibitem [{\citenamefont {Zhou}(2016)}]{Zhou2016PhysScr}%
  \BibitemOpen
  \bibfield  {author} {\bibinfo {author} {\bibfnamefont {S.-G.}\ \bibnamefont
  {Zhou}},\ }\href
  {https://iopscience.iop.org/article/10.1088/0031-8949/91/6/063008/meta}
  {\bibfield  {journal} {\bibinfo  {journal} {Phys. Scr.}\ }\textbf {\bibinfo
  {volume} {91}},\ \bibinfo {pages} {063008} (\bibinfo {year}
  {2016})}\BibitemShut {NoStop}%
\bibitem [{\citenamefont {Meng}(2016)}]{Meng2016Book}%
  \BibitemOpen
  \bibinfo {editor} {\bibfnamefont {J.}~\bibnamefont {Meng}},\ ed.,\ \href@noop
  {} {\emph {\bibinfo {title} {Relativistic Density Functional for Nuclear
  Structure}}}\ (\bibinfo  {publisher} {World Scientific},\ \bibinfo {year}
  {2016})\BibitemShut {NoStop}%
\bibitem [{\citenamefont {Shen}\ \emph {et~al.}(2019)\citenamefont {Shen},
  \citenamefont {Liang}, \citenamefont {Long}, \citenamefont {Meng},\ and\
  \citenamefont {Ring}}]{Shen2019PPNP}%
  \BibitemOpen
  \bibfield  {author} {\bibinfo {author} {\bibfnamefont {S.}~\bibnamefont
  {Shen}}, \bibinfo {author} {\bibfnamefont {H.}~\bibnamefont {Liang}},
  \bibinfo {author} {\bibfnamefont {W.~H.}\ \bibnamefont {Long}}, \bibinfo
  {author} {\bibfnamefont {J.}~\bibnamefont {Meng}}, \ and\ \bibinfo {author}
  {\bibfnamefont {P.}~\bibnamefont {Ring}},\ }\href {\doibase
  10.1016/j.ppnp.2019.103713} {\bibfield  {journal} {\bibinfo  {journal} {Prog.
  Part. Nucl. Phys.}\ }\textbf {\bibinfo {volume} {109}},\ \bibinfo {pages}
  {103713} (\bibinfo {year} {2019})}\BibitemShut {NoStop}%
\bibitem [{\citenamefont {Ren}\ and\ \citenamefont
  {Zhao}(2020)}]{Ren2020PRC(R)}%
  \BibitemOpen
  \bibfield  {author} {\bibinfo {author} {\bibfnamefont {Z.~X.}\ \bibnamefont
  {Ren}}\ and\ \bibinfo {author} {\bibfnamefont {P.~W.}\ \bibnamefont {Zhao}},\
  }\href {\doibase 10.1103/PhysRevC.102.021301} {\bibfield  {journal} {\bibinfo
   {journal} {Phys. Rev. C}\ }\textbf {\bibinfo {volume} {102}},\ \bibinfo
  {pages} {021301(R)} (\bibinfo {year} {2020})}\BibitemShut {NoStop}%
\bibitem [{\citenamefont {Ginocchio}(1997)}]{Ginocchio1997PRL}%
  \BibitemOpen
  \bibfield  {author} {\bibinfo {author} {\bibfnamefont {J.~N.}\ \bibnamefont
  {Ginocchio}},\ }\href {\doibase 10.1103/PhysRevLett.78.436} {\bibfield
  {journal} {\bibinfo  {journal} {Phys. Rev. Lett.}\ }\textbf {\bibinfo
  {volume} {78}},\ \bibinfo {pages} {436} (\bibinfo {year} {1997})}\BibitemShut
  {NoStop}%
\bibitem [{\citenamefont {Meng}\ \emph
  {et~al.}(1998{\natexlab{a}})\citenamefont {Meng}, \citenamefont
  {Sugawara-Tanabe}, \citenamefont {Yamaji}, \citenamefont {Ring},\ and\
  \citenamefont {Arima}}]{Meng1998Phys.Rev.C628}%
  \BibitemOpen
  \bibfield  {author} {\bibinfo {author} {\bibfnamefont {J.}~\bibnamefont
  {Meng}}, \bibinfo {author} {\bibfnamefont {K.}~\bibnamefont
  {Sugawara-Tanabe}}, \bibinfo {author} {\bibfnamefont {S.}~\bibnamefont
  {Yamaji}}, \bibinfo {author} {\bibfnamefont {P.}~\bibnamefont {Ring}}, \ and\
  \bibinfo {author} {\bibfnamefont {A.}~\bibnamefont {Arima}},\ }\href
  {\doibase 10.1103/PhysRevC.58.R628} {\bibfield  {journal} {\bibinfo
  {journal} {Phys. Rev. C}\ }\textbf {\bibinfo {volume} {58}},\ \bibinfo
  {pages} {R628} (\bibinfo {year} {1998}{\natexlab{a}})}\BibitemShut {NoStop}%
\bibitem [{\citenamefont {Meng}\ \emph {et~al.}(1999)\citenamefont {Meng},
  \citenamefont {Sugawara-Tanabe}, \citenamefont {Yamaji},\ and\ \citenamefont
  {Arima}}]{Meng1999PRC}%
  \BibitemOpen
  \bibfield  {author} {\bibinfo {author} {\bibfnamefont {J.}~\bibnamefont
  {Meng}}, \bibinfo {author} {\bibfnamefont {K.}~\bibnamefont
  {Sugawara-Tanabe}}, \bibinfo {author} {\bibfnamefont {S.}~\bibnamefont
  {Yamaji}}, \ and\ \bibinfo {author} {\bibfnamefont {A.}~\bibnamefont
  {Arima}},\ }\href {\doibase 10.1103/PhysRevC.59.154} {\bibfield  {journal}
  {\bibinfo  {journal} {Phys. Rev. C}\ }\textbf {\bibinfo {volume} {59}},\
  \bibinfo {pages} {154} (\bibinfo {year} {1999})}\BibitemShut {NoStop}%
\bibitem [{\citenamefont {Chen}\ \emph {et~al.}(2003)\citenamefont {Chen},
  \citenamefont {Lu}, \citenamefont {Meng}, \citenamefont {Zhang},\ and\
  \citenamefont {Zhou}}]{Chen2003CPL}%
  \BibitemOpen
  \bibfield  {author} {\bibinfo {author} {\bibfnamefont {T.~S.}\ \bibnamefont
  {Chen}}, \bibinfo {author} {\bibfnamefont {H.~F.}\ \bibnamefont {Lu}},
  \bibinfo {author} {\bibfnamefont {J.}~\bibnamefont {Meng}}, \bibinfo {author}
  {\bibfnamefont {S.~Q.}\ \bibnamefont {Zhang}}, \ and\ \bibinfo {author}
  {\bibfnamefont {S.~G.}\ \bibnamefont {Zhou}},\ }\href
  {https://iopscience.iop.org/article/10.1088/0256-307X/20/3/312/meta}
  {\bibfield  {journal} {\bibinfo  {journal} {Chin. Phys. Lett.}\ }\textbf
  {\bibinfo {volume} {20}},\ \bibinfo {pages} {358} (\bibinfo {year}
  {2003})}\BibitemShut {NoStop}%
\bibitem [{\citenamefont {Ginocchio}(2005)}]{Ginocchio2005PhysRep}%
  \BibitemOpen
  \bibfield  {author} {\bibinfo {author} {\bibfnamefont {J.~N.}\ \bibnamefont
  {Ginocchio}},\ }\href {\doibase 10.1016/j.physrep.2005.04.003} {\bibfield
  {journal} {\bibinfo  {journal} {Phys. Rep.}\ }\textbf {\bibinfo {volume}
  {414}},\ \bibinfo {pages} {165 } (\bibinfo {year} {2005})}\BibitemShut
  {NoStop}%
\bibitem [{\citenamefont {Liang}\ \emph {et~al.}(2015)\citenamefont {Liang},
  \citenamefont {Meng},\ and\ \citenamefont {Zhou}}]{Liang2015PhysRep}%
  \BibitemOpen
  \bibfield  {author} {\bibinfo {author} {\bibfnamefont {H.}~\bibnamefont
  {Liang}}, \bibinfo {author} {\bibfnamefont {J.}~\bibnamefont {Meng}}, \ and\
  \bibinfo {author} {\bibfnamefont {S.-G.}\ \bibnamefont {Zhou}},\ }\href
  {\doibase 10.1016/j.physrep.2014.12.005} {\bibfield  {journal} {\bibinfo
  {journal} {Phys. Rep.}\ }\textbf {\bibinfo {volume} {570}},\ \bibinfo {pages}
  {1} (\bibinfo {year} {2015})}\BibitemShut {NoStop}%
\bibitem [{\citenamefont {Zhou}\ \emph
  {et~al.}(2003{\natexlab{a}})\citenamefont {Zhou}, \citenamefont {Meng},\ and\
  \citenamefont {Ring}}]{Zhou2003PRL}%
  \BibitemOpen
  \bibfield  {author} {\bibinfo {author} {\bibfnamefont {S.-G.}\ \bibnamefont
  {Zhou}}, \bibinfo {author} {\bibfnamefont {J.}~\bibnamefont {Meng}}, \ and\
  \bibinfo {author} {\bibfnamefont {P.}~\bibnamefont {Ring}},\ }\href {\doibase
  10.1103/PhysRevLett.91.262501} {\bibfield  {journal} {\bibinfo  {journal}
  {Phys. Rev. Lett.}\ }\textbf {\bibinfo {volume} {91}},\ \bibinfo {pages}
  {262501} (\bibinfo {year} {2003}{\natexlab{a}})}\BibitemShut {NoStop}%
\bibitem [{\citenamefont {He}\ \emph {et~al.}(2006)\citenamefont {He},
  \citenamefont {Zhou}, \citenamefont {Meng}, \citenamefont {Zhao},\ and\
  \citenamefont {Scheid}}]{He2006EPJA}%
  \BibitemOpen
  \bibfield  {author} {\bibinfo {author} {\bibfnamefont {X.~T.}\ \bibnamefont
  {He}}, \bibinfo {author} {\bibfnamefont {S.~G.}\ \bibnamefont {Zhou}},
  \bibinfo {author} {\bibfnamefont {J.}~\bibnamefont {Meng}}, \bibinfo {author}
  {\bibfnamefont {E.~G.}\ \bibnamefont {Zhao}}, \ and\ \bibinfo {author}
  {\bibfnamefont {W.}~\bibnamefont {Scheid}},\ }\href {\doibase
  10.1140/epja/i2006-10066-0} {\bibfield  {journal} {\bibinfo  {journal} {Eur.
  Phys. J. A}\ }\textbf {\bibinfo {volume} {28}},\ \bibinfo {pages} {265}
  (\bibinfo {year} {2006})}\BibitemShut {NoStop}%
\bibitem [{\citenamefont {Koepf}\ and\ \citenamefont
  {Ring}(1989)}]{Koepf1989NPA}%
  \BibitemOpen
  \bibfield  {author} {\bibinfo {author} {\bibfnamefont {W.}~\bibnamefont
  {Koepf}}\ and\ \bibinfo {author} {\bibfnamefont {P.}~\bibnamefont {Ring}},\
  }\href {\doibase 10.1016/0375-9474(89)90532-0} {\bibfield  {journal}
  {\bibinfo  {journal} {Nucl. Phys. A}\ }\textbf {\bibinfo {volume} {493}},\
  \bibinfo {pages} {61 } (\bibinfo {year} {1989})}\BibitemShut {NoStop}%
\bibitem [{\citenamefont {Yao}\ \emph {et~al.}(2006)\citenamefont {Yao},
  \citenamefont {Chen},\ and\ \citenamefont {Meng}}]{Yao2006Phys.Rev.C24307}%
  \BibitemOpen
  \bibfield  {author} {\bibinfo {author} {\bibfnamefont {J.~M.}\ \bibnamefont
  {Yao}}, \bibinfo {author} {\bibfnamefont {H.}~\bibnamefont {Chen}}, \ and\
  \bibinfo {author} {\bibfnamefont {J.}~\bibnamefont {Meng}},\ }\href {\doibase
  10.1103/PhysRevC.74.024307} {\bibfield  {journal} {\bibinfo  {journal} {Phys.
  Rev. C}\ }\textbf {\bibinfo {volume} {74}},\ \bibinfo {pages} {024307}
  (\bibinfo {year} {2006})}\BibitemShut {NoStop}%
\bibitem [{\citenamefont {Arima}(2011)}]{Arima2011}%
  \BibitemOpen
  \bibfield  {author} {\bibinfo {author} {\bibfnamefont {A.}~\bibnamefont
  {Arima}},\ }\href {\doibase doi.org/10.1007/s11433-010-4224-6} {\bibfield
  {journal} {\bibinfo  {journal} {Sci. China Phys. Mech. Astron.}\ }\textbf
  {\bibinfo {volume} {54}},\ \bibinfo {pages} {188} (\bibinfo {year}
  {2011})}\BibitemShut {NoStop}%
\bibitem [{\citenamefont {Li}\ \emph {et~al.}(2011{\natexlab{a}})\citenamefont
  {Li}, \citenamefont {Meng}, \citenamefont {Ring}, \citenamefont {Yao},\ and\
  \citenamefont {Arima}}]{Li2011Sci.ChinaPhys.Mech.Astron.204}%
  \BibitemOpen
  \bibfield  {author} {\bibinfo {author} {\bibfnamefont {J.}~\bibnamefont
  {Li}}, \bibinfo {author} {\bibfnamefont {J.}~\bibnamefont {Meng}}, \bibinfo
  {author} {\bibfnamefont {P.}~\bibnamefont {Ring}}, \bibinfo {author}
  {\bibfnamefont {J.~M.}\ \bibnamefont {Yao}}, \ and\ \bibinfo {author}
  {\bibfnamefont {A.}~\bibnamefont {Arima}},\ }\href {\doibase
  10.1007/s11433-010-4215-7} {\bibfield  {journal} {\bibinfo  {journal} {Sci.
  China Phys. Mech. Astron.}\ }\textbf {\bibinfo {volume} {54}},\ \bibinfo
  {pages} {204} (\bibinfo {year} {2011}{\natexlab{a}})}\BibitemShut {NoStop}%
\bibitem [{\citenamefont {Li}\ \emph {et~al.}(2011{\natexlab{b}})\citenamefont
  {Li}, \citenamefont {Yao}, \citenamefont {Meng},\ and\ \citenamefont
  {Arima}}]{Li2011Prog.Theor.Phys.1185}%
  \BibitemOpen
  \bibfield  {author} {\bibinfo {author} {\bibfnamefont {J.}~\bibnamefont
  {Li}}, \bibinfo {author} {\bibfnamefont {J.~M.}\ \bibnamefont {Yao}},
  \bibinfo {author} {\bibfnamefont {J.}~\bibnamefont {Meng}}, \ and\ \bibinfo
  {author} {\bibfnamefont {A.}~\bibnamefont {Arima}},\ }\href {\doibase
  10.1143/PTP.125.1185} {\bibfield  {journal} {\bibinfo  {journal} {Prog.
  Theor. Phys.}\ }\textbf {\bibinfo {volume} {125}},\ \bibinfo {pages} {1185}
  (\bibinfo {year} {2011}{\natexlab{b}})}\BibitemShut {NoStop}%
\bibitem [{\citenamefont {Li}\ and\ \citenamefont
  {Meng}(2018)}]{Li2018Front.Phys.Beijing132109}%
  \BibitemOpen
  \bibfield  {author} {\bibinfo {author} {\bibfnamefont {J.}~\bibnamefont
  {Li}}\ and\ \bibinfo {author} {\bibfnamefont {J.}~\bibnamefont {Meng}},\
  }\href {\doibase 10.1007/s11467-018-0842-7} {\bibfield  {journal} {\bibinfo
  {journal} {Front. Phys.}\ }\textbf {\bibinfo {volume} {13}},\ \bibinfo
  {pages} {132109} (\bibinfo {year} {2018})}\BibitemShut {NoStop}%
\bibitem [{\citenamefont {K\"onig}\ and\ \citenamefont
  {Ring}(1993)}]{Konig1993PRL}%
  \BibitemOpen
  \bibfield  {author} {\bibinfo {author} {\bibfnamefont {J.}~\bibnamefont
  {K\"onig}}\ and\ \bibinfo {author} {\bibfnamefont {P.}~\bibnamefont {Ring}},\
  }\href {\doibase 10.1103/PhysRevLett.71.3079} {\bibfield  {journal} {\bibinfo
   {journal} {Phys. Rev. Lett.}\ }\textbf {\bibinfo {volume} {71}},\ \bibinfo
  {pages} {3079} (\bibinfo {year} {1993})}\BibitemShut {NoStop}%
\bibitem [{\citenamefont {Afanasjev}\ \emph {et~al.}(2000)\citenamefont
  {Afanasjev}, \citenamefont {Ring},\ and\ \citenamefont
  {Konig}}]{Afanasjev2000NPA}%
  \BibitemOpen
  \bibfield  {author} {\bibinfo {author} {\bibfnamefont {A.~V.}\ \bibnamefont
  {Afanasjev}}, \bibinfo {author} {\bibfnamefont {P.}~\bibnamefont {Ring}}, \
  and\ \bibinfo {author} {\bibfnamefont {J.}~\bibnamefont {Konig}},\ }\href
  {\doibase 10.1016/S0375-9474(00)00187-1} {\bibfield  {journal} {\bibinfo
  {journal} {Nucl. Phys. A}\ }\textbf {\bibinfo {volume} {676}},\ \bibinfo
  {pages} {196 } (\bibinfo {year} {2000})}\BibitemShut {NoStop}%
\bibitem [{\citenamefont {Afanasjev}\ and\ \citenamefont
  {Ring}(2000)}]{PhysRevC.62.031302}%
  \BibitemOpen
  \bibfield  {author} {\bibinfo {author} {\bibfnamefont {A.~V.}\ \bibnamefont
  {Afanasjev}}\ and\ \bibinfo {author} {\bibfnamefont {P.}~\bibnamefont
  {Ring}},\ }\href {\doibase 10.1103/PhysRevC.62.031302} {\bibfield  {journal}
  {\bibinfo  {journal} {Phys. Rev. C}\ }\textbf {\bibinfo {volume} {62}},\
  \bibinfo {pages} {031302(R)} (\bibinfo {year} {2000})}\BibitemShut {NoStop}%
\bibitem [{\citenamefont {Afanasjev}\ and\ \citenamefont
  {Abusara}(2010)}]{PhysRevC.82.034329}%
  \BibitemOpen
  \bibfield  {author} {\bibinfo {author} {\bibfnamefont {A.~V.}\ \bibnamefont
  {Afanasjev}}\ and\ \bibinfo {author} {\bibfnamefont {H.}~\bibnamefont
  {Abusara}},\ }\href {\doibase 10.1103/PhysRevC.82.034329} {\bibfield
  {journal} {\bibinfo  {journal} {Phys. Rev. C}\ }\textbf {\bibinfo {volume}
  {82}},\ \bibinfo {pages} {034329} (\bibinfo {year} {2010})}\BibitemShut
  {NoStop}%
\bibitem [{\citenamefont {Zhao}\ \emph
  {et~al.}(2011{\natexlab{a}})\citenamefont {Zhao}, \citenamefont {Peng},
  \citenamefont {Liang}, \citenamefont {Ring},\ and\ \citenamefont
  {Meng}}]{Zhao2011PRL}%
  \BibitemOpen
  \bibfield  {author} {\bibinfo {author} {\bibfnamefont {P.~W.}\ \bibnamefont
  {Zhao}}, \bibinfo {author} {\bibfnamefont {J.}~\bibnamefont {Peng}}, \bibinfo
  {author} {\bibfnamefont {H.~Z.}\ \bibnamefont {Liang}}, \bibinfo {author}
  {\bibfnamefont {P.}~\bibnamefont {Ring}}, \ and\ \bibinfo {author}
  {\bibfnamefont {J.}~\bibnamefont {Meng}},\ }\href {\doibase
  10.1103/PhysRevLett.107.122501} {\bibfield  {journal} {\bibinfo  {journal}
  {Phys. Rev. Lett.}\ }\textbf {\bibinfo {volume} {107}},\ \bibinfo {pages}
  {122501} (\bibinfo {year} {2011}{\natexlab{a}})}\BibitemShut {NoStop}%
\bibitem [{\citenamefont {Zhao}\ \emph
  {et~al.}(2011{\natexlab{b}})\citenamefont {Zhao}, \citenamefont {Zhang},
  \citenamefont {Peng}, \citenamefont {Liang}, \citenamefont {Ring},\ and\
  \citenamefont {Meng}}]{Zhao2011PLB}%
  \BibitemOpen
  \bibfield  {author} {\bibinfo {author} {\bibfnamefont {P.~W.}\ \bibnamefont
  {Zhao}}, \bibinfo {author} {\bibfnamefont {S.~Q.}\ \bibnamefont {Zhang}},
  \bibinfo {author} {\bibfnamefont {J.}~\bibnamefont {Peng}}, \bibinfo {author}
  {\bibfnamefont {H.~Z.}\ \bibnamefont {Liang}}, \bibinfo {author}
  {\bibfnamefont {P.}~\bibnamefont {Ring}}, \ and\ \bibinfo {author}
  {\bibfnamefont {J.}~\bibnamefont {Meng}},\ }\href {\doibase
  10.1016/j.physletb.2011.03.068} {\bibfield  {journal} {\bibinfo  {journal}
  {Phys. Lett. B}\ }\textbf {\bibinfo {volume} {699}},\ \bibinfo {pages} {181}
  (\bibinfo {year} {2011}{\natexlab{b}})}\BibitemShut {NoStop}%
\bibitem [{\citenamefont {Zhao}\ \emph
  {et~al.}(2012{\natexlab{a}})\citenamefont {Zhao}, \citenamefont {Peng},
  \citenamefont {Liang}, \citenamefont {Ring},\ and\ \citenamefont
  {Meng}}]{Zhao2012Phys.Rev.C54310}%
  \BibitemOpen
  \bibfield  {author} {\bibinfo {author} {\bibfnamefont {P.~W.}\ \bibnamefont
  {Zhao}}, \bibinfo {author} {\bibfnamefont {J.}~\bibnamefont {Peng}}, \bibinfo
  {author} {\bibfnamefont {H.~Z.}\ \bibnamefont {Liang}}, \bibinfo {author}
  {\bibfnamefont {P.}~\bibnamefont {Ring}}, \ and\ \bibinfo {author}
  {\bibfnamefont {J.}~\bibnamefont {Meng}},\ }\href {\doibase
  10.1103/PhysRevC.85.054310} {\bibfield  {journal} {\bibinfo  {journal} {Phys.
  Rev. C}\ }\textbf {\bibinfo {volume} {85}},\ \bibinfo {pages} {054310}
  (\bibinfo {year} {2012}{\natexlab{a}})}\BibitemShut {NoStop}%
\bibitem [{\citenamefont {Zhao}\ \emph {et~al.}(2015)\citenamefont {Zhao},
  \citenamefont {Itagaki},\ and\ \citenamefont {Meng}}]{Zhao2015PRL}%
  \BibitemOpen
  \bibfield  {author} {\bibinfo {author} {\bibfnamefont {P.~W.}\ \bibnamefont
  {Zhao}}, \bibinfo {author} {\bibfnamefont {N.}~\bibnamefont {Itagaki}}, \
  and\ \bibinfo {author} {\bibfnamefont {J.}~\bibnamefont {Meng}},\ }\href
  {\doibase 10.1103/PhysRevLett.115.022501} {\bibfield  {journal} {\bibinfo
  {journal} {Phys. Rev. Lett.}\ }\textbf {\bibinfo {volume} {115}},\ \bibinfo
  {pages} {022501} (\bibinfo {year} {2015})}\BibitemShut {NoStop}%
\bibitem [{\citenamefont {Wang}(2017)}]{Wang2017Phys.Rev.C54324}%
  \BibitemOpen
  \bibfield  {author} {\bibinfo {author} {\bibfnamefont {Y.~K.}\ \bibnamefont
  {Wang}},\ }\href {\doibase 10.1103/PhysRevC.96.054324} {\bibfield  {journal}
  {\bibinfo  {journal} {Phys. Rev. C}\ }\textbf {\bibinfo {volume} {96}},\
  \bibinfo {pages} {054324} (\bibinfo {year} {2017})}\BibitemShut {NoStop}%
\bibitem [{\citenamefont {Wang}(2018)}]{Wang2018Phys.Rev.64321}%
  \BibitemOpen
  \bibfield  {author} {\bibinfo {author} {\bibfnamefont {Y.~K.}\ \bibnamefont
  {Wang}},\ }\href {\doibase 10.1103/PhysRevC.97.064321} {\bibfield  {journal}
  {\bibinfo  {journal} {Phys. Rev. C}\ }\textbf {\bibinfo {volume} {97}},\
  \bibinfo {pages} {064321} (\bibinfo {year} {2018})}\BibitemShut {NoStop}%
\bibitem [{\citenamefont {Ren}\ \emph {et~al.}(2019)\citenamefont {Ren},
  \citenamefont {Zhang}, \citenamefont {Zhao}, \citenamefont {Itagaki},
  \citenamefont {Maruhn},\ and\ \citenamefont {Meng}}]{Ren2019SciChina}%
  \BibitemOpen
  \bibfield  {author} {\bibinfo {author} {\bibfnamefont {Z.~X.}\ \bibnamefont
  {Ren}}, \bibinfo {author} {\bibfnamefont {S.~Q.}\ \bibnamefont {Zhang}},
  \bibinfo {author} {\bibfnamefont {P.~W.}\ \bibnamefont {Zhao}}, \bibinfo
  {author} {\bibfnamefont {N.}~\bibnamefont {Itagaki}}, \bibinfo {author}
  {\bibfnamefont {J.~A.}\ \bibnamefont {Maruhn}}, \ and\ \bibinfo {author}
  {\bibfnamefont {J.}~\bibnamefont {Meng}},\ }\href {\doibase
  10.1007/s11433-019-9412-3} {\bibfield  {journal} {\bibinfo  {journal} {Sci.
  China Phys. Mech. Astron.}\ }\textbf {\bibinfo {volume} {62}},\ \bibinfo
  {pages} {112062} (\bibinfo {year} {2019})}\BibitemShut {NoStop}%
\bibitem [{\citenamefont {Ren}\ \emph {et~al.}(2020)\citenamefont {Ren},
  \citenamefont {Zhao}, \citenamefont {Zhang},\ and\ \citenamefont
  {Meng}}]{Ren2020NPA}%
  \BibitemOpen
  \bibfield  {author} {\bibinfo {author} {\bibfnamefont {Z.}~\bibnamefont
  {Ren}}, \bibinfo {author} {\bibfnamefont {P.}~\bibnamefont {Zhao}}, \bibinfo
  {author} {\bibfnamefont {S.}~\bibnamefont {Zhang}}, \ and\ \bibinfo {author}
  {\bibfnamefont {J.}~\bibnamefont {Meng}},\ }\href {\doibase
  https://doi.org/10.1016/j.nuclphysa.2020.121696} {\bibfield  {journal}
  {\bibinfo  {journal} {Nucl. Phys. A}\ }\textbf {\bibinfo {volume}
  {996}},\ \bibinfo {pages} {121696} (\bibinfo {year} {2020})}\BibitemShut
  {NoStop}%
\bibitem [{\citenamefont {Lalazissis}\ \emph {et~al.}(1999)\citenamefont
  {Lalazissis}, \citenamefont {Raman},\ and\ \citenamefont
  {Ring}}]{Lalazissis1999ADNDT}%
  \BibitemOpen
  \bibfield  {author} {\bibinfo {author} {\bibfnamefont {G.}~\bibnamefont
  {Lalazissis}}, \bibinfo {author} {\bibfnamefont {S.}~\bibnamefont {Raman}}, \
  and\ \bibinfo {author} {\bibfnamefont {P.}~\bibnamefont {Ring}},\ }\href
  {\doibase 10.1006/adnd.1998.0795} {\bibfield  {journal} {\bibinfo  {journal}
  {Atom. Data Nucl. Data Tabl.}\ }\textbf {\bibinfo {volume} {71}},\ \bibinfo
  {pages} {1 } (\bibinfo {year} {1999})}\BibitemShut {NoStop}%
\bibitem [{\citenamefont {Geng}\ \emph {et~al.}(2005)\citenamefont {Geng},
  \citenamefont {Toki},\ and\ \citenamefont {Meng}}]{Geng2005PTP}%
  \BibitemOpen
  \bibfield  {author} {\bibinfo {author} {\bibfnamefont {L.-S.}\ \bibnamefont
  {Geng}}, \bibinfo {author} {\bibfnamefont {H.}~\bibnamefont {Toki}}, \ and\
  \bibinfo {author} {\bibfnamefont {J.}~\bibnamefont {Meng}},\ }\href {\doibase
  10.1143/PTP.113.785} {\bibfield  {journal} {\bibinfo  {journal} {Prog. Theor.
  Phys.}\ }\textbf {\bibinfo {volume} {113}},\ \bibinfo {pages} {785} (\bibinfo
  {year} {2005})}\BibitemShut {NoStop}%
\bibitem [{\citenamefont {Afanasjev}\ \emph {et~al.}(2013)\citenamefont
  {Afanasjev}, \citenamefont {Agbemava}, \citenamefont {Ray},\ and\
  \citenamefont {Ring}}]{Afanasjev2013PLB}%
  \BibitemOpen
  \bibfield  {author} {\bibinfo {author} {\bibfnamefont {A.~V.}\ \bibnamefont
  {Afanasjev}}, \bibinfo {author} {\bibfnamefont {S.~E.}\ \bibnamefont
  {Agbemava}}, \bibinfo {author} {\bibfnamefont {D.}~\bibnamefont {Ray}}, \
  and\ \bibinfo {author} {\bibfnamefont {P.}~\bibnamefont {Ring}},\ }\href
  {\doibase 10.1016/j.physletb.2013.09.017} {\bibfield  {journal} {\bibinfo
  {journal} {Phys. Lett. B}\ }\textbf {\bibinfo {volume} {726}},\ \bibinfo
  {pages} {680 } (\bibinfo {year} {2013})}\BibitemShut {NoStop}%
\bibitem [{\citenamefont {Zhang}\ \emph
  {et~al.}(2014{\natexlab{b}})\citenamefont {Zhang}, \citenamefont {Niu},
  \citenamefont {Li}, \citenamefont {Yao},\ and\ \citenamefont
  {Meng}}]{Zhang2014FOP}%
  \BibitemOpen
  \bibfield  {author} {\bibinfo {author} {\bibfnamefont {Q.~S.}\ \bibnamefont
  {Zhang}}, \bibinfo {author} {\bibfnamefont {Z.~M.}\ \bibnamefont {Niu}},
  \bibinfo {author} {\bibfnamefont {Z.~P.}\ \bibnamefont {Li}}, \bibinfo
  {author} {\bibfnamefont {J.~M.}\ \bibnamefont {Yao}}, \ and\ \bibinfo
  {author} {\bibfnamefont {J.}~\bibnamefont {Meng}},\ }\href {\doibase
  10.1007/s11467-014-0413-5} {\bibfield  {journal} {\bibinfo  {journal} {Front.
  Phys.}\ }\textbf {\bibinfo {volume} {9}},\ \bibinfo {pages} {529} (\bibinfo
  {year} {2014}{\natexlab{b}})}\BibitemShut {NoStop}%
\bibitem [{\citenamefont {Agbemava}\ \emph {et~al.}(2014)\citenamefont
  {Agbemava}, \citenamefont {Afanasjev}, \citenamefont {Ray},\ and\
  \citenamefont {Ring}}]{Agbemava2014PRC}%
  \BibitemOpen
  \bibfield  {author} {\bibinfo {author} {\bibfnamefont {S.~E.}\ \bibnamefont
  {Agbemava}}, \bibinfo {author} {\bibfnamefont {A.~V.}\ \bibnamefont
  {Afanasjev}}, \bibinfo {author} {\bibfnamefont {D.}~\bibnamefont {Ray}}, \
  and\ \bibinfo {author} {\bibfnamefont {P.}~\bibnamefont {Ring}},\ }\href
  {\doibase 10.1103/PhysRevC.89.054320} {\bibfield  {journal} {\bibinfo
  {journal} {Phys. Rev. C}\ }\textbf {\bibinfo {volume} {89}},\ \bibinfo
  {pages} {054320} (\bibinfo {year} {2014})}\BibitemShut {NoStop}%
\bibitem [{\citenamefont {Afanasjev}\ \emph {et~al.}(2015)\citenamefont
  {Afanasjev}, \citenamefont {Agbemava}, \citenamefont {Ray},\ and\
  \citenamefont {Ring}}]{Afanasjev2015PRC}%
  \BibitemOpen
  \bibfield  {author} {\bibinfo {author} {\bibfnamefont {A.~V.}\ \bibnamefont
  {Afanasjev}}, \bibinfo {author} {\bibfnamefont {S.~E.}\ \bibnamefont
  {Agbemava}}, \bibinfo {author} {\bibfnamefont {D.}~\bibnamefont {Ray}}, \
  and\ \bibinfo {author} {\bibfnamefont {P.}~\bibnamefont {Ring}},\ }\href
  {\doibase 10.1103/PhysRevC.91.014324} {\bibfield  {journal} {\bibinfo
  {journal} {Phys. Rev. C}\ }\textbf {\bibinfo {volume} {91}},\ \bibinfo
  {pages} {014324} (\bibinfo {year} {2015})}\BibitemShut {NoStop}%
\bibitem [{\citenamefont {Lu}\ \emph {et~al.}(2015)\citenamefont {Lu},
  \citenamefont {Li}, \citenamefont {Li}, \citenamefont {Yao},\ and\
  \citenamefont {Meng}}]{Lu2015PRC}%
  \BibitemOpen
  \bibfield  {author} {\bibinfo {author} {\bibfnamefont {K.~Q.}\ \bibnamefont
  {Lu}}, \bibinfo {author} {\bibfnamefont {Z.~X.}\ \bibnamefont {Li}}, \bibinfo
  {author} {\bibfnamefont {Z.~P.}\ \bibnamefont {Li}}, \bibinfo {author}
  {\bibfnamefont {J.~M.}\ \bibnamefont {Yao}}, \ and\ \bibinfo {author}
  {\bibfnamefont {J.}~\bibnamefont {Meng}},\ }\href {\doibase
  10.1103/PhysRevC.91.027304} {\bibfield  {journal} {\bibinfo  {journal} {Phys.
  Rev. C}\ }\textbf {\bibinfo {volume} {91}},\ \bibinfo {pages} {027304}
  (\bibinfo {year} {2015})}\BibitemShut {NoStop}%
\bibitem [{\citenamefont {Pe{\~{n}}a-Arteaga}\ \emph
  {et~al.}(2016)\citenamefont {Pe{\~{n}}a-Arteaga}, \citenamefont {Goriely},\
  and\ \citenamefont {Chamel}}]{Pena-Arteaga2016EPJA}%
  \BibitemOpen
  \bibfield  {author} {\bibinfo {author} {\bibfnamefont {D.}~\bibnamefont
  {Pe{\~{n}}a-Arteaga}}, \bibinfo {author} {\bibfnamefont {S.}~\bibnamefont
  {Goriely}}, \ and\ \bibinfo {author} {\bibfnamefont {N.}~\bibnamefont
  {Chamel}},\ }\href {\doibase 10.1140/epja/i2016-16320-x} {\bibfield
  {journal} {\bibinfo  {journal} {Eur. Phys. J. A}\ }\textbf {\bibinfo {volume}
  {52}},\ \bibinfo {pages} {320} (\bibinfo {year} {2016})}\BibitemShut
  {NoStop}%
\bibitem [{\citenamefont {Xia}\ \emph {et~al.}(2018)\citenamefont {Xia},
  \citenamefont {Lim}, \citenamefont {Zhao}, \citenamefont {Liang},
  \citenamefont {Qu}, \citenamefont {Chen}, \citenamefont {Liu}, \citenamefont
  {Zhang}, \citenamefont {Zhang}, \citenamefont {Kim},\ and\ \citenamefont
  {Meng}}]{Xia2018ADNDT}%
  \BibitemOpen
  \bibfield  {author} {\bibinfo {author} {\bibfnamefont {X.~W.}\ \bibnamefont
  {Xia}}, \bibinfo {author} {\bibfnamefont {Y.}~\bibnamefont {Lim}}, \bibinfo
  {author} {\bibfnamefont {P.~W.}\ \bibnamefont {Zhao}}, \bibinfo {author}
  {\bibfnamefont {H.~Z.}\ \bibnamefont {Liang}}, \bibinfo {author}
  {\bibfnamefont {X.~Y.}\ \bibnamefont {Qu}}, \bibinfo {author} {\bibfnamefont
  {Y.}~\bibnamefont {Chen}}, \bibinfo {author} {\bibfnamefont {H.}~\bibnamefont
  {Liu}}, \bibinfo {author} {\bibfnamefont {L.~F.}\ \bibnamefont {Zhang}},
  \bibinfo {author} {\bibfnamefont {S.~Q.}\ \bibnamefont {Zhang}}, \bibinfo
  {author} {\bibfnamefont {Y.}~\bibnamefont {Kim}}, \ and\ \bibinfo {author}
  {\bibfnamefont {J.}~\bibnamefont {Meng}},\ }\href {\doibase
  10.1016/j.adt.2017.09.001} {\bibfield  {journal} {\bibinfo  {journal} {Atom.
  Data Nucl. Data Tabl.}\ }\textbf {\bibinfo {volume} {121-122}},\ \bibinfo
  {pages} {1} (\bibinfo {year} {2018})}\BibitemShut {NoStop}%
\bibitem [{\citenamefont {Yang}\ \emph
  {et~al.}(2021{\natexlab{a}})\citenamefont {Yang}, \citenamefont {Wang},
  \citenamefont {Zhao},\ and\ \citenamefont {Li}}]{Yang2021PRC}%
  \BibitemOpen
  \bibfield  {author} {\bibinfo {author} {\bibfnamefont {Y.~L.}\ \bibnamefont
  {Yang}}, \bibinfo {author} {\bibfnamefont {Y.~K.}\ \bibnamefont {Wang}},
  \bibinfo {author} {\bibfnamefont {P.~W.}\ \bibnamefont {Zhao}}, \ and\
  \bibinfo {author} {\bibfnamefont {Z.~P.}\ \bibnamefont {Li}},\ }\href
  {\doibase 10.1103/PhysRevC.104.054312} {\bibfield  {journal} {\bibinfo
  {journal} {Phys. Rev. C}\ }\textbf {\bibinfo {volume} {104}},\ \bibinfo
  {pages} {054312} (\bibinfo {year} {2021}{\natexlab{a}})}\BibitemShut
  {NoStop}%
\bibitem [{\citenamefont {Zhao}\ \emph {et~al.}(2010)\citenamefont {Zhao},
  \citenamefont {Li}, \citenamefont {Yao},\ and\ \citenamefont
  {Meng}}]{Zhao2010PRC}%
  \BibitemOpen
  \bibfield  {author} {\bibinfo {author} {\bibfnamefont {P.~W.}\ \bibnamefont
  {Zhao}}, \bibinfo {author} {\bibfnamefont {Z.~P.}\ \bibnamefont {Li}},
  \bibinfo {author} {\bibfnamefont {J.~M.}\ \bibnamefont {Yao}}, \ and\
  \bibinfo {author} {\bibfnamefont {J.}~\bibnamefont {Meng}},\ }\href {\doibase
  10.1103/PhysRevC.82.054319} {\bibfield  {journal} {\bibinfo  {journal} {Phys.
  Rev. C}\ }\textbf {\bibinfo {volume} {82}},\ \bibinfo {pages} {054319}
  (\bibinfo {year} {2010})}\BibitemShut {NoStop}%
\bibitem [{\citenamefont {Coraggio}\ \emph {et~al.}(2009)\citenamefont
  {Coraggio}, \citenamefont {Covello}, \citenamefont {Gargano}, \citenamefont
  {Itaco},\ and\ \citenamefont {Kuo}}]{Coraggio2009PPNP}%
  \BibitemOpen
  \bibfield  {author} {\bibinfo {author} {\bibfnamefont {L.}~\bibnamefont
  {Coraggio}}, \bibinfo {author} {\bibfnamefont {A.}~\bibnamefont {Covello}},
  \bibinfo {author} {\bibfnamefont {A.}~\bibnamefont {Gargano}}, \bibinfo
  {author} {\bibfnamefont {N.}~\bibnamefont {Itaco}}, \ and\ \bibinfo {author}
  {\bibfnamefont {T.}~\bibnamefont {Kuo}},\ }\href {\doibase
  https://doi.org/10.1016/j.ppnp.2008.06.001} {\bibfield  {journal} {\bibinfo
  {journal} {Prog. Part. Nucl. Phys.}\ }\textbf {\bibinfo {volume} {62}},\
  \bibinfo {pages} {135} (\bibinfo {year} {2009})}\BibitemShut {NoStop}%
\bibitem [{\citenamefont {Barrett}\ \emph {et~al.}(2013)\citenamefont
  {Barrett}, \citenamefont {Navr\'{a}til},\ and\ \citenamefont
  {Vary}}]{Barrett2013PPNP}%
  \BibitemOpen
  \bibfield  {author} {\bibinfo {author} {\bibfnamefont {B.~R.}\ \bibnamefont
  {Barrett}}, \bibinfo {author} {\bibfnamefont {P.}~\bibnamefont {Navr\'{a}til}},
  \ and\ \bibinfo {author} {\bibfnamefont {J.~P.}\ \bibnamefont {Vary}},\
  }\href {\doibase https://doi.org/10.1016/j.ppnp.2012.10.003} {\bibfield
  {journal} {\bibinfo  {journal} {Prog. Part. Nucl. Phys.}\ }\textbf {\bibinfo
  {volume} {69}},\ \bibinfo {pages} {131} (\bibinfo {year} {2013})}\BibitemShut
  {NoStop}%
\bibitem [{\citenamefont {Otsuka}\ \emph {et~al.}(2020)\citenamefont {Otsuka},
  \citenamefont {Gade}, \citenamefont {Sorlin}, \citenamefont {Suzuki},\ and\
  \citenamefont {Utsuno}}]{Otsuka2020RMP}%
  \BibitemOpen
  \bibfield  {author} {\bibinfo {author} {\bibfnamefont {T.}~\bibnamefont
  {Otsuka}}, \bibinfo {author} {\bibfnamefont {A.}~\bibnamefont {Gade}},
  \bibinfo {author} {\bibfnamefont {O.}~\bibnamefont {Sorlin}}, \bibinfo
  {author} {\bibfnamefont {T.}~\bibnamefont {Suzuki}}, \ and\ \bibinfo {author}
  {\bibfnamefont {Y.}~\bibnamefont {Utsuno}},\ }\href {\doibase
  10.1103/RevModPhys.92.015002} {\bibfield  {journal} {\bibinfo  {journal}
  {Rev. Mod. Phys.}\ }\textbf {\bibinfo {volume} {92}},\ \bibinfo {pages}
  {015002} (\bibinfo {year} {2020})}\BibitemShut {NoStop}%
\bibitem [{\citenamefont {Hoffman}\ \emph {et~al.}(2008)\citenamefont
  {Hoffman}, \citenamefont {Baumann}, \citenamefont {Bazin}, \citenamefont
  {Brown}, \citenamefont {Christian}, \citenamefont {DeYoung}, \citenamefont
  {Finck}, \citenamefont {Frank}, \citenamefont {Hinnefeld}, \citenamefont
  {Howes}, \citenamefont {Mears}, \citenamefont {Mosby}, \citenamefont {Mosby},
  \citenamefont {Reith}, \citenamefont {Rizzo}, \citenamefont {Rogers},
  \citenamefont {Peaslee}, \citenamefont {Peters}, \citenamefont {Schiller},
  \citenamefont {Scott}, \citenamefont {Tabor}, \citenamefont {Thoennessen},
  \citenamefont {Voss},\ and\ \citenamefont {Williams}}]{Hoffman2008PRL}%
  \BibitemOpen
  \bibfield  {author} {\bibinfo {author} {\bibfnamefont {C.~R.}\ \bibnamefont
  {Hoffman}}, \bibinfo {author} {\bibfnamefont {T.}~\bibnamefont {Baumann}},
  \bibinfo {author} {\bibfnamefont {D.}~\bibnamefont {Bazin}}, \bibinfo
  {author} {\bibfnamefont {J.}~\bibnamefont {Brown}}, \bibinfo {author}
  {\bibfnamefont {G.}~\bibnamefont {Christian}}, \bibinfo {author}
  {\bibfnamefont {P.~A.}\ \bibnamefont {DeYoung}}, \bibinfo {author}
  {\bibfnamefont {J.~E.}\ \bibnamefont {Finck}}, \bibinfo {author}
  {\bibfnamefont {N.}~\bibnamefont {Frank}}, \bibinfo {author} {\bibfnamefont
  {J.}~\bibnamefont {Hinnefeld}}, \bibinfo {author} {\bibfnamefont
  {R.}~\bibnamefont {Howes}}, \bibinfo {author} {\bibfnamefont
  {P.}~\bibnamefont {Mears}}, \bibinfo {author} {\bibfnamefont
  {E.}~\bibnamefont {Mosby}}, \bibinfo {author} {\bibfnamefont
  {S.}~\bibnamefont {Mosby}}, \bibinfo {author} {\bibfnamefont
  {J.}~\bibnamefont {Reith}}, \bibinfo {author} {\bibfnamefont
  {B.}~\bibnamefont {Rizzo}}, \bibinfo {author} {\bibfnamefont {W.~F.}\
  \bibnamefont {Rogers}}, \bibinfo {author} {\bibfnamefont {G.}~\bibnamefont
  {Peaslee}}, \bibinfo {author} {\bibfnamefont {W.~A.}\ \bibnamefont {Peters}},
  \bibinfo {author} {\bibfnamefont {A.}~\bibnamefont {Schiller}}, \bibinfo
  {author} {\bibfnamefont {M.~J.}\ \bibnamefont {Scott}}, \bibinfo {author}
  {\bibfnamefont {S.~L.}\ \bibnamefont {Tabor}}, \bibinfo {author}
  {\bibfnamefont {M.}~\bibnamefont {Thoennessen}}, \bibinfo {author}
  {\bibfnamefont {P.~J.}\ \bibnamefont {Voss}}, \ and\ \bibinfo {author}
  {\bibfnamefont {T.}~\bibnamefont {Williams}},\ }\href {\doibase
  10.1103/PhysRevLett.100.152502} {\bibfield  {journal} {\bibinfo  {journal}
  {Phys. Rev. Lett.}\ }\textbf {\bibinfo {volume} {100}},\ \bibinfo {pages}
  {152502} (\bibinfo {year} {2008})}\BibitemShut {NoStop}%
\bibitem [{\citenamefont {Otsuka}\ \emph {et~al.}(2010)\citenamefont {Otsuka},
  \citenamefont {Suzuki}, \citenamefont {Holt}, \citenamefont {Schwenk},\ and\
  \citenamefont {Akaishi}}]{Otsuka2010PRL}%
  \BibitemOpen
  \bibfield  {author} {\bibinfo {author} {\bibfnamefont {T.}~\bibnamefont
  {Otsuka}}, \bibinfo {author} {\bibfnamefont {T.}~\bibnamefont {Suzuki}},
  \bibinfo {author} {\bibfnamefont {J.~D.}\ \bibnamefont {Holt}}, \bibinfo
  {author} {\bibfnamefont {A.}~\bibnamefont {Schwenk}}, \ and\ \bibinfo
  {author} {\bibfnamefont {Y.}~\bibnamefont {Akaishi}},\ }\href {\doibase
  10.1103/PhysRevLett.105.032501} {\bibfield  {journal} {\bibinfo  {journal}
  {Phys. Rev. Lett.}\ }\textbf {\bibinfo {volume} {105}},\ \bibinfo {pages}
  {032501} (\bibinfo {year} {2010})}\BibitemShut {NoStop}%
\bibitem [{\citenamefont {Meng}\ and\ \citenamefont
  {Ring}(1996)}]{Meng1996PRL}%
  \BibitemOpen
  \bibfield  {author} {\bibinfo {author} {\bibfnamefont {J.}~\bibnamefont
  {Meng}}\ and\ \bibinfo {author} {\bibfnamefont {P.}~\bibnamefont {Ring}},\
  }\href {\doibase 10.1103/PhysRevLett.77.3963} {\bibfield  {journal} {\bibinfo
   {journal} {Phys. Rev. Lett.}\ }\textbf {\bibinfo {volume} {77}},\ \bibinfo
  {pages} {3963} (\bibinfo {year} {1996})}\BibitemShut {NoStop}%
\bibitem [{\citenamefont {Meng}(1998)}]{Meng1998NPA}%
  \BibitemOpen
  \bibfield  {author} {\bibinfo {author} {\bibfnamefont {J.}~\bibnamefont
  {Meng}},\ }\href {\doibase 10.1016/S0375-9474(98)00178-X} {\bibfield
  {journal} {\bibinfo  {journal} {Nucl. Phys. A}\ }\textbf {\bibinfo {volume}
  {635}},\ \bibinfo {pages} {3} (\bibinfo {year} {1998})}\BibitemShut {NoStop}%
\bibitem [{\citenamefont {Meng}\ and\ \citenamefont
  {Ring}(1998)}]{Meng1998PRL}%
  \BibitemOpen
  \bibfield  {author} {\bibinfo {author} {\bibfnamefont {J.}~\bibnamefont
  {Meng}}\ and\ \bibinfo {author} {\bibfnamefont {P.}~\bibnamefont {Ring}},\
  }\href {\doibase 10.1103/PhysRevLett.80.460} {\bibfield  {journal} {\bibinfo
  {journal} {Phys. Rev. Lett.}\ }\textbf {\bibinfo {volume} {80}},\ \bibinfo
  {pages} {460} (\bibinfo {year} {1998})}\BibitemShut {NoStop}%
\bibitem [{\citenamefont {Meng}\ \emph
  {et~al.}(2002{\natexlab{a}})\citenamefont {Meng}, \citenamefont {Toki},
  \citenamefont {Zeng}, \citenamefont {Zhang},\ and\ \citenamefont
  {Zhou}}]{Meng2002PRC(R)}%
  \BibitemOpen
  \bibfield  {author} {\bibinfo {author} {\bibfnamefont {J.}~\bibnamefont
  {Meng}}, \bibinfo {author} {\bibfnamefont {H.}~\bibnamefont {Toki}}, \bibinfo
  {author} {\bibfnamefont {J.~Y.}\ \bibnamefont {Zeng}}, \bibinfo {author}
  {\bibfnamefont {S.~Q.}\ \bibnamefont {Zhang}}, \ and\ \bibinfo {author}
  {\bibfnamefont {S.-G.}\ \bibnamefont {Zhou}},\ }\href {\doibase
  10.1103/PhysRevC.65.041302} {\bibfield  {journal} {\bibinfo  {journal} {Phys.
  Rev. C}\ }\textbf {\bibinfo {volume} {65}},\ \bibinfo {pages} {041302(R)}
  (\bibinfo {year} {2002}{\natexlab{a}})}\BibitemShut {NoStop}%
\bibitem [{\citenamefont {Zhang}\ \emph {et~al.}(2003)\citenamefont {Zhang},
  \citenamefont {Meng},\ and\ \citenamefont {Zhou}}]{Zhang2003SciChina}%
  \BibitemOpen
  \bibfield  {author} {\bibinfo {author} {\bibfnamefont {S.~Q.}\ \bibnamefont
  {Zhang}}, \bibinfo {author} {\bibfnamefont {J.}~\bibnamefont {Meng}}, \ and\
  \bibinfo {author} {\bibfnamefont {S.~G.}\ \bibnamefont {Zhou}},\ }\href
  {\doibase doi = "https://doi.org/10.1360/03yw0140"} {\bibfield  {journal}
  {\bibinfo  {journal} {Sci. China Ser. G: Phys. Ast.}\ }\textbf {\bibinfo
  {volume} {46}},\ \bibinfo {pages} {632} (\bibinfo {year} {2003})}\BibitemShut
  {NoStop}%
\bibitem [{\citenamefont {Meng}\ \emph
  {et~al.}(1998{\natexlab{b}})\citenamefont {Meng}, \citenamefont {Tanihata},\
  and\ \citenamefont {Yamaji}}]{Meng1998PLB}%
  \BibitemOpen
  \bibfield  {author} {\bibinfo {author} {\bibfnamefont {J.}~\bibnamefont
  {Meng}}, \bibinfo {author} {\bibfnamefont {I.}~\bibnamefont {Tanihata}}, \
  and\ \bibinfo {author} {\bibfnamefont {S.}~\bibnamefont {Yamaji}},\ }\href
  {\doibase 10.1016/S0370-2693(97)01386-5} {\bibfield  {journal} {\bibinfo
  {journal} {Phys. Lett. B}\ }\textbf {\bibinfo {volume} {419}},\ \bibinfo
  {pages} {1} (\bibinfo {year} {1998}{\natexlab{b}})}\BibitemShut {NoStop}%
\bibitem [{\citenamefont {Meng}\ \emph
  {et~al.}(2002{\natexlab{b}})\citenamefont {Meng}, \citenamefont {Zhou},\ and\
  \citenamefont {Tanihata}}]{Meng2002PLB}%
  \BibitemOpen
  \bibfield  {author} {\bibinfo {author} {\bibfnamefont {J.}~\bibnamefont
  {Meng}}, \bibinfo {author} {\bibfnamefont {S.~G.}\ \bibnamefont {Zhou}}, \
  and\ \bibinfo {author} {\bibfnamefont {I.}~\bibnamefont {Tanihata}},\ }\href
  {\doibase 10.1016/S0370-2693(02)01574-5} {\bibfield  {journal} {\bibinfo
  {journal} {Phys. Lett. B}\ }\textbf {\bibinfo {volume} {532}},\ \bibinfo
  {pages} {209} (\bibinfo {year} {2002}{\natexlab{b}})}\BibitemShut {NoStop}%
\bibitem [{\citenamefont {Zhang}\ \emph {et~al.}(2005)\citenamefont {Zhang},
  \citenamefont {Meng}, \citenamefont {Zhang}, \citenamefont {Geng},\ and\
  \citenamefont {Toki}}]{Zhang2005NPA}%
  \BibitemOpen
  \bibfield  {author} {\bibinfo {author} {\bibfnamefont {W.}~\bibnamefont
  {Zhang}}, \bibinfo {author} {\bibfnamefont {J.}~\bibnamefont {Meng}},
  \bibinfo {author} {\bibfnamefont {S.~Q.}\ \bibnamefont {Zhang}}, \bibinfo
  {author} {\bibfnamefont {L.~S.}\ \bibnamefont {Geng}}, \ and\ \bibinfo
  {author} {\bibfnamefont {H.}~\bibnamefont {Toki}},\ }\href {\doibase
  10.1016/j.nuclphysa.2005.02.086} {\bibfield  {journal} {\bibinfo  {journal}
  {Nucl. Phys. A}\ }\textbf {\bibinfo {volume} {753}},\ \bibinfo {pages} {106}
  (\bibinfo {year} {2005})}\BibitemShut {NoStop}%
\bibitem [{\citenamefont {Lu}\ \emph {et~al.}(2003)\citenamefont {Lu},
  \citenamefont {Meng}, \citenamefont {Zhang},\ and\ \citenamefont
  {Zhou}}]{Lyu2003EPJA}%
  \BibitemOpen
  \bibfield  {author} {\bibinfo {author} {\bibfnamefont {H.~F.}\ \bibnamefont
  {Lu}}, \bibinfo {author} {\bibfnamefont {J.}~\bibnamefont {Meng}}, \bibinfo
  {author} {\bibfnamefont {S.~Q.}\ \bibnamefont {Zhang}}, \ and\ \bibinfo
  {author} {\bibfnamefont {S.~G.}\ \bibnamefont {Zhou}},\ }\href {\doibase
  10.1140/epja/i2002-10136-3} {\bibfield  {journal} {\bibinfo  {journal} {Eur.
  Phys. J. A}\ }\textbf {\bibinfo {volume} {17}},\ \bibinfo {pages} {19}
  (\bibinfo {year} {2003})}\BibitemShut {NoStop}%
\bibitem [{\citenamefont {Zhang}\ and\ \citenamefont
  {Xia}(2016)}]{Zhang2016CPC}%
  \BibitemOpen
  \bibfield  {author} {\bibinfo {author} {\bibfnamefont {L.-F.}\ \bibnamefont
  {Zhang}}\ and\ \bibinfo {author} {\bibfnamefont {X.-W.}\ \bibnamefont
  {Xia}},\ }\href {\doibase 10.1088/1674-1137/40/5/054102} {\bibfield
  {journal} {\bibinfo  {journal} {Chin. Phys. C}\ }\textbf {\bibinfo {volume}
  {40}},\ \bibinfo {pages} {054102} (\bibinfo {year} {2016})}\BibitemShut
  {NoStop}%
\bibitem [{\citenamefont {Lim}\ \emph {et~al.}(2016)\citenamefont {Lim},
  \citenamefont {Xia},\ and\ \citenamefont {Kim}}]{Lim2016PRC}%
  \BibitemOpen
  \bibfield  {author} {\bibinfo {author} {\bibfnamefont {Y.}~\bibnamefont
  {Lim}}, \bibinfo {author} {\bibfnamefont {X.}~\bibnamefont {Xia}}, \ and\
  \bibinfo {author} {\bibfnamefont {Y.}~\bibnamefont {Kim}},\ }\href {\doibase
  10.1103/PhysRevC.93.014314} {\bibfield  {journal} {\bibinfo  {journal} {Phys.
  Rev. C}\ }\textbf {\bibinfo {volume} {93}},\ \bibinfo {pages} {014314}
  (\bibinfo {year} {2016})}\BibitemShut {NoStop}%
\bibitem [{\citenamefont {Zhou}\ \emph {et~al.}(2000)\citenamefont {Zhou},
  \citenamefont {Meng}, \citenamefont {Yamaji},\ and\ \citenamefont
  {Yang}}]{Zhou2000CPL}%
  \BibitemOpen
  \bibfield  {author} {\bibinfo {author} {\bibfnamefont {S.-G.}\ \bibnamefont
  {Zhou}}, \bibinfo {author} {\bibfnamefont {J.}~\bibnamefont {Meng}}, \bibinfo
  {author} {\bibfnamefont {S.}~\bibnamefont {Yamaji}}, \ and\ \bibinfo {author}
  {\bibfnamefont {S.-C.}\ \bibnamefont {Yang}},\ }\href {\doibase
  10.1088/0256-307x/17/10/006} {\bibfield  {journal} {\bibinfo  {journal}
  {Chin. Phys. Lett.}\ }\textbf {\bibinfo {volume} {17}},\ \bibinfo {pages}
  {717} (\bibinfo {year} {2000})}\BibitemShut {NoStop}%
\bibitem [{\citenamefont {Zhou}\ \emph {et~al.}(2010)\citenamefont {Zhou},
  \citenamefont {Meng}, \citenamefont {Ring},\ and\ \citenamefont
  {Zhao}}]{Zhou2010PRC(R)}%
  \BibitemOpen
  \bibfield  {author} {\bibinfo {author} {\bibfnamefont {S.-G.}\ \bibnamefont
  {Zhou}}, \bibinfo {author} {\bibfnamefont {J.}~\bibnamefont {Meng}}, \bibinfo
  {author} {\bibfnamefont {P.}~\bibnamefont {Ring}}, \ and\ \bibinfo {author}
  {\bibfnamefont {E.-G.}\ \bibnamefont {Zhao}},\ }\href {\doibase
  10.1103/PhysRevC.82.011301} {\bibfield  {journal} {\bibinfo  {journal} {Phys.
  Rev. C}\ }\textbf {\bibinfo {volume} {82}},\ \bibinfo {pages} {011301(R)}
  (\bibinfo {year} {2010})}\BibitemShut {NoStop}%
\bibitem [{\citenamefont {Li}\ \emph {et~al.}(2012{\natexlab{a}})\citenamefont
  {Li}, \citenamefont {Meng}, \citenamefont {Ring}, \citenamefont {Zhao},\ and\
  \citenamefont {Zhou}}]{Li2012PRC}%
  \BibitemOpen
  \bibfield  {author} {\bibinfo {author} {\bibfnamefont {L.}~\bibnamefont
  {Li}}, \bibinfo {author} {\bibfnamefont {J.}~\bibnamefont {Meng}}, \bibinfo
  {author} {\bibfnamefont {P.}~\bibnamefont {Ring}}, \bibinfo {author}
  {\bibfnamefont {E.-G.}\ \bibnamefont {Zhao}}, \ and\ \bibinfo {author}
  {\bibfnamefont {S.-G.}\ \bibnamefont {Zhou}},\ }\href {\doibase
  10.1103/PhysRevC.85.024312} {\bibfield  {journal} {\bibinfo  {journal} {Phys.
  Rev. C}\ }\textbf {\bibinfo {volume} {85}},\ \bibinfo {pages} {024312}
  (\bibinfo {year} {2012}{\natexlab{a}})}\BibitemShut {NoStop}%
\bibitem [{\citenamefont {Zhou}\ \emph
  {et~al.}(2003{\natexlab{b}})\citenamefont {Zhou}, \citenamefont {Meng},\ and\
  \citenamefont {Ring}}]{Zhou2003PRC}%
  \BibitemOpen
  \bibfield  {author} {\bibinfo {author} {\bibfnamefont {S.-G.}\ \bibnamefont
  {Zhou}}, \bibinfo {author} {\bibfnamefont {J.}~\bibnamefont {Meng}}, \ and\
  \bibinfo {author} {\bibfnamefont {P.}~\bibnamefont {Ring}},\ }\href {\doibase
  10.1103/PhysRevC.68.034323} {\bibfield  {journal} {\bibinfo  {journal} {Phys.
  Rev. C}\ }\textbf {\bibinfo {volume} {68}},\ \bibinfo {pages} {034323}
  (\bibinfo {year} {2003}{\natexlab{b}})}\BibitemShut {NoStop}%
\bibitem [{\citenamefont {Chen}\ \emph {et~al.}(2012)\citenamefont {Chen},
  \citenamefont {Li}, \citenamefont {Liang},\ and\ \citenamefont
  {Meng}}]{Chen2012PRC}%
  \BibitemOpen
  \bibfield  {author} {\bibinfo {author} {\bibfnamefont {Y.}~\bibnamefont
  {Chen}}, \bibinfo {author} {\bibfnamefont {L.}~\bibnamefont {Li}}, \bibinfo
  {author} {\bibfnamefont {H.}~\bibnamefont {Liang}}, \ and\ \bibinfo {author}
  {\bibfnamefont {J.}~\bibnamefont {Meng}},\ }\href {\doibase
  10.1103/PhysRevC.85.067301} {\bibfield  {journal} {\bibinfo  {journal} {Phys.
  Rev. C}\ }\textbf {\bibinfo {volume} {85}},\ \bibinfo {pages} {067301}
  (\bibinfo {year} {2012})}\BibitemShut {NoStop}%
\bibitem [{\citenamefont {Li}\ \emph {et~al.}(2012{\natexlab{b}})\citenamefont
  {Li}, \citenamefont {Meng}, \citenamefont {Ring}, \citenamefont {Zhao},\ and\
  \citenamefont {Zhou}}]{Li2012CPL}%
  \BibitemOpen
  \bibfield  {author} {\bibinfo {author} {\bibfnamefont {L.}~\bibnamefont
  {Li}}, \bibinfo {author} {\bibfnamefont {J.}~\bibnamefont {Meng}}, \bibinfo
  {author} {\bibfnamefont {P.}~\bibnamefont {Ring}}, \bibinfo {author}
  {\bibfnamefont {E.-G.}\ \bibnamefont {Zhao}}, \ and\ \bibinfo {author}
  {\bibfnamefont {S.-G.}\ \bibnamefont {Zhou}},\ }\href {\doibase
  10.1088/0256-307X/29/4/042101} {\bibfield  {journal} {\bibinfo  {journal}
  {Chin. Phys. Lett.}\ }\textbf {\bibinfo {volume} {29}},\ \bibinfo {pages}
  {042101} (\bibinfo {year} {2012}{\natexlab{b}})}\BibitemShut {NoStop}%
\bibitem [{\citenamefont {Sun}\ and\ \citenamefont
  {Zhou}(2021{\natexlab{a}})}]{Sun2021SciBull}%
  \BibitemOpen
  \bibfield  {author} {\bibinfo {author} {\bibfnamefont {X.-X.}\ \bibnamefont
  {Sun}}\ and\ \bibinfo {author} {\bibfnamefont {S.-G.}\ \bibnamefont {Zhou}},\
  }\href {\doibase 10.1016/j.scib.2021.07.005} {\bibfield  {journal} {\bibinfo
  {journal} {Sci. Bull.}\ }\textbf {\bibinfo {volume} {66}},\ \bibinfo {pages}
  {2072} (\bibinfo {year} {2021}{\natexlab{a}})}\BibitemShut {NoStop}%
\bibitem [{\citenamefont {Sun}\ and\ \citenamefont
  {Zhou}(2021{\natexlab{b}})}]{Sun2021arXiv}%
  \BibitemOpen
  \bibfield  {author} {\bibinfo {author} {\bibfnamefont {X.-X.}\ \bibnamefont
  {Sun}}\ and\ \bibinfo {author} {\bibfnamefont {S.-G.}\ \bibnamefont {Zhou}},\
  }\href {https://arxiv.org/abs/2107.05925} {\bibfield  {journal} {\bibinfo
  {journal} {arXiv}\ }\textbf {\bibinfo {volume} {2107.05925}} (\bibinfo {year}
  {2021}{\natexlab{b}})}\BibitemShut {NoStop}%
\bibitem [{\citenamefont {Sun}\ \emph {et~al.}(2018)\citenamefont {Sun},
  \citenamefont {Zhao},\ and\ \citenamefont {Zhou}}]{Sun2018PLB}%
  \BibitemOpen
  \bibfield  {author} {\bibinfo {author} {\bibfnamefont {X.-X.}\ \bibnamefont
  {Sun}}, \bibinfo {author} {\bibfnamefont {J.}~\bibnamefont {Zhao}}, \ and\
  \bibinfo {author} {\bibfnamefont {S.-G.}\ \bibnamefont {Zhou}},\ }\href
  {\doibase 10.1016/j.physletb.2018.08.071} {\bibfield  {journal} {\bibinfo
  {journal} {Phys. Lett. B}\ }\textbf {\bibinfo {volume} {785}},\ \bibinfo
  {pages} {530 } (\bibinfo {year} {2018})}\BibitemShut {NoStop}%
\bibitem [{\citenamefont {Sun}\ \emph {et~al.}(2020)\citenamefont {Sun},
  \citenamefont {Zhao},\ and\ \citenamefont {Zhou}}]{Sun2020NPA}%
  \BibitemOpen
  \bibfield  {author} {\bibinfo {author} {\bibfnamefont {X.-X.}\ \bibnamefont
  {Sun}}, \bibinfo {author} {\bibfnamefont {J.}~\bibnamefont {Zhao}}, \ and\
  \bibinfo {author} {\bibfnamefont {S.-G.}\ \bibnamefont {Zhou}},\ }\href
  {\doibase 10.1016/j.nuclphysa.2020.122011} {\bibfield  {journal} {\bibinfo
  {journal} {Nucl. Phys. A}\ }\textbf {\bibinfo {volume} {1003}},\ \bibinfo
  {pages} {122011} (\bibinfo {year} {2020})}\BibitemShut {NoStop}%
\bibitem [{\citenamefont {Zhang}\ \emph
  {et~al.}(2019{\natexlab{b}})\citenamefont {Zhang}, \citenamefont {Wang},\
  and\ \citenamefont {Zhang}}]{Zhang2019PRC}%
  \BibitemOpen
  \bibfield  {author} {\bibinfo {author} {\bibfnamefont {K.~Y.}\ \bibnamefont
  {Zhang}}, \bibinfo {author} {\bibfnamefont {D.~Y.}\ \bibnamefont {Wang}}, \
  and\ \bibinfo {author} {\bibfnamefont {S.~Q.}\ \bibnamefont {Zhang}},\ }\href
  {\doibase 10.1103/PhysRevC.100.034312} {\bibfield  {journal} {\bibinfo
  {journal} {Phys. Rev. C}\ }\textbf {\bibinfo {volume} {100}},\ \bibinfo
  {pages} {034312} (\bibinfo {year} {2019}{\natexlab{b}})}\BibitemShut
  {NoStop}%
\bibitem [{\citenamefont {Yang}\ \emph
  {et~al.}(2021{\natexlab{b}})\citenamefont {Yang}, \citenamefont {Kubota},
  \citenamefont {Corsi}, \citenamefont {Yoshida}, \citenamefont {Sun},
  \citenamefont {Li}, \citenamefont {Kimura}, \citenamefont {Michel},
  \citenamefont {Ogata}, \citenamefont {Yuan}, \citenamefont {Yuan},
  \citenamefont {Authelet}, \citenamefont {Baba}, \citenamefont {Caesar},
  \citenamefont {Calvet}, \citenamefont {Delbart}, \citenamefont {Dozono},
  \citenamefont {Feng}, \citenamefont {Flavigny}, \citenamefont {Gheller},
  \citenamefont {Gibelin}, \citenamefont {Giganon}, \citenamefont {Gillibert},
  \citenamefont {Hasegawa}, \citenamefont {Isobe}, \citenamefont {Kanaya},
  \citenamefont {Kawakami}, \citenamefont {Kim}, \citenamefont {Kiyokawa},
  \citenamefont {Kobayashi}, \citenamefont {Kobayashi}, \citenamefont
  {Kobayashi}, \citenamefont {Kondo}, \citenamefont {Korkulu}, \citenamefont
  {Koyama}, \citenamefont {Lapoux}, \citenamefont {Maeda}, \citenamefont
  {Marqu\'es}, \citenamefont {Motobayashi}, \citenamefont {Miyazaki},
  \citenamefont {Nakamura}, \citenamefont {Nakatsuka}, \citenamefont {Nishio},
  \citenamefont {Obertelli}, \citenamefont {Ohkura}, \citenamefont {Orr},
  \citenamefont {Ota}, \citenamefont {Otsu}, \citenamefont {Ozaki},
  \citenamefont {Panin}, \citenamefont {Paschalis}, \citenamefont {Pollacco},
  \citenamefont {Reichert}, \citenamefont {Rouss\'e}, \citenamefont {Saito},
  \citenamefont {Sakaguchi}, \citenamefont {Sako}, \citenamefont {Santamaria},
  \citenamefont {Sasano}, \citenamefont {Sato}, \citenamefont {Shikata},
  \citenamefont {Shimizu}, \citenamefont {Shindo}, \citenamefont {Stuhl},
  \citenamefont {Sumikama}, \citenamefont {Sun}, \citenamefont {Tabata},
  \citenamefont {Togano}, \citenamefont {Tsubota}, \citenamefont {Xu},
  \citenamefont {Yasuda}, \citenamefont {Yoneda}, \citenamefont {Zenihiro},
  \citenamefont {Zhou}, \citenamefont {Zuo},\ and\ \citenamefont
  {Uesaka}}]{Yang2021PRL}%
  \BibitemOpen
  \bibfield  {author} {\bibinfo {author} {\bibfnamefont {Z.~H.}\ \bibnamefont
  {Yang}}, \bibinfo {author} {\bibfnamefont {Y.}~\bibnamefont {Kubota}},
  \bibinfo {author} {\bibfnamefont {A.}~\bibnamefont {Corsi}}, \bibinfo
  {author} {\bibfnamefont {K.}~\bibnamefont {Yoshida}}, \bibinfo {author}
  {\bibfnamefont {X.-X.}\ \bibnamefont {Sun}}, \bibinfo {author} {\bibfnamefont
  {J.~G.}\ \bibnamefont {Li}}, \bibinfo {author} {\bibfnamefont
  {M.}~\bibnamefont {Kimura}}, \bibinfo {author} {\bibfnamefont
  {N.}~\bibnamefont {Michel}}, \bibinfo {author} {\bibfnamefont
  {K.}~\bibnamefont {Ogata}}, \bibinfo {author} {\bibfnamefont {C.~X.}\
  \bibnamefont {Yuan}}, \bibinfo {author} {\bibfnamefont {Q.}~\bibnamefont
  {Yuan}}, \bibinfo {author} {\bibfnamefont {G.}~\bibnamefont {Authelet}},
  \bibinfo {author} {\bibfnamefont {H.}~\bibnamefont {Baba}}, \bibinfo {author}
  {\bibfnamefont {C.}~\bibnamefont {Caesar}}, \bibinfo {author} {\bibfnamefont
  {D.}~\bibnamefont {Calvet}}, \bibinfo {author} {\bibfnamefont
  {A.}~\bibnamefont {Delbart}}, \bibinfo {author} {\bibfnamefont
  {M.}~\bibnamefont {Dozono}}, \bibinfo {author} {\bibfnamefont
  {J.}~\bibnamefont {Feng}}, \bibinfo {author} {\bibfnamefont {F.}~\bibnamefont
  {Flavigny}}, \bibinfo {author} {\bibfnamefont {J.-M.}\ \bibnamefont
  {Gheller}}, \bibinfo {author} {\bibfnamefont {J.}~\bibnamefont {Gibelin}},
  \bibinfo {author} {\bibfnamefont {A.}~\bibnamefont {Giganon}}, \bibinfo
  {author} {\bibfnamefont {A.}~\bibnamefont {Gillibert}}, \bibinfo {author}
  {\bibfnamefont {K.}~\bibnamefont {Hasegawa}}, \bibinfo {author}
  {\bibfnamefont {T.}~\bibnamefont {Isobe}}, \bibinfo {author} {\bibfnamefont
  {Y.}~\bibnamefont {Kanaya}}, \bibinfo {author} {\bibfnamefont
  {S.}~\bibnamefont {Kawakami}}, \bibinfo {author} {\bibfnamefont
  {D.}~\bibnamefont {Kim}}, \bibinfo {author} {\bibfnamefont {Y.}~\bibnamefont
  {Kiyokawa}}, \bibinfo {author} {\bibfnamefont {M.}~\bibnamefont {Kobayashi}},
  \bibinfo {author} {\bibfnamefont {N.}~\bibnamefont {Kobayashi}}, \bibinfo
  {author} {\bibfnamefont {T.}~\bibnamefont {Kobayashi}}, \bibinfo {author}
  {\bibfnamefont {Y.}~\bibnamefont {Kondo}}, \bibinfo {author} {\bibfnamefont
  {Z.}~\bibnamefont {Korkulu}}, \bibinfo {author} {\bibfnamefont
  {S.}~\bibnamefont {Koyama}}, \bibinfo {author} {\bibfnamefont
  {V.}~\bibnamefont {Lapoux}}, \bibinfo {author} {\bibfnamefont
  {Y.}~\bibnamefont {Maeda}}, \bibinfo {author} {\bibfnamefont {F.~M.}\
  \bibnamefont {Marqu\'es}}, \bibinfo {author} {\bibfnamefont {T.}~\bibnamefont
  {Motobayashi}}, \bibinfo {author} {\bibfnamefont {T.}~\bibnamefont
  {Miyazaki}}, \bibinfo {author} {\bibfnamefont {T.}~\bibnamefont {Nakamura}},
  \bibinfo {author} {\bibfnamefont {N.}~\bibnamefont {Nakatsuka}}, \bibinfo
  {author} {\bibfnamefont {Y.}~\bibnamefont {Nishio}}, \bibinfo {author}
  {\bibfnamefont {A.}~\bibnamefont {Obertelli}}, \bibinfo {author}
  {\bibfnamefont {A.}~\bibnamefont {Ohkura}}, \bibinfo {author} {\bibfnamefont
  {N.~A.}\ \bibnamefont {Orr}}, \bibinfo {author} {\bibfnamefont
  {S.}~\bibnamefont {Ota}}, \bibinfo {author} {\bibfnamefont {H.}~\bibnamefont
  {Otsu}}, \bibinfo {author} {\bibfnamefont {T.}~\bibnamefont {Ozaki}},
  \bibinfo {author} {\bibfnamefont {V.}~\bibnamefont {Panin}}, \bibinfo
  {author} {\bibfnamefont {S.}~\bibnamefont {Paschalis}}, \bibinfo {author}
  {\bibfnamefont {E.~C.}\ \bibnamefont {Pollacco}}, \bibinfo {author}
  {\bibfnamefont {S.}~\bibnamefont {Reichert}}, \bibinfo {author}
  {\bibfnamefont {J.-Y.}\ \bibnamefont {Rouss\'e}}, \bibinfo {author}
  {\bibfnamefont {A.~T.}\ \bibnamefont {Saito}}, \bibinfo {author}
  {\bibfnamefont {S.}~\bibnamefont {Sakaguchi}}, \bibinfo {author}
  {\bibfnamefont {M.}~\bibnamefont {Sako}}, \bibinfo {author} {\bibfnamefont
  {C.}~\bibnamefont {Santamaria}}, \bibinfo {author} {\bibfnamefont
  {M.}~\bibnamefont {Sasano}}, \bibinfo {author} {\bibfnamefont
  {H.}~\bibnamefont {Sato}}, \bibinfo {author} {\bibfnamefont {M.}~\bibnamefont
  {Shikata}}, \bibinfo {author} {\bibfnamefont {Y.}~\bibnamefont {Shimizu}},
  \bibinfo {author} {\bibfnamefont {Y.}~\bibnamefont {Shindo}}, \bibinfo
  {author} {\bibfnamefont {L.}~\bibnamefont {Stuhl}}, \bibinfo {author}
  {\bibfnamefont {T.}~\bibnamefont {Sumikama}}, \bibinfo {author}
  {\bibfnamefont {Y.~L.}\ \bibnamefont {Sun}}, \bibinfo {author} {\bibfnamefont
  {M.}~\bibnamefont {Tabata}}, \bibinfo {author} {\bibfnamefont
  {Y.}~\bibnamefont {Togano}}, \bibinfo {author} {\bibfnamefont
  {J.}~\bibnamefont {Tsubota}}, \bibinfo {author} {\bibfnamefont {F.~R.}\
  \bibnamefont {Xu}}, \bibinfo {author} {\bibfnamefont {J.}~\bibnamefont
  {Yasuda}}, \bibinfo {author} {\bibfnamefont {K.}~\bibnamefont {Yoneda}},
  \bibinfo {author} {\bibfnamefont {J.}~\bibnamefont {Zenihiro}}, \bibinfo
  {author} {\bibfnamefont {S.-G.}\ \bibnamefont {Zhou}}, \bibinfo {author}
  {\bibfnamefont {W.}~\bibnamefont {Zuo}}, \ and\ \bibinfo {author}
  {\bibfnamefont {T.}~\bibnamefont {Uesaka}},\ }\href {\doibase
  10.1103/PhysRevLett.126.082501} {\bibfield  {journal} {\bibinfo  {journal}
  {Phys. Rev. Lett.}\ }\textbf {\bibinfo {volume} {126}},\ \bibinfo {pages}
  {082501} (\bibinfo {year} {2021}{\natexlab{b}})}\BibitemShut {NoStop}%
\bibitem [{\citenamefont {Sun}(2021)}]{Sun2021PRC}%
  \BibitemOpen
  \bibfield  {author} {\bibinfo {author} {\bibfnamefont {X.-X.}\ \bibnamefont
  {Sun}},\ }\href {\doibase 10.1103/PhysRevC.103.054315} {\bibfield  {journal}
  {\bibinfo  {journal} {Phys. Rev. C}\ }\textbf {\bibinfo {volume} {103}},\
  \bibinfo {pages} {054315} (\bibinfo {year} {2021})}\BibitemShut {NoStop}%
\bibitem [{\citenamefont {Zhang}\ \emph {et~al.}(2020)\citenamefont {Zhang},
  \citenamefont {Cheoun}, \citenamefont {Choi}, \citenamefont {Chong},
  \citenamefont {Dong}, \citenamefont {Geng}, \citenamefont {Ha}, \citenamefont
  {He}, \citenamefont {Heo}, \citenamefont {Ho}, \citenamefont {In},
  \citenamefont {Kim}, \citenamefont {Kim}, \citenamefont {Lee}, \citenamefont
  {Lee}, \citenamefont {Li}, \citenamefont {Luo}, \citenamefont {Meng},
  \citenamefont {Mun}, \citenamefont {Niu}, \citenamefont {Pan}, \citenamefont
  {Papakonstantinou}, \citenamefont {Shang}, \citenamefont {Shen},
  \citenamefont {Shen}, \citenamefont {Sun}, \citenamefont {Sun}, \citenamefont
  {Tam}, \citenamefont {Thaivayongnou}, \citenamefont {Wang}, \citenamefont
  {Wong}, \citenamefont {Xia}, \citenamefont {Yan}, \citenamefont {Yeung},
  \citenamefont {Yiu}, \citenamefont {Zhang}, \citenamefont {Zhang},\ and\
  \citenamefont {Zhou}}]{Zhang2020PRC}%
  \BibitemOpen
  \bibfield  {author} {\bibinfo {author} {\bibfnamefont {K.}~\bibnamefont
  {Zhang}}, \bibinfo {author} {\bibfnamefont {M.-K.}\ \bibnamefont {Cheoun}},
  \bibinfo {author} {\bibfnamefont {Y.-B.}\ \bibnamefont {Choi}}, \bibinfo
  {author} {\bibfnamefont {P.~S.}\ \bibnamefont {Chong}}, \bibinfo {author}
  {\bibfnamefont {J.}~\bibnamefont {Dong}}, \bibinfo {author} {\bibfnamefont
  {L.}~\bibnamefont {Geng}}, \bibinfo {author} {\bibfnamefont {E.}~\bibnamefont
  {Ha}}, \bibinfo {author} {\bibfnamefont {X.}~\bibnamefont {He}}, \bibinfo
  {author} {\bibfnamefont {C.}~\bibnamefont {Heo}}, \bibinfo {author}
  {\bibfnamefont {M.~C.}\ \bibnamefont {Ho}}, \bibinfo {author} {\bibfnamefont
  {E.~J.}\ \bibnamefont {In}}, \bibinfo {author} {\bibfnamefont
  {S.}~\bibnamefont {Kim}}, \bibinfo {author} {\bibfnamefont {Y.}~\bibnamefont
  {Kim}}, \bibinfo {author} {\bibfnamefont {C.-H.}\ \bibnamefont {Lee}},
  \bibinfo {author} {\bibfnamefont {J.}~\bibnamefont {Lee}}, \bibinfo {author}
  {\bibfnamefont {Z.}~\bibnamefont {Li}}, \bibinfo {author} {\bibfnamefont
  {T.}~\bibnamefont {Luo}}, \bibinfo {author} {\bibfnamefont {J.}~\bibnamefont
  {Meng}}, \bibinfo {author} {\bibfnamefont {M.-H.}\ \bibnamefont {Mun}},
  \bibinfo {author} {\bibfnamefont {Z.}~\bibnamefont {Niu}}, \bibinfo {author}
  {\bibfnamefont {C.}~\bibnamefont {Pan}}, \bibinfo {author} {\bibfnamefont
  {P.}~\bibnamefont {Papakonstantinou}}, \bibinfo {author} {\bibfnamefont
  {X.}~\bibnamefont {Shang}}, \bibinfo {author} {\bibfnamefont
  {C.}~\bibnamefont {Shen}}, \bibinfo {author} {\bibfnamefont {G.}~\bibnamefont
  {Shen}}, \bibinfo {author} {\bibfnamefont {W.}~\bibnamefont {Sun}}, \bibinfo
  {author} {\bibfnamefont {X.-X.}\ \bibnamefont {Sun}}, \bibinfo {author}
  {\bibfnamefont {C.~K.}\ \bibnamefont {Tam}}, \bibinfo {author} {\bibnamefont
  {Thaivayongnou}}, \bibinfo {author} {\bibfnamefont {C.}~\bibnamefont {Wang}},
  \bibinfo {author} {\bibfnamefont {S.~H.}\ \bibnamefont {Wong}}, \bibinfo
  {author} {\bibfnamefont {X.}~\bibnamefont {Xia}}, \bibinfo {author}
  {\bibfnamefont {Y.}~\bibnamefont {Yan}}, \bibinfo {author} {\bibfnamefont
  {R.~W.-Y.}\ \bibnamefont {Yeung}}, \bibinfo {author} {\bibfnamefont {T.~C.}\
  \bibnamefont {Yiu}}, \bibinfo {author} {\bibfnamefont {S.}~\bibnamefont
  {Zhang}}, \bibinfo {author} {\bibfnamefont {W.}~\bibnamefont {Zhang}}, \ and\
  \bibinfo {author} {\bibfnamefont {S.-G.}\ \bibnamefont {Zhou}} (\bibinfo
  {collaboration} {DRHBc Mass Table Collaboration}),\ }\href {\doibase
  10.1103/PhysRevC.102.024314} {\bibfield  {journal} {\bibinfo  {journal}
  {Phys. Rev. C}\ }\textbf {\bibinfo {volume} {102}},\ \bibinfo {pages}
  {024314} (\bibinfo {year} {2020})}\BibitemShut {NoStop}%
\bibitem [{\citenamefont {Pan}\ \emph {et~al.}(2019)\citenamefont {Pan},
  \citenamefont {Zhang},\ and\ \citenamefont {Zhang}}]{Pan2019IJMPE}%
  \BibitemOpen
  \bibfield  {author} {\bibinfo {author} {\bibfnamefont {C.}~\bibnamefont
  {Pan}}, \bibinfo {author} {\bibfnamefont {K.}~\bibnamefont {Zhang}}, \ and\
  \bibinfo {author} {\bibfnamefont {S.}~\bibnamefont {Zhang}},\ }\href
  {\doibase 10.1142/S0218301319500824} {\bibfield  {journal} {\bibinfo
  {journal} {Int. J. Mod. Phys. E}\ }\textbf {\bibinfo {volume} {28}},\
  \bibinfo {pages} {1950082} (\bibinfo {year} {2019})}\BibitemShut {NoStop}%
\bibitem [{\citenamefont {In}\ \emph {et~al.}(2021)\citenamefont {In},
  \citenamefont {Papakonstantinou}, \citenamefont {Kim},\ and\ \citenamefont
  {Hong}}]{In2021IJMPE}%
  \BibitemOpen
  \bibfield  {author} {\bibinfo {author} {\bibfnamefont {E.~J.}\ \bibnamefont
  {In}}, \bibinfo {author} {\bibfnamefont {P.}~\bibnamefont
  {Papakonstantinou}}, \bibinfo {author} {\bibfnamefont {Y.}~\bibnamefont
  {Kim}}, \ and\ \bibinfo {author} {\bibfnamefont {S.-W.}\ \bibnamefont
  {Hong}},\ }\href {\doibase 10.1142/S0218301321500099} {\bibfield  {journal}
  {\bibinfo  {journal} {Int. J. Mod. Phys. E}\ }\textbf {\bibinfo {volume}
  {30}},\ \bibinfo {pages} {2150009} (\bibinfo {year} {2021})}\BibitemShut
  {NoStop}%
\bibitem [{\citenamefont {Zhang}\ \emph {et~al.}(2021)\citenamefont {Zhang},
  \citenamefont {He}, \citenamefont {Meng}, \citenamefont {Pan}, \citenamefont
  {Shen}, \citenamefont {Wang},\ and\ \citenamefont {Zhang}}]{Zhang2021PRC(L)}%
  \BibitemOpen
  \bibfield  {author} {\bibinfo {author} {\bibfnamefont {K.}~\bibnamefont
  {Zhang}}, \bibinfo {author} {\bibfnamefont {X.}~\bibnamefont {He}}, \bibinfo
  {author} {\bibfnamefont {J.}~\bibnamefont {Meng}}, \bibinfo {author}
  {\bibfnamefont {C.}~\bibnamefont {Pan}}, \bibinfo {author} {\bibfnamefont
  {C.}~\bibnamefont {Shen}}, \bibinfo {author} {\bibfnamefont {C.}~\bibnamefont
  {Wang}}, \ and\ \bibinfo {author} {\bibfnamefont {S.}~\bibnamefont {Zhang}},\
  }\href {\doibase 10.1103/PhysRevC.104.L021301} {\bibfield  {journal}
  {\bibinfo  {journal} {Phys. Rev. C}\ }\textbf {\bibinfo {volume} {104}},\
  \bibinfo {pages} {L021301} (\bibinfo {year} {2021})}\BibitemShut {NoStop}%
\bibitem [{\citenamefont {Pan}\ \emph {et~al.}(2021)\citenamefont {Pan},
  \citenamefont {Zhang}, \citenamefont {Chong}, \citenamefont {Heo},
  \citenamefont {Ho}, \citenamefont {Lee}, \citenamefont {Li}, \citenamefont
  {Sun}, \citenamefont {Tam}, \citenamefont {Wong}, \citenamefont {Yeung},
  \citenamefont {Yiu},\ and\ \citenamefont {Zhang}}]{Pan2021PRC}%
  \BibitemOpen
  \bibfield  {author} {\bibinfo {author} {\bibfnamefont {C.}~\bibnamefont
  {Pan}}, \bibinfo {author} {\bibfnamefont {K.~Y.}\ \bibnamefont {Zhang}},
  \bibinfo {author} {\bibfnamefont {P.~S.}\ \bibnamefont {Chong}}, \bibinfo
  {author} {\bibfnamefont {C.}~\bibnamefont {Heo}}, \bibinfo {author}
  {\bibfnamefont {M.~C.}\ \bibnamefont {Ho}}, \bibinfo {author} {\bibfnamefont
  {J.}~\bibnamefont {Lee}}, \bibinfo {author} {\bibfnamefont {Z.~P.}\
  \bibnamefont {Li}}, \bibinfo {author} {\bibfnamefont {W.}~\bibnamefont
  {Sun}}, \bibinfo {author} {\bibfnamefont {C.~K.}\ \bibnamefont {Tam}},
  \bibinfo {author} {\bibfnamefont {S.~H.}\ \bibnamefont {Wong}}, \bibinfo
  {author} {\bibfnamefont {R.~W.-Y.}\ \bibnamefont {Yeung}}, \bibinfo {author}
  {\bibfnamefont {T.~C.}\ \bibnamefont {Yiu}}, \ and\ \bibinfo {author}
  {\bibfnamefont {S.~Q.}\ \bibnamefont {Zhang}},\ }\href {\doibase
  10.1103/PhysRevC.104.024331} {\bibfield  {journal} {\bibinfo  {journal}
  {Phys. Rev. C}\ }\textbf {\bibinfo {volume} {104}},\ \bibinfo {pages}
  {024331} (\bibinfo {year} {2021})}\BibitemShut {NoStop}%
\bibitem [{\citenamefont {He}\ \emph {et~al.}(2021)\citenamefont {He},
  \citenamefont {Wang}, \citenamefont {Zhang},\ and\ \citenamefont
  {Shen}}]{He2021CPC}%
  \BibitemOpen
  \bibfield  {author} {\bibinfo {author} {\bibfnamefont {X.-T.}\ \bibnamefont
  {He}}, \bibinfo {author} {\bibfnamefont {C.}~\bibnamefont {Wang}}, \bibinfo
  {author} {\bibfnamefont {K.-Y.}\ \bibnamefont {Zhang}}, \ and\ \bibinfo
  {author} {\bibfnamefont {C.-W.}\ \bibnamefont {Shen}},\ }\href {\doibase
  10.1088/1674-1137/ac1b99} {\bibfield  {journal} {\bibinfo  {journal} {Chin.
  Phys. C}\ }\textbf {\bibinfo {volume} {45}},\ \bibinfo {pages} {101001}
  (\bibinfo {year} {2021})}\BibitemShut {NoStop}%
\bibitem [{\citenamefont {Kucharek}\ and\ \citenamefont
  {Ring}(1991)}]{Kucharek1991ZPA}%
  \BibitemOpen
  \bibfield  {author} {\bibinfo {author} {\bibfnamefont {H.}~\bibnamefont
  {Kucharek}}\ and\ \bibinfo {author} {\bibfnamefont {P.}~\bibnamefont
  {Ring}},\ }\href
  {https://link.springer.com/content/pdf/10.1007/BF01282930.pdf} {\bibfield
  {journal} {\bibinfo  {journal} {Z. Phys. A}\ }\textbf {\bibinfo {volume}
  {339}},\ \bibinfo {pages} {23} (\bibinfo {year} {1991})}\BibitemShut
  {NoStop}%
\bibitem [{\citenamefont {Ring}\ and\ \citenamefont
  {Schuck}(1980)}]{Peter1980Book}%
  \BibitemOpen
  \bibfield  {author} {\bibinfo {author} {\bibfnamefont {P.}~\bibnamefont
  {Ring}}\ and\ \bibinfo {author} {\bibfnamefont {P.}~\bibnamefont {Schuck}},\
  }\href@noop {} {\emph {\bibinfo {title} {The Nuclear Many-body Problem}}}\
  (\bibinfo  {publisher} {Springer-Verlag, Berlin},\ \bibinfo {year}
  {1980})\BibitemShut {NoStop}%
\bibitem [{\citenamefont {Zhao}\ \emph
  {et~al.}(2012{\natexlab{b}})\citenamefont {Zhao}, \citenamefont {Song},
  \citenamefont {Sun}, \citenamefont {Geissel},\ and\ \citenamefont
  {Meng}}]{Zhao2012Phys.Rev.C64324}%
  \BibitemOpen
  \bibfield  {author} {\bibinfo {author} {\bibfnamefont {P.~W.}\ \bibnamefont
  {Zhao}}, \bibinfo {author} {\bibfnamefont {L.~S.}\ \bibnamefont {Song}},
  \bibinfo {author} {\bibfnamefont {B.}~\bibnamefont {Sun}}, \bibinfo {author}
  {\bibfnamefont {H.}~\bibnamefont {Geissel}}, \ and\ \bibinfo {author}
  {\bibfnamefont {J.}~\bibnamefont {Meng}},\ }\href {\doibase
  10.1103/PhysRevC.86.064324} {\bibfield  {journal} {\bibinfo  {journal} {Phys.
  Rev. C}\ }\textbf {\bibinfo {volume} {86}},\ \bibinfo {pages} {064324}
  (\bibinfo {year} {2012}{\natexlab{b}})}\BibitemShut {NoStop}%
\bibitem [{\citenamefont {Angeli}\ and\ \citenamefont
  {Marinova}(2013)}]{Angeli2013ADNDT}%
  \BibitemOpen
  \bibfield  {author} {\bibinfo {author} {\bibfnamefont {I.}~\bibnamefont
  {Angeli}}\ and\ \bibinfo {author} {\bibfnamefont {K.}~\bibnamefont
  {Marinova}},\ }\href {\doibase 10.1016/j.adt.2011.12.006} {\bibfield
  {journal} {\bibinfo  {journal} {Atom. Data Nucl. Data Tabl.}\ }\textbf
  {\bibinfo {volume} {99}},\ \bibinfo {pages} {69 } (\bibinfo {year}
  {2013})}\BibitemShut {NoStop}%
\bibitem [{\citenamefont {Li}\ \emph {et~al.}(2021)\citenamefont {Li},
  \citenamefont {Luo},\ and\ \citenamefont {Wang}}]{Li2021ADNDT}%
  \BibitemOpen
  \bibfield  {author} {\bibinfo {author} {\bibfnamefont {T.}~\bibnamefont
  {Li}}, \bibinfo {author} {\bibfnamefont {Y.}~\bibnamefont {Luo}}, \ and\
  \bibinfo {author} {\bibfnamefont {N.}~\bibnamefont {Wang}},\ }\href {\doibase
  10.1016/j.adt.2021.101440} {\bibfield  {journal} {\bibinfo  {journal} {Atom.
  Data Nucl. Data Tabl.}\ }\textbf {\bibinfo {volume} {140}},\ \bibinfo {pages}
  {101440} (\bibinfo {year} {2021})}\BibitemShut {NoStop}%
\bibitem [{\citenamefont {Li}\ \emph {et~al.}(2014)\citenamefont {Li},
  \citenamefont {Long}, \citenamefont {Margueron},\ and\ \citenamefont {{Van
  Giai}}}]{Li2014PLB}%
  \BibitemOpen
  \bibfield  {author} {\bibinfo {author} {\bibfnamefont {J.~J.}\ \bibnamefont
  {Li}}, \bibinfo {author} {\bibfnamefont {W.~H.}\ \bibnamefont {Long}},
  \bibinfo {author} {\bibfnamefont {J.}~\bibnamefont {Margueron}}, \ and\
  \bibinfo {author} {\bibfnamefont {N.}~\bibnamefont {{Van Giai}}},\ }\href
  {\doibase 10.1016/j.physletb.2014.03.031} {\bibfield  {journal} {\bibinfo
  {journal} {Phys. Lett. B}\ }\textbf {\bibinfo {volume} {732}},\ \bibinfo
  {pages} {169 } (\bibinfo {year} {2014})}\BibitemShut {NoStop}%
\bibitem [{\citenamefont {Tian}\ \emph {et~al.}(2009)\citenamefont {Tian},
  \citenamefont {Ma},\ and\ \citenamefont {Ring}}]{Tian2009PLB}%
  \BibitemOpen
  \bibfield  {author} {\bibinfo {author} {\bibfnamefont {Y.}~\bibnamefont
  {Tian}}, \bibinfo {author} {\bibfnamefont {Z.}~\bibnamefont {Ma}}, \ and\
  \bibinfo {author} {\bibfnamefont {P.}~\bibnamefont {Ring}},\ }\href {\doibase
  https://doi.org/10.1016/j.physletb.2009.04.067} {\bibfield  {journal}
  {\bibinfo  {journal} {Physics Letters B}\ }\textbf {\bibinfo {volume}
  {676}},\ \bibinfo {pages} {44 } (\bibinfo {year} {2009})}\BibitemShut
  {NoStop}%
\bibitem [{\citenamefont {Dong}\ \emph {et~al.}(2018)\citenamefont {Dong},
  \citenamefont {Zhang}, \citenamefont {Zuo}, \citenamefont {Gu}, \citenamefont
  {Wang},\ and\ \citenamefont {Sun}}]{Dong2018PRC(R)}%
  \BibitemOpen
  \bibfield  {author} {\bibinfo {author} {\bibfnamefont {J.~M.}\ \bibnamefont
  {Dong}}, \bibinfo {author} {\bibfnamefont {Y.~H.}\ \bibnamefont {Zhang}},
  \bibinfo {author} {\bibfnamefont {W.}~\bibnamefont {Zuo}}, \bibinfo {author}
  {\bibfnamefont {J.~Z.}\ \bibnamefont {Gu}}, \bibinfo {author} {\bibfnamefont
  {L.~J.}\ \bibnamefont {Wang}}, \ and\ \bibinfo {author} {\bibfnamefont
  {Y.}~\bibnamefont {Sun}},\ }\href {\doibase 10.1103/PhysRevC.97.021301}
  {\bibfield  {journal} {\bibinfo  {journal} {Phys. Rev. C}\ }\textbf {\bibinfo
  {volume} {97}},\ \bibinfo {pages} {021301(R)} (\bibinfo {year}
  {2018})}\BibitemShut {NoStop}%
\bibitem [{\citenamefont {Dong}\ \emph {et~al.}(2019)\citenamefont {Dong},
  \citenamefont {Shang}, \citenamefont {Zuo}, \citenamefont {Niu},\ and\
  \citenamefont {Sun}}]{Dong2019NPA}%
  \BibitemOpen
  \bibfield  {author} {\bibinfo {author} {\bibfnamefont {J.}~\bibnamefont
  {Dong}}, \bibinfo {author} {\bibfnamefont {X.}~\bibnamefont {Shang}},
  \bibinfo {author} {\bibfnamefont {W.}~\bibnamefont {Zuo}}, \bibinfo {author}
  {\bibfnamefont {Y.}~\bibnamefont {Niu}}, \ and\ \bibinfo {author}
  {\bibfnamefont {Y.}~\bibnamefont {Sun}},\ }\href {\doibase
  https://doi.org/10.1016/j.nuclphysa.2019.01.003} {\bibfield  {journal}
  {\bibinfo  {journal} {Nucl. Phys. A}\ }\textbf {\bibinfo {volume} {983}},\
  \bibinfo {pages} {133} (\bibinfo {year} {2019})}\BibitemShut {NoStop}%
\bibitem [{\citenamefont {Wang}\ \emph
  {et~al.}(2021{\natexlab{b}})\citenamefont {Wang}, \citenamefont {Sun},\ and\
  \citenamefont {Zhou}}]{Wang2021CPC}%
  \BibitemOpen
  \bibfield  {author} {\bibinfo {author} {\bibfnamefont {X.-Q.}\ \bibnamefont
  {Wang}}, \bibinfo {author} {\bibfnamefont {X.-X.}\ \bibnamefont {Sun}}, \
  and\ \bibinfo {author} {\bibfnamefont {S.-G.}\ \bibnamefont {Zhou}},\ }\href
  {\doibase 10.1088/1674-1137/ac3904} {\bibfield  {journal} {\bibinfo
  {journal} {Chin. Phys. C}\ } (\bibinfo {year} {2021}{\natexlab{b}}),\
  10.1088/1674-1137/ac3904}\BibitemShut {NoStop}%
\bibitem [{\citenamefont {Geng}\ \emph {et~al.}(2006)\citenamefont {Geng},
  \citenamefont {Meng}, \citenamefont {Toki}, \citenamefont {Long},\ and\
  \citenamefont {Shen}}]{Geng2006CPL}%
  \BibitemOpen
  \bibfield  {author} {\bibinfo {author} {\bibfnamefont {L.-S.}\ \bibnamefont
  {Geng}}, \bibinfo {author} {\bibfnamefont {J.}~\bibnamefont {Meng}}, \bibinfo
  {author} {\bibfnamefont {H.}~\bibnamefont {Toki}}, \bibinfo {author}
  {\bibfnamefont {W.-H.}\ \bibnamefont {Long}}, \ and\ \bibinfo {author}
  {\bibfnamefont {G.}~\bibnamefont {Shen}},\ }\href {\doibase
  10.1088/0256-307x/23/5/021} {\bibfield  {journal} {\bibinfo  {journal} {Chin.
  Phys. Lett.}\ }\textbf {\bibinfo {volume} {23}},\ \bibinfo {pages} {1139}
  (\bibinfo {year} {2006})}\BibitemShut {NoStop}%
\bibitem [{\citenamefont {Pritychenko}\ \emph {et~al.}(2016)\citenamefont
  {Pritychenko}, \citenamefont {Birch}, \citenamefont {Singh},\ and\
  \citenamefont {Horoi}}]{Pritychenko2016ADNDT}%
  \BibitemOpen
  \bibfield  {author} {\bibinfo {author} {\bibfnamefont {B.}~\bibnamefont
  {Pritychenko}}, \bibinfo {author} {\bibfnamefont {M.}~\bibnamefont {Birch}},
  \bibinfo {author} {\bibfnamefont {B.}~\bibnamefont {Singh}}, \ and\ \bibinfo
  {author} {\bibfnamefont {M.}~\bibnamefont {Horoi}},\ }\href {\doibase
  10.1016/j.adt.2015.10.001} {\bibfield  {journal} {\bibinfo  {journal} {Atom.
  Data Nucl. Data Tabl.}\ }\textbf {\bibinfo {volume} {107}},\ \bibinfo {pages}
  {1 } (\bibinfo {year} {2016})}\BibitemShut {NoStop}%
\bibitem [{\citenamefont {Ma}\ \emph {et~al.}(2020)\citenamefont {Ma},
  \citenamefont {Su}, \citenamefont {Liu}, \citenamefont {Ren}, \citenamefont
  {Xu},\ and\ \citenamefont {Gao}}]{Ma2020PRC}%
  \BibitemOpen
  \bibfield  {author} {\bibinfo {author} {\bibfnamefont {Y.}~\bibnamefont
  {Ma}}, \bibinfo {author} {\bibfnamefont {C.}~\bibnamefont {Su}}, \bibinfo
  {author} {\bibfnamefont {J.}~\bibnamefont {Liu}}, \bibinfo {author}
  {\bibfnamefont {Z.}~\bibnamefont {Ren}}, \bibinfo {author} {\bibfnamefont
  {C.}~\bibnamefont {Xu}}, \ and\ \bibinfo {author} {\bibfnamefont
  {Y.}~\bibnamefont {Gao}},\ }\href {\doibase 10.1103/PhysRevC.101.014304}
  {\bibfield  {journal} {\bibinfo  {journal} {Phys. Rev. C}\ }\textbf {\bibinfo
  {volume} {101}},\ \bibinfo {pages} {014304} (\bibinfo {year}
  {2020})}\BibitemShut {NoStop}%
\bibitem [{\citenamefont {Dong}\ \emph {et~al.}(2021)\citenamefont {Dong},
  \citenamefont {An}, \citenamefont {Lu},\ and\ \citenamefont
  {Geng}}]{Dong2021arXiv}%
  \BibitemOpen
  \bibfield  {author} {\bibinfo {author} {\bibfnamefont {X.-X.}\ \bibnamefont
  {Dong}}, \bibinfo {author} {\bibfnamefont {R.}~\bibnamefont {An}}, \bibinfo
  {author} {\bibfnamefont {J.-X.}\ \bibnamefont {Lu}}, \ and\ \bibinfo {author}
  {\bibfnamefont {L.-S.}\ \bibnamefont {Geng}},\ }\href
  {https://arxiv.org/abs/2109.09626} {\bibfield  {journal} {\bibinfo  {journal}
  {arXiv}\ }\textbf {\bibinfo {volume} {2109.09626}} (\bibinfo {year}
  {2021})}\BibitemShut {NoStop}%
\bibitem [{\citenamefont {Terasaki}\ \emph {et~al.}(2006)\citenamefont
  {Terasaki}, \citenamefont {Zhang}, \citenamefont {Zhou},\ and\ \citenamefont
  {Meng}}]{Terasaki2006PRC}%
  \BibitemOpen
  \bibfield  {author} {\bibinfo {author} {\bibfnamefont {J.}~\bibnamefont
  {Terasaki}}, \bibinfo {author} {\bibfnamefont {S.~Q.}\ \bibnamefont {Zhang}},
  \bibinfo {author} {\bibfnamefont {S.~G.}\ \bibnamefont {Zhou}}, \ and\
  \bibinfo {author} {\bibfnamefont {J.}~\bibnamefont {Meng}},\ }\href {\doibase
  10.1103/PhysRevC.74.054318} {\bibfield  {journal} {\bibinfo  {journal} {Phys.
  Rev. C}\ }\textbf {\bibinfo {volume} {74}},\ \bibinfo {pages} {054318}
  (\bibinfo {year} {2006})}\BibitemShut {NoStop}%
\bibitem [{\citenamefont {Yao}\ \emph {et~al.}(2010)\citenamefont {Yao},
  \citenamefont {Meng}, \citenamefont {Ring},\ and\ \citenamefont
  {Vretenar}}]{Yao2010Phys.Rev.C44311}%
  \BibitemOpen
  \bibfield  {author} {\bibinfo {author} {\bibfnamefont {J.~M.}\ \bibnamefont
  {Yao}}, \bibinfo {author} {\bibfnamefont {J.}~\bibnamefont {Meng}}, \bibinfo
  {author} {\bibfnamefont {P.}~\bibnamefont {Ring}}, \ and\ \bibinfo {author}
  {\bibfnamefont {D.}~\bibnamefont {Vretenar}},\ }\href {\doibase
  10.1103/PhysRevC.81.044311} {\bibfield  {journal} {\bibinfo  {journal} {Phys.
  Rev. C}\ }\textbf {\bibinfo {volume} {81}},\ \bibinfo {pages} {044311}
  (\bibinfo {year} {2010})}\BibitemShut {NoStop}%
\bibitem [{\citenamefont {Yao}\ \emph {et~al.}(2011)\citenamefont {Yao},
  \citenamefont {Mei}, \citenamefont {Chen}, \citenamefont {Meng},
  \citenamefont {Ring},\ and\ \citenamefont
  {Vretenar}}]{Yao2011Phys.Rev.C14308}%
  \BibitemOpen
  \bibfield  {author} {\bibinfo {author} {\bibfnamefont {J.~M.}\ \bibnamefont
  {Yao}}, \bibinfo {author} {\bibfnamefont {H.}~\bibnamefont {Mei}}, \bibinfo
  {author} {\bibfnamefont {H.}~\bibnamefont {Chen}}, \bibinfo {author}
  {\bibfnamefont {J.}~\bibnamefont {Meng}}, \bibinfo {author} {\bibfnamefont
  {P.}~\bibnamefont {Ring}}, \ and\ \bibinfo {author} {\bibfnamefont
  {D.}~\bibnamefont {Vretenar}},\ }\href {\doibase 10.1103/PhysRevC.83.014308}
  {\bibfield  {journal} {\bibinfo  {journal} {Phys. Rev. C}\ }\textbf {\bibinfo
  {volume} {83}},\ \bibinfo {pages} {014308} (\bibinfo {year}
  {2011})}\BibitemShut {NoStop}%
\bibitem [{\citenamefont {Pei}\ \emph {et~al.}(2007)\citenamefont {Pei},
  \citenamefont {Xu}, \citenamefont {Lin},\ and\ \citenamefont
  {Zhao}}]{Pei2007PRC}%
  \BibitemOpen
  \bibfield  {author} {\bibinfo {author} {\bibfnamefont {J.~C.}\ \bibnamefont
  {Pei}}, \bibinfo {author} {\bibfnamefont {F.~R.}\ \bibnamefont {Xu}},
  \bibinfo {author} {\bibfnamefont {Z.~J.}\ \bibnamefont {Lin}}, \ and\
  \bibinfo {author} {\bibfnamefont {E.~G.}\ \bibnamefont {Zhao}},\ }\href
  {\doibase 10.1103/PhysRevC.76.044326} {\bibfield  {journal} {\bibinfo
  {journal} {Phys. Rev. C}\ }\textbf {\bibinfo {volume} {76}},\ \bibinfo
  {pages} {044326} (\bibinfo {year} {2007})}\BibitemShut {NoStop}%
\bibitem [{\citenamefont {Reinhard}(1988)}]{Reinhard1988ZPA}%
  \BibitemOpen
  \bibfield  {author} {\bibinfo {author} {\bibfnamefont {P.~G.}\ \bibnamefont
  {Reinhard}},\ }\href {\doibase 10.1007/BF01290231} {\bibfield  {journal}
  {\bibinfo  {journal} {Z. Phys. A}\ }\textbf {\bibinfo {volume} {329}},\
  \bibinfo {pages} {257} (\bibinfo {year} {1988})}\BibitemShut {NoStop}%
\end{thebibliography}

\newpage

\section*{Explanation of Tables}

\textbf{Table II. Ground-state properties of even-even nuclei calculated by the DRHBc theory}

\begin{tabular}{@{}p{0.0in}p{1.5in}p{4.5in}@{}}
& $Z$ & Proton number\\
& $N$ & Neutron number\\
& $A$ & Mass number\\
& $E^{\mathrm{cal}}_{\mathrm{b}}$ & Binding energy from DRHBc calculations\\
& $E^{\mathrm{cal}}_{\mathrm{b}+\mathrm{rot}}$ & Binding energy plus rotational correction energy from DRHBc calculations, which is suggested to be compared with the experimental value\\
& $E^{\mathrm{exp}}_{\mathrm{b}}$ & Binding energy from experimental data\\
& $S_{2n}$ & Two-neutron separation energy\\
& $S_{2p}$ & Two-proton separation energy\\
& $R_{n}$ & Neutron root-mean-square radius \\
& $R_{p}$ & Proton root-mean-square radius \\
& $R_{m}$ & Matter root-mean-square radius \\
& $R^{\mathrm{cal}}_{\mathrm{ch}}$ & Charge radius from DRHBc calculations \\
& $R^{\mathrm{exp}}_{\mathrm{ch}}$ & Charge radius from experimental data\\
& $\beta_{2n}$ & Neutron quadrupole deformation \\
& $\beta_{2p}$ & Proton quadrupole deformation \\
& $\beta_{2}$ & Matter quadrupole deformation \\
& $\lambda_{n}$ & Neutron Fermi surface\\
& $\lambda_{p}$ & Proton Fermi surface\\
& $\sigma$ & rms deviations for binding energies and charge radii for each isotopic chain \\
\\
\end{tabular}

\textbf{Note: Since PC-PK1 is a non-linear density functional, it encounters high density instability in $^{24,26}$Mg and $^{26,28}$Si, similar to that in $^{12}$C by NL1~\cite{Reinhard1988ZPA}.}

\newpage
\begin{landscape}
\pagestyle{empty}
\begin{longtable}{rrrrrrrrrrrrrrrrr}
\caption{Ground-state properties.}  \\

\hline \hline
$A$ & $N$ & $E_{\mathrm{b}}^{\mathrm{cal}}$ & $E_{\mathrm{b}+\mathrm{rot}}^{\mathrm{cal}}$ & $E_{\mathrm{b}}^{\mathrm{exp}}$ & $S_{2n}$ & $S_{2p}$ & $R_n$ & $R_p$ & $R_m$ & $R_{\mathrm{ch}}^{\mathrm{cal}}$ & $R_{\mathrm{ch}}^{\mathrm{exp}}$ & $\beta_{2n}$ & $\beta_{2p}$ & $\beta_2$ & $\lambda_n$ & $\lambda_p$ \\
 &  & (MeV) & (MeV) & (MeV) & (MeV) & (MeV) & (fm) & (fm) & (fm) & (fm) & (fm) &  &  &  & (MeV) & (MeV) \\
\hline \endfirsthead

\caption{(continued)} \\
\hline
$A$ & $N$ & $E_{\mathrm{b}}^{\mathrm{cal}}$ & $E_{\mathrm{b}+\mathrm{rot}}^{\mathrm{cal}}$ & $E_{\mathrm{b}}^{\mathrm{exp}}$ & $S_{2n}$ & $S_{2p}$ & $R_n$ & $R_p$ & $R_m$ & $R_{\mathrm{ch}}^{\mathrm{cal}}$ & $R_{\mathrm{ch}}^{\mathrm{exp}}$ & $\beta_{2n}$ & $\beta_{2p}$ & $\beta_2$ & $\lambda_n$ & $\lambda_p$ \\
 &  & (MeV) & (MeV) & (MeV) & (MeV) & (MeV) & (fm) & (fm) & (fm) & (fm) & (fm) &  &  &  & (MeV) & (MeV) \\
\hline \endhead

\multicolumn{16}{r}{(Continued on next page)} \\
\endfoot

\multicolumn{16}{r}{(End)} \\
\endlastfoot

\multicolumn{8}{l}{$Z=8$ (O)}     \\ 
12 & 4 & 59.70 & 59.70 & 58.58 &  & \underline{-2.37} & 2.329 & 2.915 & 2.734 & 3.022 &  & 0.000 & 0.000 & 0.000 & -19.36 & -2.15  \\
14 & 6 & 101.16 & 101.16 & 98.73 & 41.46 & 1.49 & 2.250 & 2.639 & 2.480 & 2.758 &  & 0.000 & 0.000 & 0.000 & -16.55 & -1.18  \\
16 & 8 & 127.28 & 127.28 & 127.62 & 26.11 & 20.45 & 2.626 & 2.650 & 2.638 & 2.768 & 2.699 & 0.000 & 0.000 & 0.000 & -11.42 & -7.78  \\
18 & 10 & 140.97 & 140.97 & 139.81 & 13.69 & 27.18 & 2.806 & 2.642 & 2.734 & 2.760 & 2.773 & 0.000 & 0.000 & 0.000 & -6.67 & -11.63  \\
20 & 12 & 152.43 & 152.43 & 151.37 & 11.47 & 33.79 & 2.960 & 2.645 & 2.838 & 2.763 &  & 0.000 & 0.000 & 0.000 & -5.70 & -15.15  \\
22 & 14 & 162.45 & 162.45 & 162.03 & 10.02 & 40.10 & 3.105 & 2.653 & 2.949 & 2.771 &  & 0.000 & 0.000 & 0.000 & -4.88 & -18.31  \\
24 & 16 & 171.07 & 171.07 & 168.95 & 8.62 & 45.83 & 3.267 & 2.659 & 3.078 & 2.777 &  & 0.000 & 0.000 & 0.000 & -3.33 & -20.24  \\
\end{longtable}
\end{landscape}

%

\end{document}